\RequirePackage{fix-cm}
\documentclass[natbib]{svjour3}         
\usepackage{floatrow}

\smartqed  

\usepackage[normalem]{ulem}

\newcount\listnorom
\newcommand\listromanDE{\global\advance \listnorom by 1
{\lowercase\expandafter{(\romannumeral\listnorom)}\ }}

\newcount\listnumber
\newcommand\listDE{\global\advance \listnumber by 1
{\lowercase\expandafter{(\number\listnumber)}\ }}

\newcount\Inum
\newcount\IInum
\newcount\IIInum
\newcount\IVnum
\def\I{\global\multiply\IInum by 0 \global\multiply\IIInum by 0
            \global\multiply\IVnum by 0 \global\advance \Inum by 1
            {\the\Inum. }}
\def\II{\global\multiply\IIInum by 0\global\multiply\IVnum by 0
       \global\advance \IInum by 1 {\the\Inum.\the\IInum. }}
\def\III{\global\multiply\IVnum by 0\global\advance \IIInum by 1
            {\the\Inum.\the\IInum.\the\IIInum. }}
\def\IV{\global\advance \IVnum by 1
            {\the\IVnum. }}
\usepackage{graphicx}
\usepackage{epsf}
\usepackage{amssymb}
\usepackage{amsmath}
\usepackage{mathptmx}
\usepackage{placeins}
\usepackage{color}
\usepackage{multirow}
 \journalname{SSRv}
\newcommand{\Msun}{\mbox{$M_{\odot}\;$}}

\newcommand{\kmps}{km s$^{-1}$}

\newcommand{\xx}[1]{\!\times\!10^{#1}}

\newcommand{\be}{\begin{equation}}
\newcommand{\ee}{\end{equation}}
\newcommand{\beq}{\begin{eqnarray}}
\newcommand{\eeq}{\end{eqnarray}}
\newcommand\subsun[1]{{$_{\normalsize\odot}$}}
\newcommand{\ergs}{erg~s$^{-1}$}   
\newcommand{\kms}{~km~s$^{-1}$}

\definecolor{pink}{rgb}{0.91, 0.67, 0.81}
\definecolor{violet}{rgb}{0.93, 0.51, 0.93}

\def\asr{Adv. Space Res.}%
\def\aj{AJ}%
\def\aj{Astron.~J.}%
\def\arnps{Annu.~Rev.~Nucl.~Part.~Sci.}%
\def\araa{Annu.~Rev.~Astron.~Astrophys.}%
\def\apj{Astrophys.~J.}%
\def\apjl{Astrophys. J. Lett.}%
\def\apjs{Astrophys.~J.~Suppl.}%
\def\aspp{Astropart.~Phys.}
\def\ao{Appl.~Opt.}%
\def\apss{Astrophys. Space Sci.}%
\def\aap{Astron.~Astrophys.}%
\def\aapr{ Astron.~Astrophys.~ Rev.}%
          \def\jetp {Sov. Phys. JETP}
\def\jcap{J. Cosmol. Astropart. Phys.}%
\def\mnras{Mon. Not. R. Astron. Soc.}%
\def\prd{Phys.~Rev.~D}%
\def\prl{Phys.~Rev.~Lett.}%
\def\ssr{Space~Sci.~Rev.}%
\def\nat{Nature}%
\def\memsai{Mem.~Soc.~Astron.~Italiana}%
\def\procspie{Proc.~SPIE}%
\def \rpp {Rep.~Prog.~Phys.}
\def\lsim{\;\raise0.3ex\hbox{$<$\kern-0.75em\raise-1.1ex\hbox{$\sim$}}\;}
\def\gsim{\;\raise0.3ex\hbox{$>$\kern-0.75em\raise-1.1ex\hbox{$\sim$}}\;}

\def\lsim{\;\raise0.3ex\hbox{$<$\kern-0.75em\raise-1.1ex\hbox{$\sim$}}\;}
\def\gsim{\;\raise0.3ex\hbox{$>$\kern-0.75em\raise-1.1ex\hbox{$\sim$}}\;}

\def\kms{\rm ~km~s^{-1}}

\def \kms {\rm ~km~s^{-1}}

\def\lsim{\;\raise0.3ex\hbox{$<$\kern-0.75em\raise-1.1ex\hbox{$\sim$}}\;}
\def\gsim{\;\raise0.3ex\hbox{$>$\kern-0.75em\raise-1.1ex\hbox{$\sim$}}\;}

\definecolor{purple}{rgb}{0.63, 0.36, 0.94}

\def\kms{\rm ~km~s^{-1}}

\begin{document}
\title{High-energy particles and radiation in star-forming regions
}
\titlerunning{}

\author{Andrei~M.~Bykov \and Alexandre~Marcowith \and Elena~Amato \and Maria~E.~Kalyashova \and J.~M.~Diederik~Kruijssen   \and Eli~Waxman} 

\authorrunning{A.~M.~Bykov et al.} 

\institute{A.~M.~Bykov \at
           Ioffe Institute, 194021, St. Petersburg, Russia;\\ 
           \email{byk@astro.ioffe.ru} \and A.~Marcowith  \at Laboratoire Univers et Particules de Montpellier CNRS/Universit\'e de Montpellier,
Place E. Bataillon, 34095 Montpellier, France \and  E.~Amato \at INAF - Osservatorio Astrofisico di Arcetri, Largo E. Fermi, 5, 50125, Firenze, Italy; \email{amato@arcetri.astro.it}  \and M.~E.~Kalyashova \at
Ioffe Institute, 194021, St. Petersburg, Russia \and J.~M.~D.~Kruijssen \at
          Astronomisches Rechen-Institut, Zentrum f\"{u}r Astronomie der Universit\"{a}t Heidelberg, M\"onchhofstr. 12-14, 69120 Heidelberg, Germany; \email{kruijssen@uni-heidelberg.de}  \and E.~Waxman \at Weizmann Institute of Science, 76100, Rehovot, Israel}

\date{Received: \dots / Accepted: \dots}

\maketitle

\begin{abstract}
Non-thermal particles and high-energy radiation can play a role in the dynamical processes in star-forming regions and provide an important piece of the multiwavelength observational picture of their structure and components.      
Powerful stellar winds and supernovae in compact clusters of massive stars and 
OB associations are known to be  favourable sites of high-energy particle acceleration 
and sources of non-thermal radiation and neutrinos. 
Namely, young massive stellar clusters are likely sources of the PeV (petaelectronvolt) regime cosmic rays (CRs). They can also be responsible for the cosmic ray composition, e.g., $^{22}$Ne/$^{20}$Ne anomalous isotopic ratio in CRs.
Efficient particle acceleration can be accompanied by super-adiabatic amplification of the fluctuating magnetic fields in the systems converting a part of kinetic power of the winds and supernovae into the magnetic energy through the CR-driven instabilities. 
The escape and CR propagation in the vicinity of the sources are affected by the non-linear CR feedback. These effects are expected to be important in starburst galaxies, which produce high-energy neutrinos and gamma-rays.
We give a brief review of the theoretical models and observational data
on high-energy particle acceleration 
and their radiation in star-forming regions with young stellar population.   

 \keywords{Star-forming regions \and Cosmic rays \and OB associations \and Young massive star clusters \and Starburst galaxies }
%
\end{abstract}


\section{Introduction} \label{section:intro}
Star formation is the key phenomenon in the galactic environment connecting together all 
of the important constituents from molecular gas to magnetic fields and cosmic rays 
in a close
relationship. Multiwavelength observations of star-forming regions 
both in the Milky Way and in starburst galaxies revealed broadband non-thermal radiation 
from radio to gamma rays, indicating the presence of relativistic particles and their interactions with matter, radiation and magnetic fields 
\citep[see e.g.][]{Condon1992ARAA,2009Sci...326.1080A,2012ApJ...755..164A,NGC253_HESS_Fermi18}. Cosmic rays (CRs) are expected to be efficiently produced by strong fast winds of young massive stars and core-collapse supernovae in star-forming regions (SFRs) \citep[see e.g.][]{Bykov2014,Lingenfelter2018,2018ARNPS..68..377T}.

Low-energy CRs accelerated in these sources can penetrate deep into molecular clouds providing ionization and heating \citep[see for details][]{2020SSRv..216...29P}. Self-regulated propagation of CRs escaping from the sources may form CR halos or drive the winds in the vicinity of star-forming regions \citep[see e.g.][]{ipavich1975,1991A&A...245...79B,2017ApJ...834..208R,2018ApJ...856..112F,2019A&A...626A..85O}. 
 The CR pressure gradients around the compact clusters or groups of young massive stars with ages of about 5-10 Myrs can affect then the star formation in 50-100 pc vicinity of the CR source. The analysis by \citet{2018A&A...612A..50B} of the ages of OB stars in Cygnus X, derived from isochrones, revealed a spatial clustering of the groups of different ages: young (0–5 Myr), intermediate (5–10 Myr), and old ($\geq  10$ Myr). It may suggest that the massive star formation has proceeded in the Cygnus region. However, the distances derived with Gaia DR2 \citep{2019MNRAS.484.1838B} indicated that  Cygnus OB2 group is farther away than previously estimated and it may be spatially more separated from the foreground group in the line of sight. 
 Current models are zooming into the star-forming galaxies with high-resolution cosmological simulations accounting for mechanical, radiative, magnetic and cosmic-ray feedback from massive stars \citep[see e.g.][]{2019FrASS...6....7K,2019MNRAS.tmp.2993H,2019MNRAS.490.1271H,2019A&A...626A..85O}.

Radiation produced by the accelerated particles in star-forming regions provides important information on the distribution of matter and magnetic fields. At the same time, the increasing sensitivity of high-energy neutrino telescopes and upcoming gamma-ray facilities is opening a new window on the processes of cosmic ray interactions in star-forming galaxies \citep{LW06,2014PhRvD..90d3005A,2017nacs.book...33W,IceCube7years}. 
We review below some recent observations of star-forming regions and the future observational perspective as well as specific physical mechanisms of particle acceleration in these objects.

The plan of the paper is as follows.
In \S~2, we give an overview of the sources of high-energy particles and radiation in star-forming regions and starburst galaxies. Particle acceleration in OB associations and compact clusters of young massive stars can be the reason of high-energy radiation from these objects. The high and ultra-high energy CRs accelerated in the starburst galaxies are considered as the important sources of the energetic neutrinos observed by the {\sl IceCube Observatory}.   
In \S~3, we present time-dependent models of relativistic particle acceleration and CR spectra formation by large-scale magnetohydrodynamic (MHD) plasma motions with multiple shocks produced by fast winds of massive stars and supernovae in SFRs. Accelerated cosmic rays escaping the accelerators may carry away a substantial fraction of the mechanical power luminosity. Therefore, the non-linear effects of CR-induced turbulence may affect the CR propagation near the sources. The non-linear CR propagation models are discussed in \S~4. Young massive star clusters may significantly contribute into the observed spectrum of galactic CRs above PeV energy regime and can be responsible for some observed composition anomalies. We discuss in \S~5 the enhanced ratio of $^{22}$Ne/$^{20}$Ne isotopes which is observed in low-energy CRs and the expected anisotropy of 100 PeV protons accelerated in compact clusters of young massive stars.
The observational perspective of multiwavelength observations of the star-forming regions and starburst galaxies is discussed in \S~6. 

%

\section{Overview of the sources of high-energy particles and radiation associated with star-forming regions}
\label{section:overview}

Clustered star formation as a result of the evolution of massive molecular clouds has profound effect on the evolution of the interstellar medium \citep[e.g.][]{1988ApJ...324..776M,1990ApJ...354..483H,2005ARA&A..43..337C} and turbulent magnetic fields \citep[e.g.][]{2008ApJ...680..362H,2019Galax...7...45S} as well as on the production and evolution of high-energy cosmic rays \citep[e.g.][]{Bykov2014,Lingenfelter2018}.  

\citet{2007IAUS..237..106O} used a
definition of a superbubble (SB) which was introduced by You-Hua Chu in the 1980s as a shell structure
originating from multiple stars, rather than a definition related to the system size. This is certainly well-motivated given the vast differences in the shell sizes \citep[e.g.][]{1997MNRAS.289..570O}. On the other hand in the context of CR acceleration sources we have to distinguish the superbubbles produced by OB associations from the compact young massive star clusters (YMSCs)  where the winds of early-type stars are collected in a very compact core of a pc size.

\subsection{Cosmic rays and gamma-ray emission in OB associations}
\label{section:OBstars}
%


%

%
%


  {Most massive stars in Milky Way-like galaxies formed in loose OB associations \citep[see below and e.g.][]{DK12,ward19}. Supernovae and powerful stellar winds in OB associations are the likely sources of CRs \citep[e.g.,][]{cm83,Bykov2001,Bykov2014, Parizot2004, Binns2005,2015ARA&A..53..199G,Lingenfelter2018,2018SSRv..214...41B,Aharonian2019, 2019ApJS..245...30L}.

  An excess of hard-spectrum TeV gamma-ray emission was revealed by the atmospheric Cherenkov telescope HEGRA \citep{AharonianEtal2002} from the field located in the direction of the Cygnus OB2. An explanation proposed for this excess was that the observed gamma-ray emission could be connected with TeV particles, accelerated on multiple shocks from SNe and massive stars' winds in the Cyg OB2 association, which is considered to be one of the richest OB associations in our Galaxy, containing hundreds of massive stars.}


 {Cygnus OB2 is embedded within the wider region of star formation, Cygnus X, which is one of the brightest sources at radio wavelengths. The region is $\sim 200$ pc size and is $\sim 1.4$ kpc far away from the solar system.
}
{
In the recent {\sl Fermi Large Area Telescope} survey of the star-forming region Cygnus X, an extended gamma-ray source was discovered, now known as the Cygnus cocoon. \citet{AckermannSB2011} associated the source with a superbubble filled with accelerated particles. The hard emission from the source extends to a $\sim 50$\,pc wide region and has a flux of $(5.8 \pm 0.9)\times 10^{-8}$ ph cm$^{-2}$ s$^{-1}$ in the
$1-100$\,GeV range. At the estimated distance of the spatially coincident superbubble Cyg OB2 this corresponds to a gamma-ray luminosity $L_{\gamma}=(9 \pm 2)\times 10^{34}$ erg s$^{-1}$, which is less than 1 \% of the total kinetic power of Cyg OB2 massive stars' winds.}

{There is evidence that Cygnus OB2 was formed out of a giant molecular complex --- the progenitor of the present-day Cygnus X. Based on red supergiants observations in near infrared, \citet{ComeronEtal2016} investigated that the star formation in the Cygnus OB2/Cygnus X region started at least 20 Myrs ago, which is well before the last star-forming burst that caused, $\sim 3$ Myrs ago, the appearance of the compact group of early-type stars, now determining the behavior of the OB association. 
The mechanical power from SN explosions depends on the age of an OB association. \citet{Martinea10} reproduced stellar population of Cyg OB2 and gave an estimate of its kinetic luminosity over the 3 Myr of its life: $\sim 4\xx{38}$\,\ergs.
}

 {More recently, \citet{Katsuta2017} reported about a study of extended gamma-ray emission from the star-forming region G25.0+0.0 detected by {\sl Fermi}-LAT, the possible second case of a gamma-ray detection from a SFR. The hard GeV emission from the G25 region is estimated to have about 10 times larger gamma-ray luminosity than the Cygnus cocoon, which means that the efficiencies of particle acceleration in this source should also be much higher. A problem is that at the estimated distance of about 8 kpcs G25.0+0.0 is in a highly obscured region, which makes comprehensive studies of its stellar content very difficult.} 

{The spectra of extended gamma-ray emission of both the Cygnus cocoon and G25.0+0.0 SFR in the GeV range are best described with a power law with the same photon index of $\sim 2.1$. }

%
%
%

%
OB associations can be formed on timescales $\gsim 10$ Myr with the numerous assembled SNe \citep[e.g.][]{efremov98,kawamura09,longmore14,kruijssen19,chevance20}. The energy and momentum injection from stellar groups and subgroups within the association were studied in the framework of population synthesis with stellar evolution models \citep{2009A&A...504..531V,2010A&A...520A..51V}. They emphasized that the presence of groups/subgroups of different ages makes it possible to keep at relatively high level the stellar wind contribution into the mechanical luminosity of the system over $\sim 10$ Myr time intervals. The emergence of a superbubble with an internal structure consisting of shells and filaments was demonstrated by \citet{2013A&A...550A..49K,2014A&A...566A..94K}. They followed the evolutionary tracks of three massive stars produced by \citet{2005A&A...429..581M} and used 3D-hydrodynamical simulations with account of radiative cooling and photo-electric heating of the optically thin gas. In the hydrodynamical model the energy injected into the superbubble by supernovae has completely dissipated in $\sim$ 1 Myr time interval. The feedback  of magnetic fields, possibly amplified in superbubbles by turbulent dynamo and CR-driven instabilities, still needs to be studied, as well as the effect of CRs, which are likely evacuating away a fraction of the released mechanical luminosity, on the superbubble dynamics. 

Massive stars and supernovae, which form superbubbles in galaxies, produce unstable isotopes such as $^{60}$Fe and $^{26}$Al, which are gamma-ray line emitters at the characteristic energies of 1.173, 1.332, and 1.809 MeV. The sensitivity of the gamma-ray spectrometer {SPI INTEGRAL} has allowed the significant detection of the combined two lines of $^{60}$Fe: the morphology of the emission suggests that its origin is likely diffuse, rather than resulting from the combination of a few point sources  \citep{2019arXiv191207874W}. The 1.809 MeV line from the decay of $^{26}$Al was detected from the Cygnus region centered on the position of the Cyg OB2, while for the $^{60}$Fe lines from the same region only an upper limit was derived  \citep{2009A&A...506..703M}. The width of the detected $^{26}$Al line implies turbulent velocities below 200 $\kms$ for the line emitting material.  

On a smaller scale, 3D-hydrodynamical simulations were performed for a compact colliding-wind binary systems \citep[see e.g.][]{2010MNRAS.403.1657P}. These systems are known to be sources of non-thermal radiation  \citep[see e.g.][]{2013A&A...558A..28D,2019RLSFN.tmp....3R}. Different aspects of diffusive shock acceleration of relativistic particles in colliding winds of massive stars were discussed in \citep{2019ApJ...871...55G,2019arXiv191205299P} while the acceleration of very high energy CRs by a supernova shock colliding with a fast wind produced by a massive star, or a compact cluster, was studied by \citet{BEGO2015MNRAS}.  

{Recently {\sl High Altitude Water Cherenkov Observatory} (HAWC)  collaboration presented preliminary results of TeV gamma-ray emission from the Cygnus cocoon \citep{HAWC_Cygnus_19}. They reported the existence of a significant TeV counterpart of the cocoon region, described by a power law spectrum with index $\sim 2.6$. At GeV energies they obtained a harder spectrum, with an index of $\sim 2.1$, which well agrees with the findings of earlier spectral studies of the Cygnus cocoon in the GeV range. These authors suggested a hadronic origin of the observed radiation which points to CR acceleration in the Cygnus cocoon to, at least, few hundred TeV. From an energy budget calculation, the CR acceleration efficiency in the Cygnus OB2 is found to be of $\sim 0.1 \%$.} 

{Deeper studies of the Cygnus OB2 are expected in the near future. Nuclear interactions of CRs accelerated in Cygnus cocoon may lead to high-energy neutrino production in Cygnus X. \citet{Yoast-Hull2017} calculated that this neutrino flux can be large enough to be observable with the {\sl IceCube Observatory}. Apart from { HAWC} recent study, sensitive and high-resolution observations are planned with {\sl Cherenkov Telescope
Array} in the energy band from tens of GeV to hundreds of TeV. At the lower energies  it is expected that the next generation MeV-GeV instruments \citep{GRIPS12,ASTROGAM18}  will provide a comprehensive view of Cygnus OB2 and Cygnus X.}
\\
\\
\\

\subsection{Massive star clusters as sources of cosmic rays and gamma-ray emission}

\subsubsection{CR acceleration in young massive star clusters
}\label{section:CRYMSC}

Similar to superbubbles, young massive star clusters contain large populations of massive stars and hence a high SN explosion rate. In galaxies like the Milky Way, the total cluster mass can reach $10^5 \Msun$ \citep[e.g.][]{adamo15,reinacampos17}, but, unlike OB associations, all the stars are concentrated in $\sim 1~\mathrm{pc}$ radius with the star density in a cluster  $\gsim 10^{3} M_{\odot}~\mathrm{pc}^{-3}$, thus, YMSCs represent prominent and uniquely compact sites of massive star formation. YMSCs in the Galaxy are observed in almost all energy bands. Some of them are found very close to the solar system \citep{Kuhn, Getman2018}.  

Spectroscopic data show that O-, B- and Wolf-Rayet stars have powerful winds, with velocities of $(1-3)\times 10^3$ \kmps. The total kinetic power of OB winds in a cluster is typically $\sim 5 \times 10^{38}$\,\ergs.
Colliding shocks from massive star winds can be the site of effective particle acceleration up to TeV energies. Moreover, strong cluster winds can be formed in active star formation regions like YMSCs (see \citealt{chev_clegg85} for the modeling of starburst galaxies and nuclei). 


Supernovae in massive star clusters are likely to be powerful CR accelerators: acceleration efficiency and maximum energy are expected to be enhanced with respect to the standard values derived for SN remnant shocks, thanks to the fact that here the SN blast wave may interact with fast OB star winds.
Indeed it was shown by \cite{Bykov2001} 
that multiple shocks can accelerate CRs with efficiencies up to 30 \%. Particle acceleration in SN shocks colliding with a fast wind was studied quantitatively by \citet{Bykov2014,BEGO2015MNRAS}  with the help of non-linear modeling. That calculation showed that the energies of protons, accelerated in a system of SN-wind shocks, can reach hundreds of PeV, which far exceeds the expected maximum energy of Fermi acceleration in an isolated supernova remnant (SNR). 
The particle spectrum in such a system may be very hard, which can increase the efficiency of high-energy neutrino production and contribute to the neutrino events detected by the {\sl IceCube Observatory}. 

%
%
Thus, YMSCs are potential sources of galactic CRs at all energies, except for UHECRs (particles with energies in excess of $10^{18}$ eV) which are considered to be of extragalactic origin. There are a number of OB associations and clusters of massive stars of ages in the range of a few millions of years in the vicinity of the solar system  \citep[see e.g.][]{2012A&A...547A..97A,2015A&A...584A..26B,2015ApJ...808...23H}: their contribution can affect both the spectrum and composition of CRs observed at the Earth.

The distribution of massive star formation over YMSCs and OB associations is set by the `cluster formation efficiency' $\Gamma$, which is defined as the fraction of stars that form in bound clusters and varies strongly with the host galaxy properties \citep[see e.g.][and references therein]{DK12,2018ASSL..424...91A}. Galaxies with higher gas pressures (manifesting itself in terms of a higher star formation rate surface density) have larger $\Gamma$. Moreover, the cluster formation efficiency can vary by up to a factor of 3--4 within an individual galaxy, depending on its radial gas surface density profile. Therefore, YMSCs are important sources of CRs in starburst galaxies, where $\Gamma\sim50\%$ can be reached. Such starburst galaxies have been predicted to be important sources of high-energy neutrinos \citep{LW06,2014JCAP...09..043T}.    


\subsubsection{ Gamma rays from massive star clusters}
\label{section:GRYMSC}

Gamma-ray emission from YMSCs is considered to be the result of cluster kinetic energy conversion to CRs. Diffuse GeV gamma-ray emission has been detected by {\sl Fermi}-LAT from the compact clusters NGC3603, Westerlund 2, along with Cygnus OB2, while the H.E.S.S. Collaboration discovered TeV gamma-ray emission near one of the most massive star clusters in the Galaxy -- Westerlund 1. On the basis of {\sl Fermi}-LAT and H.E.S.S. data \cite{Aharonian2019,AharonianLincei2019} investigated the spectral and spatial distribution of CRs in the vicinity of Westerlund 1, Westerlund 2 and Cygnus OB2, very powerful clusters with luminosities above 10$^{38}$ \ergs. \cite{Aharonian2019} reported that in these clusters the CR density declines as $r^{-1}$, which indicates that relativistic particles are continuously injected into the interstellar medium (ISM) during a long ($\gtrsim10^6$ yr) period of time. The authors pointed out that this is an argument in favour of CR acceleration by stellar winds of massive stars, because maintaining the observed continuous particle acceleration by SNRs would require a very high rate of SNe in a cluster ($\sim 1$ SN per $10^3$ yr), which is not very realistic.

Notice, however, that the earlier census of a possible GeV emission from another sample of YMSCs did not lead to any significant gamma-ray signal \citep{Maurin2016}. In that paper the authors analysed the {\sl Fermi} telescope observations of YMSCs associated with the Rosette and Orion Nebulae (of luminosities below 10$^{37}$ \ergs), NGC2175,  NGC2467 and some others. The conclusion is that less than 10\% of the stellar wind luminosity is supplied to the relativistic particles in these objects. Some clusters even show acceleration efficiency of less than 1\%. The {\sl Cherenkov Telescope Array} may be able to clarify the situation and to constrain the capability of a given YMSC to be a site of gamma-ray emission.

\subsection{High-energy phenomena in the Galactic Center and the Central Molecular Zone}
\label{section:CMZ}

The Central Molecular Zone (CMZ) of the Milky Way is the region surrounding the Galactic Centre (GC), extending a few $\times$ 100~pc away from Sgr~A$^\star$. The CMZ is an extreme environment \citep{morris96}, exhibiting e.g.\ gas pressures, densities, velocity dispersions, temperatures, cosmic ray ionisation rates, and magnetic field strengths that are factors of several to orders of magnitude higher than in the galactic disc \citep[e.g.][]{crocker10,ao13,ginsburg16,henshaw16,krieger17,kauffmann17,yusefzadeh19}. These conditions are similar to those seen in vigorously star-forming galaxies at high redshift \citep{kruijssen13}. The CMZ hosts 90~percent of the dense molecular gas in the Milky Way, but contributes only a few percent of the galactic star formation rate \citep[e.g.][]{longmore13,barnes17}. The star formation history of the central 150 pc of the Milky Way is likely not quasi-continuous but has a number of intense star formation periods with the latest one  at about several tens of millions of years ago \citep[see e.g.][]{2019NatAs.tmp....4N}.  

Importantly in the context of this review, the CMZ is extremely bright in gamma rays \citep[e.g.][]{aharonian06,su10,2018Galax...6...55L}. Analysing 10 years of H.E.S.S. data obtained with arcminute angular resolution from the region surrounding the GC, \citet{HESS2016} found the presence of the extended very high energy gamma-ray emission implying the CR profile which is peaked towards the GC and compatible with an $R^{-1}$ law. This can be expected from a stationary point-like source at the GC injecting accelerated CRs. The authors concluded that the apparent lack of the spectral cut-off in the  H.E.S.S. data likely indicates the acceleration of the PeV protons within the central 10~pc of the Milky Way. A scenario of PeV CRs acceleration in the vicinity of massive black hole in the GC was discussed by \citet{Aharonian2005} \citep[see also] []{GuoEtal2017}.

Another possible class of the sources of CRs with energies above PeV can be associated with YMSCs like Arches and Quintuplet \citep{Bykov2014}. Together with the Young Nuclear Cluster of age 6$\pm$2~Myr, very close to Sgr A$^\star$ \citep{2018ASSL..424...69L}, these three young massive star clusters have masses of about 10$^4$~\Msun and reside within $\sim30$~pc from the GC in projection (although the Arches and Quintuplet are likely displaced along the line of sight, such that their total galactocentric radius is $\sim 60$~pc, see \citealt{stolte14} and \citealt{kruijssen15}). The Young Nuclear Cluster contains more than 100 hot stars including 23 WR-type stars. The scenario that high-energy CR acceleration takes place in the YMSCs located near the GC is further supported by recent detection of gamma-ray emission from YMSCs Westerlund 1 and 2 \citep{Aharonian2019}. Recently, \citet{2018A&A...612A...9H} tested the various contributions to the total gamma-ray emission detected by  H.E.S.S. They showed that the emission correlated with the dense gas phases covers the full extent of the CMZ. They suggested that it is a single diffuse component and its flux is about half of the total diffuse emission flux from the region.

Explanations for the high-energy CRs and gamma-ray emission in the GC and CMZ are varied \citep[e.g.][]{su10,hooper11b,abazajian14} and range from supernovae \citep[e.g.][]{2017MNRAS.467.4622J} to activity from Sgr~A$^\star$ \citep[e.g.][]{chernyakova11}, pulsars \citep[e.g.][]{brandt15,bartels16}, and dark matter annihilation \citep[e.g.][]{hooper11}. Distinguishing between these scenarios requires detailed, 7-dimensional maps (position, velocity, and time) of the molecular gas distribution, as well as sites of massive star formation and supernovae \citep{2018Galax...6...55L}. Only with a detailed model for the energy injection from astrophysical sources it is possible to model how each of the above mechanisms manifests itself in observable tracers \citep[e.g.][]{crocker12}. Major recent efforts have made important steps in mapping the CMZ's structure \citep[e.g.][]{molinari11,kruijssen15,henshaw16}, mass flows \citep[e.g.][]{sormani18,sormani19}, energy cycle and balance \citep[e.g.][]{crocker12,kruijssen14} and star formation history \citep[e.g.][]{barnes17,krumholz17}, but this remains an area of active research. We refer the reader to \citet{2018Galax...6...55L} for further details.

\subsection{High-energy phenomena in starburst galaxies}
The most significant point in the Northern hemisphere from the {\sl IceCube} scanning of the
sky is coincident with the Seyfert II galaxy NGC 1068 \citep{2019arXiv191008488I}.  
Superwinds of starburst galaxies were suggested as possible sites of high-energy CRs acceleration. Ultra-high-energy cosmic rays (UHECR) acceleration up to 10$^{11}$ GeV by superwinds of starburst galaxies was discussed by \citet{PhysRevD.97.063010}, while more conservative maximum energies (10$^{7}$ GeV for protons) were obtained in the model by  
\citet{2019arXiv191207969R}. The latter are consistent with the CR energies accelerated in YMSCs \citep{Bykov2014}, which are expected to be numerous in the starburst galaxies. \citet{NGC253_20} discussed a possible role of a massive black hole with a relatively low accretion rate as UHECR accelerator in the nucleus of starburst galaxy NGC 253.   The origin of UHECR is possibly associated with the relativistic shocks in the outflows of compact objects \citep[see e.g.][]{2018NPPP..297..267L} rather than with the superwinds of the starburst galaxies, while further studies are needed.  
We discuss in brief in \S\ref{CSFS} results of modeling of high-energy CR acceleration by supernovae exploding in the vicinity of a fast wind. The spectrum of CR protons accelerated at the colliding shock flows produced by a supernova shock at the free expansion stage and a fast wind may extend to 100 PeV in the starburst galaxy environment with high magnetic field. This spectrum has a specific shape with an upturn at TeV-PeV regime as it is shown in Fig.~\ref{CSF_Sp}. This increases the energetic efficiency of the production of the high-energy neutrinos in such systems.

Recently, \citet{PhysRevLett.124.101102} performed the studies of cross-correlation between gamma rays and the mass distribution probed by weak gravitational lensing. They detected the cross-correlation signal with hard energy spectrum of best-fit spectral index $\alpha=1.81^{+0.20}_{-0.24}$. We assume that a population of star-forming galaxies where CR acceleration is dominated by supernovae in clusters of young massive stars may provide hard enough spectra of gamma-ray radiation and thus may help to understand the obtained results.

The isotropy and flavor content of the {\sl IceCube} detected neutrinos, and the coincidence, within current uncertainties, of the 50~TeV to 2~PeV flux and the spectrum with the Waxman-Bahcall bound, suggest a cosmological origin of the neutrinos, related to the sources of UHECR with energies $>10^{10}$~GeV. The most natural explanation of the UHECR and neutrino signals is that both are produced by the same population of cosmological sources, producing CRs (likely protons) at a similar rate, $E^2d\dot{n}/dE\propto E^{0}$, over the [$10^{6}$~GeV,$10^{11}$~GeV] energy range, and residing in "calorimetric" environments, like galaxies with high star formation rate, in which $E/Z<100$~PeV CRs lose much of their energy to pion production \citep[see discussion in][]{2017nacs.book...33W}. 

In the recent work, \citet{Peretti2020} argued that starburst galaxies nuclei can account for the diffuse neutrino flux above 200 TeV,  while producing  $\lsim$ 40\% of the extragalactic diffuse gamma-ray background. They find that the neutrino flux below 200 TeV from the starburst population is expected to be lower than the observed one.  

A tenfold increase in the effective mass of the {\sl IceCube} detector at $\gtrsim100$~TeV is required in order to significantly improve the accuracy of current measurements, to enable the detection of a few bright nearby starburst "calorimeters", and to open the possibility of identifying the CR sources embedded within the calorimeters, by associating neutrinos with photons accompanying transient events responsible for their generation. Source identification and a large neutrino sample may enable one to use astrophysical neutrinos to constrain new physics models.

%


\section{Models of particle acceleration by large-scale magnetohydrodynamic turbulence with multiple shocks in star-forming regions}

A kinetic energy release from the fast winds of young massive stars and multiple supernovae within the bubble created by a stellar
association or compact stellar cluster may exceed  $10^{38}$ erg s$^{-1}$. 
The kinetic energy is released in the form of the supersonic and the super-Alfv\'enic 
flows accompanied by shocks. The strong primary shocks interacting with density inhomogeneities of different scales produce multiple weak secondary shocks and   
large-scale flows and the broad spectra of magnetohydrodynamic fluctuations with frozen-in magnetic fields. 
Vortex electric fields induced by the large-scale motions of highly
conductive plasma result in the non-equilibrium distribution of
the charged nuclei producing  high-energy tails. Because of the presence of multiple strong shocks, particle
distribution is highly intermittent. Statistical description
of the intermittent systems is a challenging problem.
The intermittent character of the system with the multiple shocks can be accounted by the kinetic equation approach with an appropriate averaging over the multiple characteristic scales \citep{BT93}.

The distribution function
$N({\mbox{\boldmath $r$}},p,t)$
of non-thermal nuclei
averaged over an ensemble of turbulent motions and shocks
satisfies the kinetic equation
\begin{equation}\label{KE:lowE}
      \frac{\partial N}{\partial t} -
       \frac{\partial}{\partial r_{\alpha}} \: \chi_{\alpha \beta} \:
       \frac{\partial N}{\partial r_{\beta}}  =
       G1  \hat{L} N +
      \frac{1}{p^2} \: \frac{\partial}{\partial p} \: p^4 D \:
      \frac{\partial N}{\partial p} + A {\hat{L}}^2 N +
      2B \hat{L} \hat{P} N + F_{j}(p).
\end{equation}

The source term $ F_{j}(p)$ is determined by injection
of the nuclei of a type $j$.
The integro-differential operators $\hat{L}$ and $\hat{P}$ are given by

\begin{equation}
      \hat{L}= \frac{1}{3p^2} \: \frac{\partial}{\partial p} \:
      p^{3-\gamma} \: \int_{0}^{p} {\rm d}p' \: {p'}^\gamma
      \frac{\partial}{\partial p'} \;;~~~~~
      \hat{P}= \frac{p }{3} \: \frac{\partial}{\partial p} \, .
\end{equation}

The kinetic coefficients satisfy the renormalization equations:

$$
  \chi = \kappa + {1 \over 3} \int { d^3 {\bf k} \, d\omega \over (2\pi)^4 }
  \left[ {2T+S \over i\omega+k^2\chi }
        -{2k^2\chi S \over \left( i\omega + k^2\chi\right)^2 } \right] \, ,
$$

$$
  D={\chi \over 9} \int { d^3 {\bf k} \,  d\omega \over (2\pi)^4 }\;
   {k^4 S(k,\omega) \over \omega^2 + k^4 \chi^2 } \, ,
$$

$$
 A = \chi \int { d^3 {\bf k} \, d\omega \over (2\pi)^4 } \;
   {k^4 \tilde{\phi}(k,\omega) \over \omega^2 + k^4 \chi^2 }\, ,
$$

$$
  G2 = \chi \int { d^3 {\bf k}\, d\omega \over (2\pi)^4 }\;
  {k^4 \tilde{\mu}(k,\omega) \over \omega^2 + k^4 \chi^2 }\, .
$$

Here $G1 = ( 1/\tau_{sh}+ G2)$.  The kinetic coefficients $G1,\kappa,D, A$ and $G2$ are determined by 
the statistical moments of MHD turbulence  and in particular by the correlation tensor of bulk velocities of plasma $ <\mathbf{u_{\alpha}}(t',\mathbf{r'})\cdot \mathbf{u_{\beta}}(t,\mathbf{r})>$. It is convenient to express the kinetic coefficients through the Fourier  components of the correlation tensor  introducing its transverse $T(k,\omega)$ and the longitudinal  $S(k,\omega)$ components \citep[see e.g.][]{1971sfmm.book.....M} as well as the  cross correlations between velocity jumps on shock fronts
and rarefaction motions in between given by $\tilde{\phi}(k,\omega)$ and  $\tilde{\mu}(k,\omega)$. We assume here the presence of a broad range of magnetic fluctuations with the spectrum $dB_k^2/dk \propto k^{-\nu}$, where the index $1 \leq \nu \leq 2$ where the energy containing (maximum) scale $l_{\mathrm{corr}}$ is the same as the energy containing scale of the kinetic energy of plasma motions initiated by the fast winds and shocks. Apart from the locally anisotropic Kolmogorov-type cascading also the CR-driven instabilities may be important in some appropriate scales for building the MHD fluctuation spectra. It is also important to have in mind that the Landau damping of compressible modes in the hot gas may affect the spectral properties of MHD turbulence in superbubble.
With available computer capabilities it is not easy to model the properties of MHD turbulence because of the very broad dynamic range which spans of about ten decades from tens of parsecs down to the particle gyroradii scales.
Therefore, we are using here some parametrizations assuming a simplified picture of the magnetic turbulence which certainly requires dedicated studies.

 In this context we call turbulent motions the long-wavelength if their wavelengths are much larger than the gyroradii of CRs. 
The long-wavelength supersonic and super-Alfv\'enic turbulence between the shocks is characterized by the mean square velocity amplitude $u_{\star} = \sqrt{<u^2>}$  and the correlation length $l_{\mathrm{corr}}$.  There are two different regimes of CR transport and acceleration in such systems.  These depend on the relation of the characteristic correlation time of the strong long-wavelength turbulence  $\tau_{\mathrm{corr}} \sim l_{\mathrm{corr}}/u_{\star}$  and the CR diffusion time over the turbulence correlation length $\tau_{\mathrm{diff}}(p) \sim l_{\mathrm{corr}}^2/\kappa(p)$. 
If the CR diffusion over the correlation length is slow and the ratio 
$\eta(p) = \tau_{\mathrm{corr}}/\tau_{\mathrm{diff}}(p)$ is large then the CR transport is determined mainly by the turbulent advection. This regime is typical for CRs with low energies if their gyroradii are  in resonance with the magnetic fluctuations. While in the case of the large diffusion coefficients $\eta(p) <$1.   

We define the momentum  $p_{\star}$ which approximate (not exactly)  the transition between the regimes by equation from  $\eta(p_{\star}) = 1$. Then using the expression for the mean free path of CRs due to scattering by the resonant magnetic fluctuations (of scales comparable to the CR gyroradius $R_H(p)$) given by \citet{1985crim.book.....T}      
\begin{equation}
\Lambda(p) = G(\nu)\cdot l_{\mathrm{corr}}\cdot  \left [\frac{R_H(p)}{l_{\mathrm{corr}}}\right]^{2 - \nu}
\end{equation}   
one can get 
\begin{equation}
\eta(p) \approx  \frac{c}{u_{\star}} \cdot \left[\frac{R_H(p)}{l_{\mathrm{corr}}}\right]^{2-\nu}. 
\end{equation}
This results in a strong dependence of the transition momentum $p_{\star}$ on the large-scale turbulent velocity amplitude $u_{\star}$  as $p_{\star} \propto u_{\star}^{1/(2-\nu)}$ for $\nu > 3/2$.  From the definition of $p_{\star}$ one can get the following estimation for $\nu =1.7$  
\begin{equation}\label{estar}
\epsilon_{\star} \approx 0.5\, {\rm GeV}\, \left[\frac{B}{10\, \mathrm{ \mu G}}\right]\cdot \left[\frac{l_{\mathrm{corr}}}{10\, {\rm pc}}\right] \cdot \left[\frac{u_{\star}}{1,000\, {\rm km ~ s^{-1}}}\right]^{3.33} 
\end{equation}

One should note here that the Fermi-type CR acceleration mechanism described above
directly used the kinetic energy of large-scale plasma motions as the source of free energy for the accelerated CRs \citep[see for a review][]{2019PhRvD..99h3006L}. The resonant magnetic fluctuations providing CR scattering needed for the mechanism to operate have scales much shorter and the energy densities much smaller than the energy containing long-wavelength motions produced by the fast stellar winds and supernovae shocks.   

\begin{center}
\begin{figure}
\includegraphics [width=115mm] {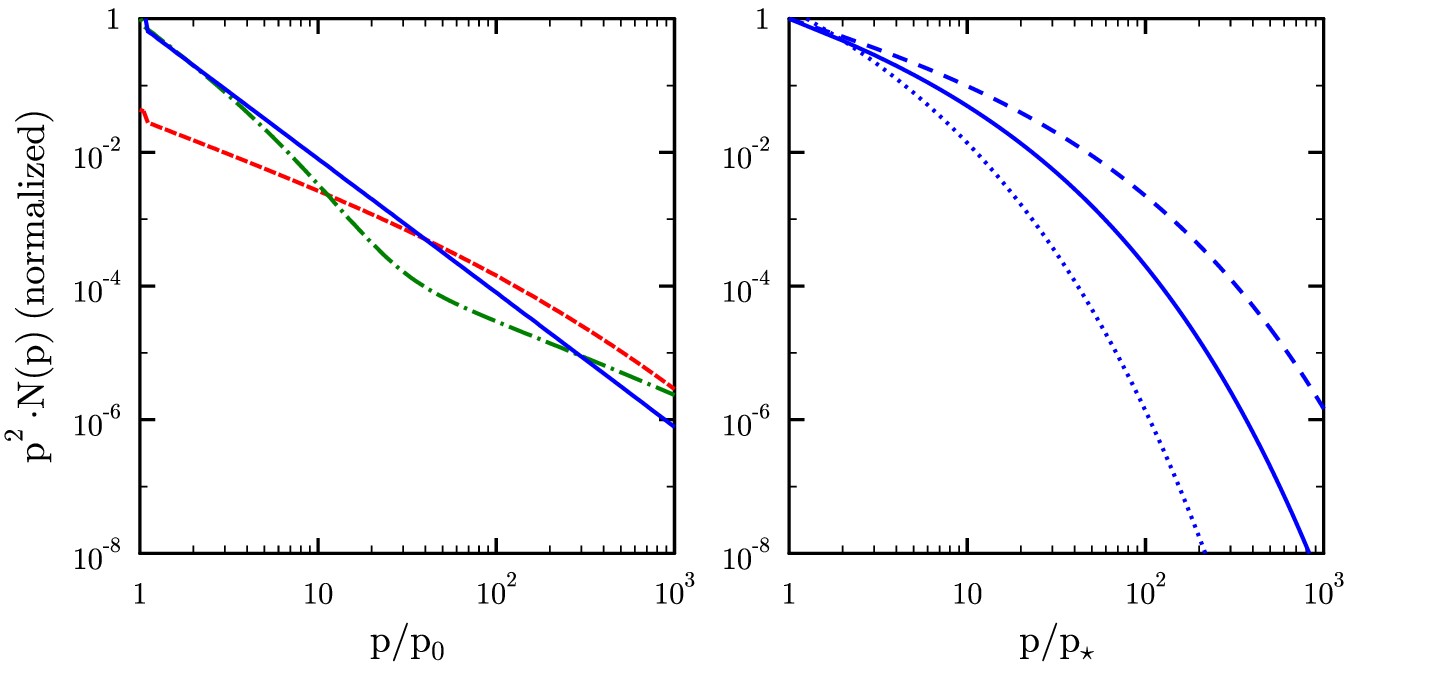}
\caption{The model spectra of CR ions accelerated by shock ensemble and large-scale MHD plasma motions in the systems with multiple winds of young massive stars and supernovae. Left panel shows temporal evolution of CR spectrum from the injection momentum $p_0$ in the advection dominated regime where $\eta(p)>$ 1 limited by the CR momentum $p_{\star}$. The dashed curve is the CR spectrum after 4 acceleration times while the dot dashed  and the solid curves are respectively after 6 and 10 acceleration times.The acceleration time is about 3$\times 10^5$ years in this case \citep[see][]{Bykov2014}.  
Right panel illustrates the asymptotic CR spectra at the high-energy end $\eta(p)<$ 1. The spectral index of MHD turbulence inside the system was fixed at $\nu =1.7$. The curves are given for the different ratios of $l_{\mathrm{corr}}/L$ with parameter $\Delta$ =2 (dashed line), 3 (solid line) and 5 (dotted line).}
\label{SB_CR}
\end{figure}
\end{center}

The test particle calculations of the kinetic equations in the low-energy regime 
 $p < p_{\star}$ by  \citet{Bykov92} provided a very hard spectrum of 
 accelerated particles which would absorb a substantial fraction
of the available kinetic energy  after a few acceleration times.
Therefore, the account for the backreaction of the accelerated particles on the turbulence is necessary to model the particle spectrum. 
A model, where the kinetic equations described above were supplied 
with the energy conservation equation for the total system including the shock turbulence and the accelerated CR particles, demonstrated a temporal evolution of initially narrow particle spectrum around the initial momentum $p_0$ to a hard and then softer spectrum in the energy interval  $p_0 < p < p_{\star}$ \citep{Bykov2001,Bykov2014} illustrated in the left panel of Fig. \ref{SB_CR}.

CR particles at high energies  $p > p_{\star}$ have $c \Lambda(p) > u_{\star} l_{\rm mathcorr}$ which imply that the characteristic diffusion scales of CRs accelerated at shock are larger than the mean distances between the shocks. Therefore, the high-energy CRs are rather homogeneously distributed within the system contrary to the highly intemittent distribution of CRs with  $p < p_{\star}$ described by Eq.(\ref{KE:lowE}) as discussed above. 

The distribution function of the high-energy CRs of $p > p_{\ast}$ can be obtained from the Fokker-Planck type equation which contains only the terms  dominating in Eq.(\ref{KE:lowE}) in the high-energy regime:

\begin{equation}\label{KE:highE}
      \frac{\partial N}{\partial t} -
       \frac{\partial}{\partial r_{\alpha}} \: \kappa_{\alpha \beta}(p) \:
       \frac{\partial N}{\partial r_{\beta}}  =
      \frac{1}{p^2} \: \frac{\partial}{\partial p} \: p^2 D \:
      \frac{\partial N}{\partial p} ,
\end{equation}
where the parallel spatial diffusion coefficient $\kappa(p) = v \Lambda(p)/3$. The momentum diffision coefficient is determined by 
\begin{equation}
  \label{MDiff}
  \begin{split}
    D(p) =  \frac{p^2}{9} \int_{}^{} dt'd\mathbf{r'}  <\nabla
    \cdot \mathbf{u}(t',\mathbf{r'})\cdot \nabla
    \cdot \mathbf{u}(t,\mathbf{r})> \rm{G}(t',\mathbf{r'};t,\mathbf{r}). 
  \end{split}
\end{equation}
Here $\rm{G}(t',\mathbf{r'};t,\mathbf{r})$ is the Green function described the diffussion propagation of GRs in space with the diffusion coefficient $\kappa(p)$. CR acceleration described by Eq.(\ref{MDiff}) corresponds to {\sl Fermi} acceleration by the large-scale motions of plasma with the wavelengths larger than the CR mean free paths $\Lambda(p)$ \citep{BT93,2001AstL...27..625B}. In the systems  where the root mean square (r.m.s.) amplitude of the large-scale MHD motions $u_{\star}$ is supersonic and super-Alfv\'enic this non-resonant CR acceleration should be dominating over the stochastic acceleration by the resonant MHD modes. The stochastic resonant acceleration of CRs \citep[see e.g.][]{1968JETP...26..821T,Schlickeiser02,2019ApJ...879...66T} is important in the systems with the subsonic turbulence. 

One should have in mind that the Eq.(\ref{KE:highE}) can be used to describe the CR distribution asymptotically in the high-energy regime and the matching its solutions with the low-energy solutions of Eq.(\ref{KE:lowE}) needs some care. The CR disribution function above $p_{\star}$ satisfying the Eq.(\ref{KE:highE}) and the condition of CR escape in the system boundary is
\begin{equation}
  \label{eq:D}
N(p) = A_0 (p/p_{\star})^{-(\nu +1)/2} K_{\emph a}(Y) ,   
\end{equation}
where the modified Bessel function (Macdonald function) $ K_{\emph a}(Y)$ has the argument  $Y = (p/p_{\star})^{2-\nu} \Delta$ and the index ${\emph a} = (\nu +1)/|4 - 2\nu|$ (for $\nu \neq 2$). Here $\Delta = K \cdot \kappa_{\star}/u_{\star} L$, where $L$ is the characteristic size of the system and  the numerical factor $K$ depends on the shape of the superbubble and the boundary conditions for the CR distribution function there. Realistic boundary conditions should be determined by the energy dependent escape of the accelerated CRs through the supershell surrounding the superbubble which require a dedicated study of its ionization state and MHD turbulence growth and  damping there. Namely, the flux of escaping CRs which depends on the CR acceleration efficiency and the scale size of accelerator may be high enough at some evolution stages to produce CR-driven turbulence providing the effects of CR self-confinement. Such the effect was discussed recently in the case of supernova remnants \citep[see e.g.][]{2018AdSpR..62.2731A,Brahimi20}.

The cut-off momentum obeys the relation
\begin{equation}
p_{\rm c} \propto p_{\star} \cdot \Delta^{1/(\nu - 2)} 
\end{equation} 
based on the Macdonald function behaviour when $p>p_{\star}$. Therefore, 
\begin{equation}\label{pcrit}
p_{\rm c} \propto p_{\star} (L/l_{\mathrm corr})^{1/(2 - \nu)}.
\end{equation}
This indicates rather a strong dependence $p_{\rm c}$ on $L$ if the spectral indexes of MHD turbulence are $5/3 \leq \nu <2$ (i.e. the Kolmogorov-type or steeper). 

The asymptotics of the Macdonald function at the large arguments $Y \gg 1 $ is $K_{\emph a}(Y) \approx \sqrt{\frac{\pi}{2Y}}\exp{(-Y)}$.
Therefore, for the large momenta of CRs $p > p_{\rm c}$ the spectrum Eq.(\ref{eq:D}) has exponential 
asymptotics 
\begin{equation}\label{HE asympt}
N(p) \propto (p/p_{\rm c})^{-3/2} \exp{[-(p/p_{\rm c})^{(2-\nu)}]},
\end{equation}

The results discussed above are valid for MHD turbulence of index $\nu <2$. Note that MHD turbulence with $\nu = 2$ can be realized in some interval of wavenumbers in the systems with multiple weak shocks which can be the secondary shocks produced by interactions of the primary strong supernova shocks propagating through inhomogeneous matter in a superbubble \citep{1987Ap&SS.138..341B}. The mean free paths of CRs with gyroradii exceeding the mean distance between the weak shocks are  energy independent.         

It follows from the Eq.(\ref{estar}) that the CR spectrum in superbubble given by the solid curve in the left panel of Fig. \ref{SB_CR} may extend to GeV regime for magnetic fields $ B > 10$ $ \mu$G inside the the accelerator and high enough plasma bulk velocities $\gsim 1,000 \kms$ which may occur either at the early stages of a relatively compact cluster of young massive stars with fast winds or at some stages dominated by multiple rather frequent supernovae. 
Strong shock waves produced by supernovae propagating through the hot gas may substantially amplify turbulent magnetic fields inside the superbubble over a time scale $\sim 10$ million years. CR-driven instabilities can transfer up to 10\% of the shock ram pressure into the fluctuating magnetic fields \citep[e.g.][]{Bykov2014}. This allows us to expect the magnetic turbulence of r.m.s. amplitudes up to about 50 $\mu$G at some evolutionary stages of the superbubbles.      
This may be the case in the Cygnus Cocoon \citep{AckermannSB2011} which was discussed above. In the YMSCs like Westerlund 1 and 2 the values of the r.m.s. magnetic fields can be substantially higher.

The CR spectra shown in Fig. \ref{SB_CR} can explain both {\sl Fermi} data \citep{AckermannSB2011} (solid curve at the left panel) and the {HAWC} measurement \citep{HAWC_Cygnus_19} (right panel) in the scenario with efficient CR acceleration. In the case of not efficient enough superbubbles only the rather steep spectra illustrated in the right panel of Fig. \ref{SB_CR} are expected. The winds of massive stars and supernovae in the nearby region of star formation has produced a low-density ionized gas cavern of about 200 pc size called the Orion-Eridanus superbubble. Detailed study of the superbubble revealed the presence of a hot gas with a range of temperatures above 10$^6$ K with a pressure in excess of the local surroundings and the magnetic fields up to 15 $\mu$G  \citep[][]{OrionEridanusAA19}. Analysis of gamma-ray observations of the Orion-Eridanus superbubble  with the {\sl Fermi} observatory \citep[cf][]{OrionEridanusCR} found that the gas gamma-ray emissivity spectrum there is consistent with the average spectrum that was measured in the local interstellar medium in a contrast to that in the Cygnus cocoon. In the context of the CR acceleration model discussed above the strong dependence of the characteristic momentum $p_{\star}$ on $u_{\star}$ for the magnetic fluctuation spectra of indexes $\nu \geq 5/3$ may imply that the r.m.s. velocity there is $u_{\star} \leq 500 \kms$. This is not the only possible reason, still lack of any information on the magnetic turbulence in superbubbles
forced us to rely on rather simple approximations which in some cases are justified by the direct measurements of the magnetic turbulence in the solar wind \citep[see e.g.][]{2005ppfa.book.....K,2016LNP...928.....B}. 

\subsection{Cosmic Ray Pevatrons from supernovae in compact star clusters}\label{CSFS}
The CR acceleration mechanism of the Fermi type in superbubbles discussed above used the mechanical energy supplied by the fast winds of massive stars and the ejecta of supernovae over a few million years. With the conversion efficiency about 10\% or slightly higher a superbubble can transfer about 10$^{52}$ ergs to the CRs. It can produce the time evolving  CR ion spectra up to characteristic momentum $p_{\rm c}$ which scales with the ratio of the system size to the turbulence correlation length according to  Eq.~\ref{pcrit}. This can provide multi TeV ions under a wide range of parameters, but to reach the ion energies above PeV the turbulence spectral index $\nu \rightarrow$ 2 is required. The spectral index $\nu$ close to 2 within some dynamic range of wavenumbers is expected in the models of turbulence which is dominated by multiple weak shocks \citep{1987Ap&SS.138..341B}.    

\begin{center}
\begin{figure}
\includegraphics [width=100mm]{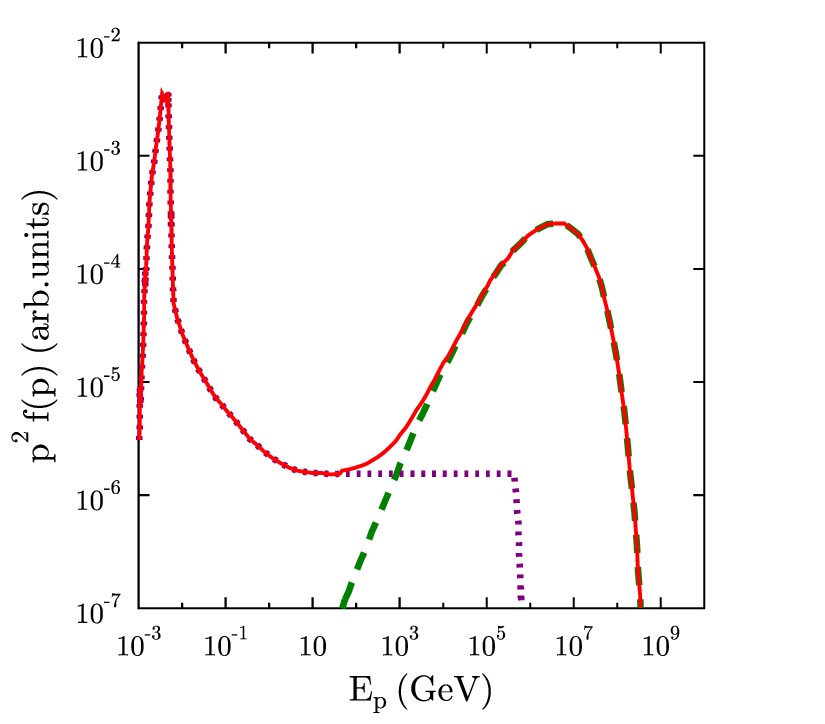}
\caption{A model of instant spectrum of CR ions accelerated by a supernova shock of speed about 10,000 $\kms$ colliding with a fast wind of a young massive star or the collective wind of YMSC. Here, $f(p)=p^2 N(p)$, where $N(p)$ is particle distribution function in the phase space defined earlier. The acceleration time is about 400  years in this case \citep[see][]{BEGO2015MNRAS}. The dotted line shows the CR spectrum produced by an isolated supernovae expanding in a homogeneous ISM. The dashed curve illustrated 
the contribution of the CR accelerated at the stage of colliding flows - supernova '
shock colliding with the fast wind of a speed of $\sim$ 3,000 $\kms$, and the 
solid curve shows the total spectrum just at the maximal acceleration phase which lasts 
a few hundred years only.}
\label{CSF_Sp}
\end{figure}
\end{center}

However, CR ion can be accelerated above PeV energies by supernovae exploding in YMSCs 
like Westerlund 1 \citep{Bykov2014,BEGO2015MNRAS}. In Fig.~\ref{CSF_Sp} we illustrate a spectrum of CR ions accelerated at the colliding shock flows produced by high velocity supernova shock at the free expansion stage colliding with a fast wind of either another massive star in the YMSC or the collective wind of the compact cluster. The shock collision stage typically lasts less than a thousand years (depending on the size of the compact cluster) which is very short comparing to the acceleration time of CRs in the extended superbubble discussed above.
However, the maximum energies of CRs accelerated by a supernova exploding in YMSC under favorable conditions can be well above PeV. 

Currently operating Cherenkov telescopes  H.E.S.S,. MAGIC, VERITAS provide spectral measurements of gamma-ray sources up to 100 TeV. H.E.S.S. observations detected high energy emission possibly associated with some clusters of young massive stars, namely Westerlund 1 and 2, Cl*1806-20  \citep[][]{2018A&A...612A..11H,2018A&A...612A...1H}. LHAASO (Large High Altitude Air Shower Observatory) \citep{2019arXiv190502773B} has a performance designed for sensitive observations of gamma-rays up to PeV energies thus opening a good perspective for  studies of the galactic Pevatrons. 

We discussed above the CR spectra inside the accelerators. The Fermi-type acceleration mechanism requires that accelerated particles are confined within the acceleration site. This means that the spectra of CRs which escaped the accelerator differ from CR spectra inside. If the CR accelerator is highly efficient the electric current of the escaping CRs may drive instabilities which amplify magnetic turbulence in the ambient ISM. Therefore, the non-linear effects of CRs backreaction may control the CR propagation in the vicinity    
of the sources \citep[see e.g.][]{Ptuskin2008,Malkov2013,bbmo13,Brahimi20}. We shall discuss these effects below.  

\section{Cosmic ray propagation near their sources}

\subsection{Supernova remnants and superbubbles}
We discussed above particle acceleration mechanisms in star-forming regions with multiple winds from the massive early-type stars and core-collapsed supernovae. Isolated supernova remnants were for a long time considered as possible sources of CRs \citep{BZ34,Ginzburg1964,Axford81,Berezinski90,Drury2001,2018SSRv..214...41B}. CR are  accelerated there likely by diffusive shock acceleration (DSA), see \citet{1977ICRC...11..132A,Kry77, Bell78a, Bell1978b, Blandford1978, Marcowith2016}. The particles, at least in the adiabatic phases of blast evolution, are injected from the heated shocked gas by thermal leakage \citep{Gieseler2000}. High-energy CRs tend to cover large distances ahead of the shock being responsible for magnetic field amplification \citep{Bell1978b,BE87,Blandford2007, Bell2004,SchureEtal2012}.

 Different scenarios of CR escape from SNR have been discussed. First, particles of high enough energy can escape by the "geometrical" losses, as their mean free path tends to exceed the SNR radius \citep[e.g.][]{Berezhko1994}, these particles have a decreasing probability to come back to the shock front \citep[see e.g.][]{Drury2011}.  \citet{BellEtal2013} argued that the current carried by the highest energy CR is the main free energy source available for magnetic field amplification at the shock front. The maximum CR energy in this scenario is fixed by the minimum amount of charge by unit of area of the SNR forward shock (see also \citet{Zira08}). In all cases, the highest energy CRs very likely escape first. Then, as these particles carry a large amount of pressure, even if they start to free stream in the ISM, they still carry enough free energy to amplify other magnetic perturbations around the CR source leading to production of the time and energy-dependent CR halos \citep{Ptuskin2008}. The issue now is to evaluate the typical time of existence and the extent of these transient structures. 
 
\begin{center}
\begin{figure}
\includegraphics [width=100mm]{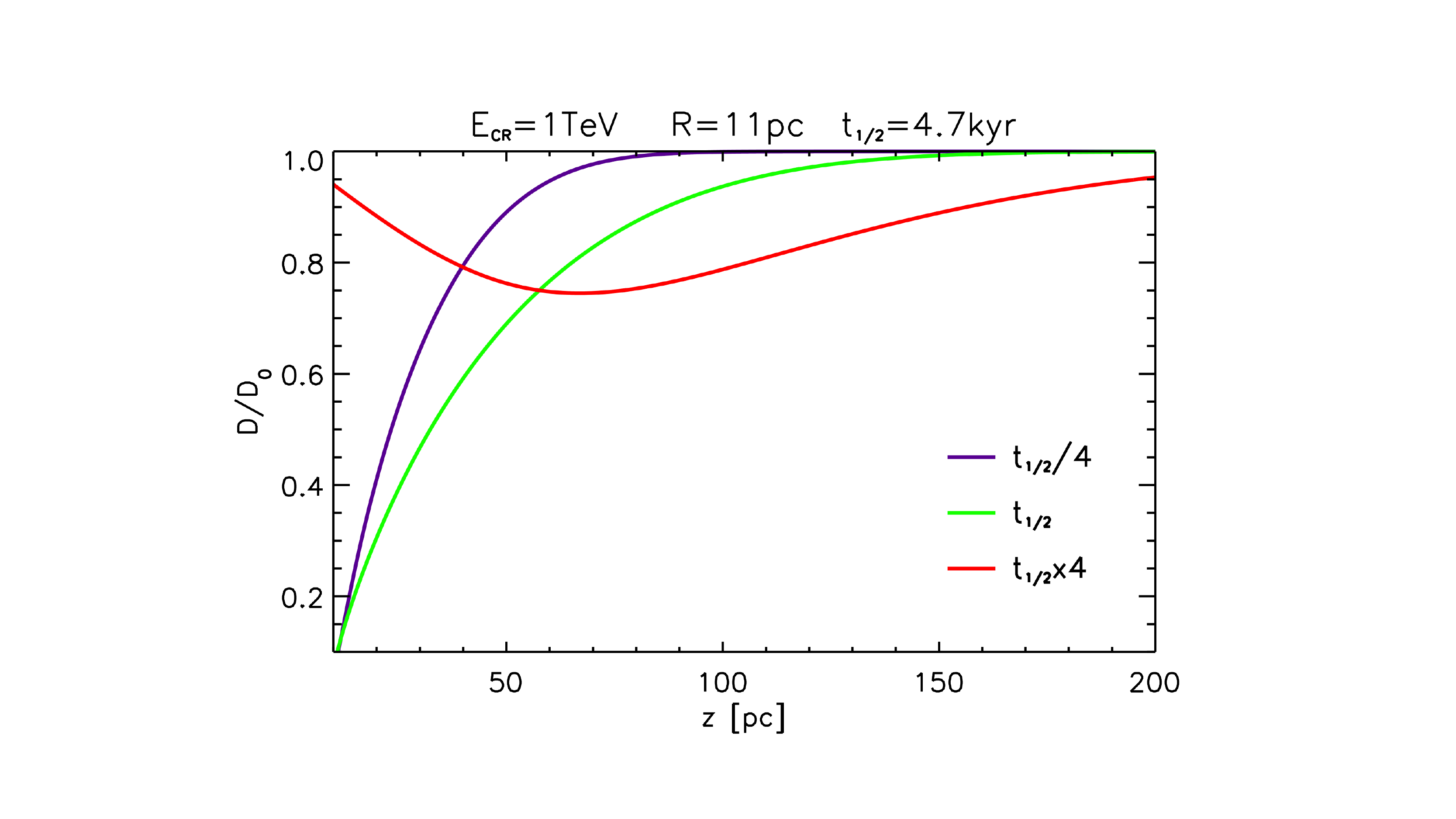}
\caption{Time evolution at different distances from the CR cloud of the ratio of the self-generated CR diffusion coefficient D to the standard background CR diffusion coefficient $D_0$ for particles with a kinetic energy of 1 TeV in the case of a SNR propagating in the WIM phase. The figure shows the escaping radius and the time $t_{1/2}$ after which the CR pressure in the cloud as decreased by a factor 2. From \citet{Nava2016}.}
\label{Fig:N16}
\end{figure}
\end{center}

 One way to investigate this problem is to decouple the acceleration and escape problems (which is an approximation, see \citet{Telezinsky2012}), by considering each young SNR to be surrounded by a CR cloud composed of particles  escaped from the shock \citep{Malkov2013}. These authors demonstrate that the escape can be treated in 1D at a good approximation level: CRs escaping from the cloud follow the local background magnetic field lines up to a distance where the field line wandering tends to dominate the CR transport which hence becomes 3D \citep{Nava2013}. Now the way CRs propagate around a SNR depends strongly on the properties of the ambient ISM gas. The ISM has an impact on CR losses and magnetic perturbations propagation and damping. A series of works has successively investigated the propagation in hot ionized ISM \citep{Dangelo2016, Nava2019}, warm ISM phases \citep{Nava2016, Dangelo2016}, and atomic and molecular phases of ISM \citep{Brahimi20}. In ionized media the main damping processes for perturbations with wavelength in resonance with CR gyromotion are the linear and non-linear Landau damping and the effect of CR self-generated perturbations interaction with background turbulence \citep{Farmer2004}, sometimes referred as the turbulent damping \citep{Lazarian2016}, whereas ion-neutral collisions dominate in partially ionized media \citep{Xu2016} at least for large enough wavenumbers \citep{Brahimi20}. Particle propagation is investigated by solving a system of two coupled equations for the CR pressure $P_{\rm CR}(E)$ and the self-generated wave magnetic energy $I(k)$ \citep{Nava2016}:
 \begin{equation}
    \label{eq:CRs}
    	{\partial P_{\rm CR} \over \partial t} + V_{\rm A} {\partial P_{\rm CR} \over \partial z} = 
        {\partial \over \partial z} \left( D {\partial {P_{\rm CR}} \over \partial z} \right) \ ,
    \end{equation}
    
    \begin{equation}
    \label{eq:waves}
    	{\partial I \over \partial t} + V_{\rm A} {\partial I \over \partial z} = 2 \left(\Gamma_{\rm growth}  - \Gamma_{\rm d}\right) I + Q \ . 
    \end{equation}
In this study particles generate resonant perturbations at wavenumber $k= 1/r_{\rm g}$, where $r_{\rm g}$ is the CR gyroradius through a resonant streaming instability with a growth rate $\Gamma_{\rm growth} \propto \partial P_{\rm CR}/ \partial z$. The ISM phase-dependent damping rate is $\Gamma_{\rm d}$. Q is the background turbulence injection rate. Finally, the CR diffusion coefficient is $D \sim vr_{\rm L}/I(k)$, {\sl v} is the particle speed. The background galactic magnetic field is supposed to lie in the {\sl z-} direction. The second left-hand side terms in Eqs. \ref{eq:CRs}, \ref{eq:waves} describe the streaming of CRs at the local Alfv\'en speed. CRs are injected by evaluating the cloud size $a_{\rm esc}$ at which half of the initial CR pressure is escaped. Among all possible solutions for $a_{\rm esc}$ \citet{Nava2016} selected the one which corresponds to the radius of the SNR dynamics imposed by the local ISM \citep{Truelove1999, Cioffi1988}. We illustrate in Fig. \ref{Fig:N16} the time evolution of the diffusion coefficient $D$ induced by the self-generated turbulence in the case the SNR propagates in the warm ionized phase (WIM) of the ISM \citep{Nava2016}. We can see the important confinement effect due to the self-generated turbulence at distances $<$ 50 pc around the CR cloud for timescales to about 5 kyrs. Once the time and energy-dependent diffusion coefficient $D$ is calculated, it is easy to deduce the grammage of the particles due to their self-confinement \citep{Dangelo2016, Nava2019, Brahimi20} and the gamma-ray emission expected from CR halos \citep{Dangelo2018}. 
The extended galactic superbubbles are surrounded by supershells and may interact with the progenitor molecular clouds. The ionisation state and the heating of the gas in the supershells may change over the lifetime of the supebubble since the lifetime of the most powerful ionizing stars is shorter than the SB lifetime. This may regulate the regimes of the CR escape and the CR halo formation.\\
We here give rough estimates of the diffusion coefficient parallel to the background magnetic field in the SB and YMSC cases. We use a procedure similar as the one adopted in \citet{Commercon19} to evaluate the diffusion coefficient of GeV CRs produced by self-generated turbulence around SNR in atomic and molecular phases. We consider a CR source which releases a CR pressure $P_{\rm CR, s}= {\rm f} P_{\rm CR , b}$, f ($>1$) times larger than the background CR pressure at a given energy E over a typical scale corresponding to a fraction g ($<1$) of source size $L_{\rm s}$. Hereafter we fix $P_{\rm CR, b}= 1~\rm{eV/cm^3}$. The level of resonant self-generated turbulence $I(k)$ is fixed by balancing the wave growth rate due to the resonant streaming instability $\Gamma_{\rm g}$ and a specific phase-dependent damping rate $\Gamma_{\rm d}$. If $V_{\rm A}$ is the local Alfv\'en speed and $W_{\rm B}$ is the ambient magnetic energy density we find
\begin{equation}\label{eq:Ik}
    I(k)= {V_{\rm A} P_{\rm CR}' \over 2 \Gamma_{\rm d} W_{\rm B}} \ ,
\end{equation}
where $P_{\rm CR}'= (P_{\rm CR,s}-P_{\rm CR,b})/{\rm g} L_{\rm s}$. Using the quasi-linear theory of CR transport \citep{Schlickeiser02} we deduce the CR mean free path along the background magnetic field lines 
\begin{equation}
    \lambda_\parallel \simeq {8 \over \pi} \left({W_{\rm B} \Gamma_{\rm d} \over P_{\rm CR}' V_{\rm a} }\right) r_{\rm g} \ .
\end{equation}
In number this gives:
\begin{equation}
    \lambda_{\parallel} \simeq 30~\rm{pc} \times \left({g \over f-1}\right) \Gamma_{\rm d,yr^{-1}} L_{\rm s, pc} \sqrt{n_{\rm g, cc}} E_{\rm TeV} \ . 
\end{equation}
The gas density $n_{\rm g}$ is in $\rm{cm^{-3}}$ units, the CR kinetic energy E in TeV units and all scales are in parsec units. The damping rate $\Gamma_{\rm d}$ is yr$^{-1}$ units. The final result is not dependent on the background magnetic field strength. 
The solutions depend on the dominant damping process. In SBs we consider CRs are injected in the shell of cold gas surrounding these structures. In YMSCs CRs are preferentially injected in HII regions. We discuss below the expressions for $\Gamma_{\rm d}$ in each of these media
\begin{itemize}
    \item Superbubbles: CRs released from SBs encounter first HI shells. There two types of damping may be relevant 1) ion-neutral collisions 2) turbulent damping due to the interaction of self-generated turbulence with large-scale injected turbulence. Ion-neutral damping is the fastest process in the cold neutral medium\footnote{especially if the injection scale of the turbulence is large as it is the case here with $L \sim L_{\rm s} \sim 100$ pc, see table \ref{T:mfp}.} \citep{Brahimi20} with $\Gamma_{\rm d}(\rm{E}=1~\rm{TeV}) \sim 10^{-4}~\rm{yr}^{-1}$. This value will be retained
    in the supershell.
    \item Young massive stellar clusters: CRs released from YMSCs first encounter HII regions of dense molecular gas ionized by the U.V. radiation from young massive stars. The main damping process in HII regions is the turbulent damping \citep{Farmer2004, Lazarian2016}. We consider the following parameters: $n_{\rm g} =1000~\rm{cm^{-3}}$, the magnetic field $B= 10~\mu$G  and gas temperature $T \sim 10^4$ K \citep{Maurin2016}. HII regions show irregular structures in optical images \citep{1995ApJ...454..316M} possibly connected to turbulent motions. The numerical modeling of this turbulence finds rms speed $v_{\rm t}$ of the turbulent velocity of the order $\sim 10$ \kmps, slightly supersonic \citep{2014MNRAS.445.1797M}. Owing to the previous choice of parameters we deduce that HII regions are rather super-Alfv\'enic with an Alfv\'enic Mach number $M_{\rm a}=v_{\rm t}/v_{\rm A} \sim 10$. The injection scale of the turbulence is assumed to be of the order of the size of HII regions, $L \sim 10$ pc, this scale is much larger the TeV CR Larmor radius, hence the damping rate is $\Gamma_{\rm d} = (v_{\rm A}/L) M_{\rm a}^{3/2} (r_{\rm g}/L)^{1/2}$. We find $\Gamma_{\rm d}(\rm{E=1~TeV}) \sim 2 \times 10^{-3}~\rm{yr^{-1}}$. 
    \item In both cases non-linear Landau damping which has a rate \citep{Nava2019}
    \[
    \Gamma_{\rm NLD} \sim 5.5 \times10^{-5}~\rm{yr^{-1}} \left({T_{\rm K} I(k) \over E_{\rm TeV} B_{\rm 10\mu\rm{G}}}\right) \ ,
    \] is found smaller than the other damping rates as soon as $I(k) < 0.1$, which has to be the case because all the calculation derived above has been performed in the quasi-linear limit. The ambient temperature T is calculated in Kelvin units. 
\end{itemize}
In Table \ref{T:mfp} we report the values of $\lambda_\parallel$ for different gas parameters at 1 TeV. In the table we also derive the ratio $r_\lambda= \lambda_\parallel/\lambda_{\rm b}$ of the parallel mean free path of CR in the self-generated turbulence to the mean free path of CR deduced from the secondary to primary elemental abundances obtained from direct observations. As a reference we take  $\lambda_{\rm b} \sim 10~\rm{pc} \times E_{\rm TeV}^{0.5}$. \\
In the SB case because of the low damping rate, we find $\lambda_\parallel \sim 1~\rm{pc} \times g/(f-1)$. Using a conservative value $g =1$ and $f=10$ \citep{Bykov92} we find that CR mean free path in the supershell drops below $\lambda_{\rm d}$ and the shell size (taken here as $\sim 0.1 L_{s} \simeq 10$ pc). In this framework, CRs are expected to be well confined in HI shells surrounding CR active SBs over timescales of the order of 1 Myrs. However, the mean free path is very sensitive to $f$, a small CR pressure contrast ($f \rightarrow 1$) produces a large mean free path.\\
In the YMSC case, because of the interaction with background turbulence the damping of self-generated perturbations is slightly more severe in HII regions. We consider a typical source size of 1 pc. At last we find $\lambda_\parallel \sim 2$ \rm{pc}$ \times g/(f-1)$. Using $g = 1$ and $f \sim 300$ (cite) we find again a mean free path smaller than the typical size of the HII region.\\
In both cases using Eq. \ref{eq:Ik} we had a consistency check that $I(k) \ll 1$, so the quasi-linear theory framework used to obtain these results is valid.\\
In summary, due to large overpressure and modest damping rates the regions surrounding SBs and YMSCs may harbor a halo of freshly injected CRs confined in these structures over a typical timescale of $t_{\rm diff} \sim 3 L_{\rm shell/HII}^2/(\lambda_\parallel c) $. This timescale is however rather short (in the range $10^4-10^5$ yrs) because the size of HI shells and HII regions is about 10 pc. The longest diffusion times are found in the case of large HII regions and may reach 1 Myrs if the CR pressure contrast is large enough. Any assessment of these crude estimates require more refine modeling based on solving Eqs \ref{eq:CRs} and \ref{eq:waves}. 

\begin{center}
\begin{table}
\begin{tabular}{c|c|c|c|c|c|c}
 Source type & $n_{\rm g,cc}$ & $L_{\rm s, pc}$ & f & g & $\lambda_\parallel$, pc & $r_\lambda$ \\
 \hline
SBs     & 10 & 100 & 10 & 1 & 0.1 & 0.01 \\
\hline
YMSCs    & 1000 & 1 & 300 & 1 & $6~10^{-3}$ & $6~10^{-4}$\\  
\hline
\end{tabular}
\caption{CR parallel mean paths at 1 TeV in SBs and YMSC environments. Values of the parameters are deduced from \citet{Bykov92}.}
\label{T:mfp}
\end{table}
\end{center}


\subsection{Cosmic ray propagation in atomic and molecular interstellar medium}
These particular media of the ISM are not -- a priori -- expected to much confine CRs above 100 MeV because of the effect ion-neutral collisions in damping magnetohydrodynamic waves which support the spatial transport of CRs \citep{Cesarsky78}. However, close to a CR source if the CR overpressure is high enough or the CR current is strong enough CRs can trigger instabilities over times short enough to compensate over the ion-neutral damping time. This is especially true at CR energies beyond a few 100 GeV-1 TeV as in these energy regimes the MHD waves in resonance with the particles have a pulsation low enough to be in the coupled regime where ions and neutrals move together \citep{Xu2016, Brahimi20}. \\
\citet{Xu2016} considered only the effect of background MHD turbulence injected at scales larger than the typical size of the phase under question. The CR mean free path is found to be strongly reduced in the range 1-100 TeV in molecular clouds because of both gyro-resonant and transit-time damping interactions with MHD waves. \\
\citet{Brahimi20} using the CR cloud model (see previous section) considered the case of self-generated turbulence composed of resonant modes only produced by the resonant instability. There the CR overpressure imposed by the presence of a nearby SNR is large enough to compensate over the ion-neutral damping effect. This overpressure is even larger in these type of phases because the CR cloud is less extended with respect to more dilute phases. Hence, even if the escape time is shorter (typically a fraction to a few kyrs), the timescale on which the diffusion coefficient reduces by an order of magnitude
can still be substantial (typically 10-100 kyrs). \\
\citet{Inoue19} proposed an alternative mechanism for TeV CR confinement. In the configuration of a SNR in interaction with a molecular cloud the current produced by escaping CRs which penetrate into the cloud is found to be strong enough to trigger the non-resonant streaming instability. This effect can induce a spectral modification by magnetic field amplification of CR and gamma-ray spectra around molecular clouds.

\section{The specific features of cosmic rays accelerated in young massive stellar clusters}
Young massive star clusters
may contribute into the observed spectrum and composition of the galactic cosmic rays both at low and high energies.
We discuss below the expected ratio of $^{22}$Ne/$^{20}$Ne in CRs accelerated in YMSCs to understand the well-known neon isotopic composition anomaly in CRs. We also examine the constraints from the measured CR anisotropy on the scenario of CRs accelerated well above PeV by supernovae in YMSCs.         
\subsection{Massive star clusters as possible sources of  $^{22}$Ne-enriched cosmic rays}
\subsubsection{Composition anomalies of cosmic rays}

 The elemental and isotopic abundances in different astrophysical environments are the important pieces of information, which can clarify high-energy processes in space.
Solar isotopic abundances are well-known from the investigation of C1 carbonaceous chondrite meteoritic abundances \citep{Lodders2003}.
Chemical composition of the galactic cosmic rays has been studied in a number of experiments: { IMP-7} \citep{Garcia1979}, { ISEE-3} \citep{Wiedenbeck1981},{ Voyager} \citep{Lukasiak1994}, { ACE-CRIS} \citep{Binns2005} and others. It was shown that the isotopic abundances of cosmic rays are mostly similar with the solar system values. Nonetheless, there are several differences: $^{12}$C/$^{16}$O, $^{22}$Ne/$^{20}$Ne,$^{58}$Fe/$^{56}$Fe. {  ACE-CRIS} measurements have resulted in a comprehensive review, where this deviation was quantitatively studied: it was shown that in CRs $^{22}$Ne/$^{20}$Ne=$0.387 \pm 0.027$, while in the solar system $^{22}$Ne/$^{20}$Ne=0.07. It corresponds the $5.3 \pm 0.3$ $^{22}$Ne overabundance. The $^{22}$Ne/$^{20}$Ne ratio was measured in energy range $84\leq$E/M$\leq 273$ MeV/nucleon \citep{Binns2005}.  Finding a reason of this overabundance is a long-standing issue, which is called the neon problem.

\subsubsection{Cosmic ray neon isotopic problem}

The discovery of the neon problem was followed with the number of suggested solutions (e.g., \cite {Woosley1981, Reeves1978, Olive1982}). Basically, there is only one astrophysical object, which can produce significantly more $^{22}$Ne than $^{20}$Ne --- the carbon (WC) sequence of Wolf-Rayet stars. That was originally suggested by  \cite{Casse1982}. Carbon stage is usually one of the latest, pre-supernova stages in star evolution.  During WC stage almost all $^{14}$N converts to $^{22}$Ne through the chain of reactions  $^{14}\mathrm{N}(\alpha, \gamma)$ $ ^{18}\mathrm{F}(e^+ \nu) $ $^{18}\mathrm{O} (\alpha, \gamma)$ $^{22}\mathrm{Ne}$. The authors estimated that $^{22}$Ne/$^{20}$Ne isotopic ratio in the wind of WC star is 120 times as high as solar, so their contribution to the CR production can be 2$\%$ to satisfy the measured value.


It is obvious that to solve the neon problem, the source of a part of CRs should be connected with WR stars. But the main question is how and where CRs from WR winds are accelerated to high energies. Several suggestions were proposed so far.  \cite{Higdon2003} suggested that $^{22}$Ne-enriched CRs are originated and accelerated in galactic superbubbles by shockwaves from stellar winds themselves and supernovae, which explode at the end of O- and B-stars life. In the superbubble paradigm, there is good agreement with the observational data, if 20 \% of galactic CRs are of superbubble origin. 

\citet{Prantzos2012} argued with this point of view and suggested another mechanism of acceleration of $^{22}$Ne-rich CRs. According to him, when the forward shock of a supernova runs through its own pre-supernova wind, it accelerates, among other matter, the material from WR winds. 

It is probable that in the measured energy range CRs rich with $^{22}$Ne can originate in WR winds in YMSCs. The modeling of particle acceleration on collective wind termination shock and SNe shocks for bound and loose clusters was performed by \citet{Gupta2019}. At the same time, in the observed energy band cosmic rays can be accelerated only on multiple shocks from 
massive stars in the cluster. Calculations and estimates provided below show that suggested sources can produce the observed Ne isotopic ratio and CR flux near Earth.
\subsubsection {Modeling of $^{22}$Ne/$^{20}$Ne in cosmic rays from YMSCs}

The modeling of stellar nucleosynthesis is currently well-developed, but still requires high computational capacities. The most advanced models so far were introduced by {\sl Geneva} \citep{Ekstrom2012, Georgy2012} and {\sl Frascati} \citep{Limongi2018} scientific groups. Their results include $^{22}$Ne and  $^{20}$Ne yields and  mass loss rates for different star initial masses as a function of time. Integrated with initial mass function (IMF) from 15 $M_{\odot}$ to 120 $M_{\odot}$, these yields provide the desired neon isotopic ratio in the YMSC. 

\begin{center}
\begin{figure}
\includegraphics [width=115mm] {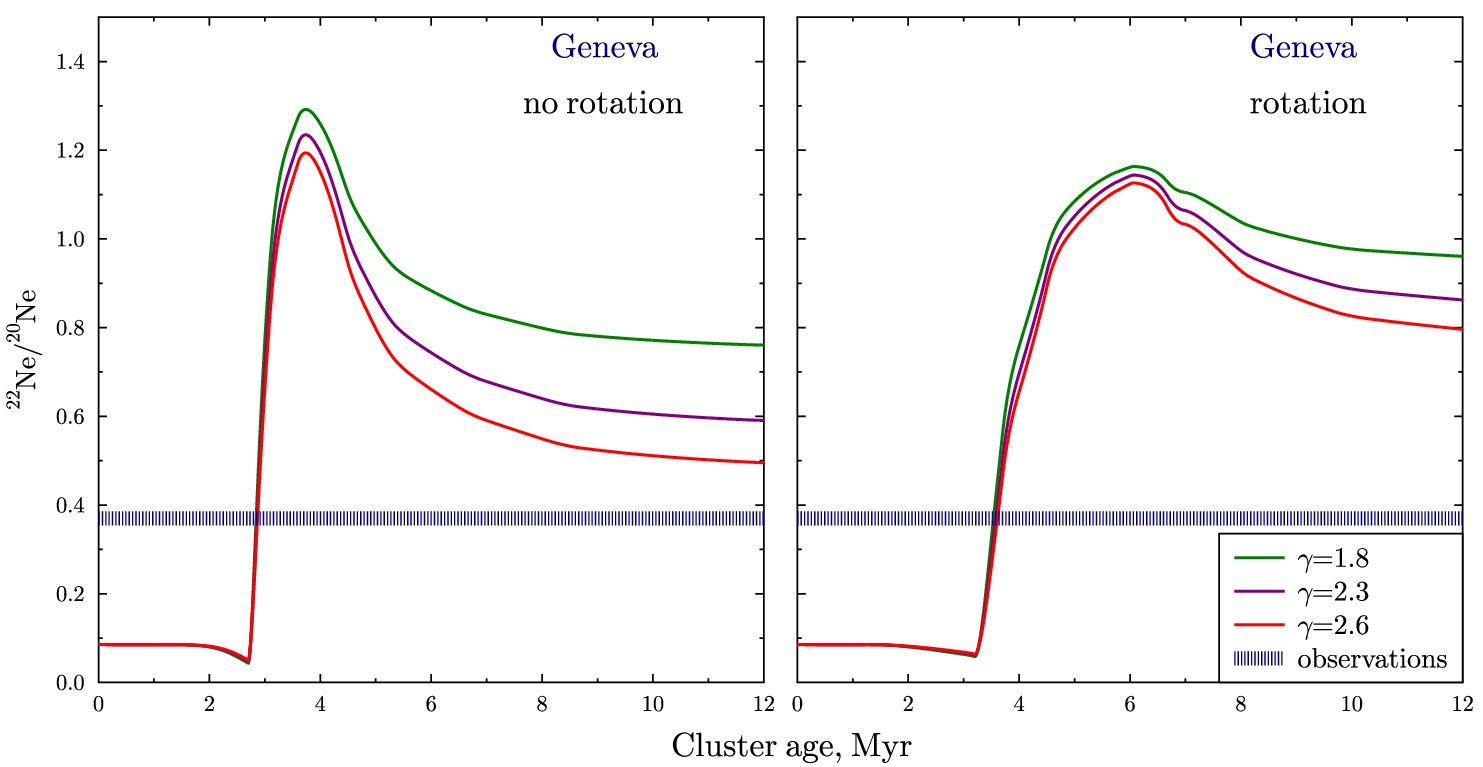}
\caption{$^{22}$Ne/$^{20}$Ne isotopic ratio produced by YMSC as a function of cluster age based on models of \textit{Geneva} group. The velocity of rotation in rotating models is $v=0.4 v_{crit}$, where $v_{crit} =\sqrt{2 GM/3R}$ ($G$ is the gravitational constant, $M$ is the star mass, $R$ is the star polar
radius) \citep{Ekstrom2012}}
\label{fig:NeGe}
\end{figure}
\end{center}

\begin{center}
\begin{figure}[ht]
\includegraphics [width=115mm] {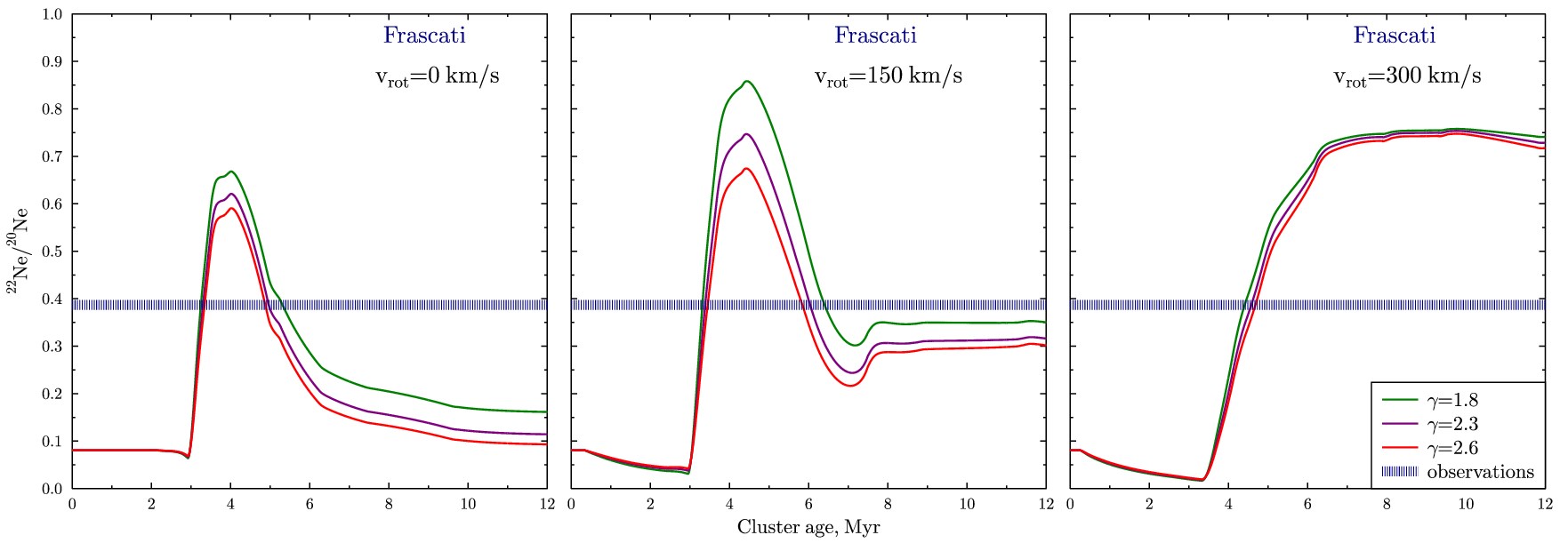}
\caption{$^{22}$Ne/$^{20}$Ne isotopic ratio produced by YMSC as a function of cluster age based on models of \textit{Frascati} group}
\label{fig:NeFr}
\end{figure}
\end{center}

There is a rationale to examine the dependence of the $^{22}$Ne/$^{20}$Ne ratio from the parameters, such as IMF power law index $\gamma$ (there is evidence that in some YMSCs it can be significantly different from Salpeter's value  -- see, e.g., \citet{Hosek2019, Lim2013}) and the velocity of rotation (it can influence dramatically the evolution of the star). The results of the calculations are presented in Fig. \ref{fig:NeGe}, Fig. \ref{fig:NeFr}.

Inspection of Fig.\ref{fig:NeGe}, Fig.\ref{fig:NeFr} shows that $^{22}$Ne/$^{20}$Ne ratio increases visibly after $\sim 3$ Myr, when the most massive stars become WRs. The flattening of IMF leads to increase in the fraction of massive stars and, as a result, to growth of $^{22}$Ne/$^{20}$Ne ratio. The rotation also affects the neon isotopic ratio: it allows less massive stars ($\sim 30  M_{\odot}$) to become WR stars with the total increase in $^{22}$Ne/$^{20}$Ne.

It is seen from Figs. \ref{fig:NeGe}, \ref{fig:NeFr} that depending on parameters and the cluster age, the $^{22}$Ne/$^{20}$Ne ratio can be both well above and below the observed value. Spin rates of the massive O- and WR stars in the Galaxy are found to be $\sim 100$ \kmps\ for single stars and $\sim 200-400$ \kmps\ for binaries \citep{Penny1996, Howarth1997, Shara2017}. This results in higher amount of $^{22}$Ne generating in a cluster according to model predictions.  We assume then that massive star winds in YMSCs should contribute $30-50 \%$ of galactic CRs in 100 MeV-1 GeV energy band.

\citet{Seo2018} provided estimates of total luminosity of all massive stars stellar winds in the Galaxy: $L_w \approx 1.1 \times 10^{41}$ erg s$^{-1}$, based on the average supernova explosion rate in the Galaxy. 
From the estimation of the typical YMSC mass and the number of clusters in our Galaxy, it follows that the fraction of all massive stars that are members of YMSCs is $\alpha=0.09$. 
Then the total wind luminosity of the massive clusters members in the Galaxy is $P_{cl} \approx 10^{40}$ erg s$^{-1}$.

The fraction of wind energy, which is converted to the galactic cosmic rays acceleration, can reach $\eta=0.3$, according to \cite{Bykov2001}. The most common value is $\eta=0.1$. The estimation of the CR flux near Earth is based on $\eta$, cluster wind luminosity, diffusion coefficient in the Galaxy, average distance to a cluster and gives differential flux at GeV energies $J/E^2=7 \cdot 10^{2} \: [\rm {GeV \:  m^2 \:  sr \:  s]^{-1}}$.

The observed flux of cosmic ray protons at Earth is $J_o/E^2 \approx 2 \cdot 10^3 $ $
 [\rm {GeV \:  m^2 \:  sr \:  s]^{-1}}$. It corresponds well with the suggestion that massive stars winds in YMSCs are the sources of $30-50 \%$ of GeV cosmic rays.


\subsection{Anisotropy of PeV regime cosmic rays from YMSCs}

Colliding shock flows from supernovae with velocities of $10^4$ \kmps\ in YMSCs allow proton acceleration up to hundreds of PeV, and heavy ions up to even higher energies (see Sect. \ref{section:CRYMSC}).
The validity of YMSCs as sources of 100 PeV regime CRs can be examined by modeling the observed characteristics of CRs arriving to the solar system: fluxes and anisotropies. It is very important how particles propagate through the Galaxy from YMSCs to Earth and how they behave in the galactic magnetic field (GMF). Such modeling was performed by  \cite{Bykov2019}. 

The galactic magnetic field has complex structure, including regular and stochastic components. The modeling of the GMF is mainly based on observations of pulsars rotation measures (RMs) and dispersion measures (DMs). The most recent investigation of the spectrum and magnitude of stochastic part of the GMF belongs to \cite{Han2017}, while regular GMF was reproduced with accuracy in the latest models of \cite{Pshirkov}, \cite{Jansson1}. 

\begin{figure}

    \floatbox[{\capbeside\thisfloatsetup{capbesideposition={right,top},capbesidewidth=4cm}}]{figure}[\FBwidth]
{\caption{Anisotropy vs. time for 100 PeV protons. The dark blue curve is for CR scattering model without the regular galactic magnetic field. Pale blue curve shows the effect of including the regular galactic field of \citet{Jansson1}; the poor statistics are due to computational restrictions. See \cite{Bykov2019} for details}\label{fig:anis}}
{\includegraphics[width=70mm]{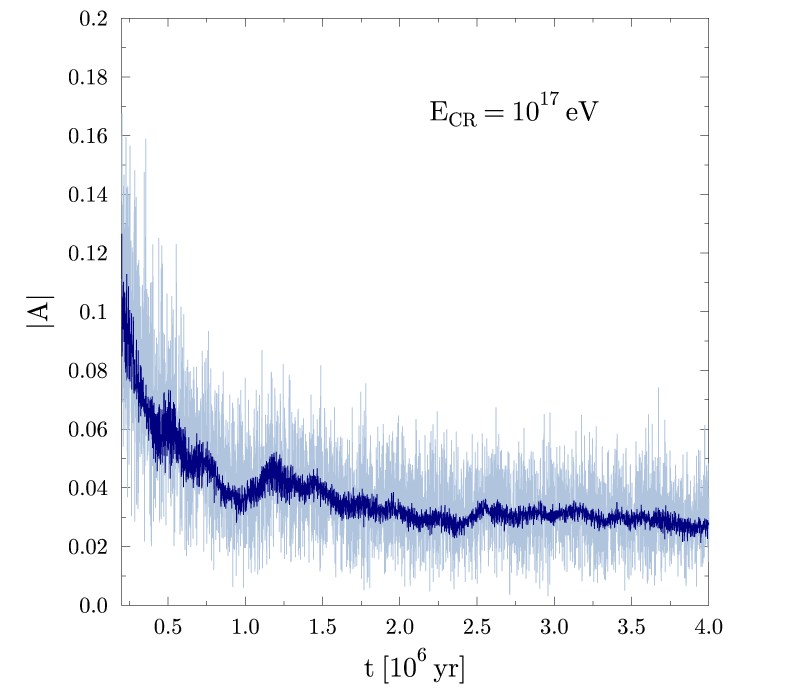}}
 

%

\end{figure}

\begin{figure}

    \floatbox[{\capbeside\thisfloatsetup{capbesideposition={right,top},capbesidewidth=4cm}}]{figure}[\FBwidth]
{\caption{Flux at Earth assuming the efficiency of converting of the SN energy to the kinetic energy of 100 PeV protons is $\chi=0.015 \%$. The pale blue band corresponds to the observed $10^{17}$\,eV CR flux near Earth.}\label{fig:flux}}
{\includegraphics[width=70mm]{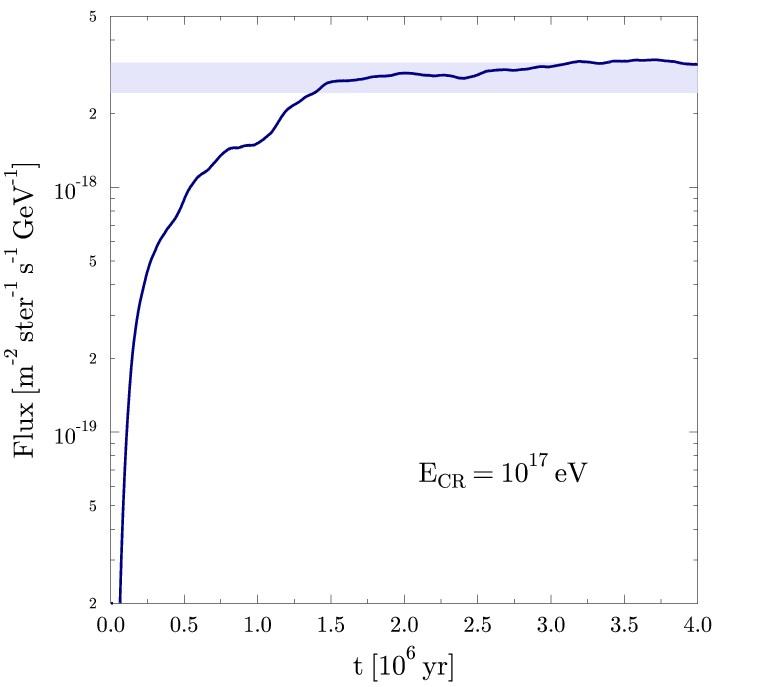}}
 

%

\end{figure}

\citet{Bykov2019} provided calculation of the diffusion coefficient in the GMF and, using Monte Carlo technique, finds anisotropies and fluxes of CR protons from the YMSCs. The results of the modeling are shown in Figs. \ref{fig:anis},\ref{fig:flux}.

The observed dipole anisotropy for 100 PeV CRs is $\lsim 0.01$ and the model anisotropy of particles from YMSCs is $|A| \simeq 0.03$, which means that $1/3$ of PeV CRs can originate in young massive star clusters. The other $2/3$ is the isotropic flux of extragalactic origin. Modeling also shows (Fig. \ref{fig:flux}) that the observed flux of 100 PeV CRs at Earth can be achieved even for very small acceleration efficiency $\chi=0.015 \%$. Thus, YMSCs are the very probable galactic sources of cosmic rays with energies of hundreds of PeV.

\section{The observational perspective}
The new perspective of high spatial resolution observations of star-forming regions in the Local Universe will be opened with the infrared { James Webb Space Telescope} (JWST), expected to launch in 2021. JWST scientific aims include revealing key issues of star formation as well as deep investigations of young stellar clusters and massive star populations. 

Observational study of non-thermal processes discussed previously in this work is an important part in scientific programs of recent and planned X-ray and gamma-ray observatories. In the Sect.\ref{subsection:Xrayfac} we give a brief review of X-ray observational prospects of star-forming regions. 

 Gamma-ray line  spectroscopy is potentially very informative concerning the stellar nucleosynthesis processes    and requires sensitive observations at MeV energies. Currently available observations at energies above 0.3 MeV  obtained with the {\sl Gamma Ray Observatory} and the {\sl International Gamma-Ray Astrophysics Laboratory} (INTEGRAL) provided maps of the Milky Way made with 511 keV annihilation line and $^{26}$Al decay line \citep{2019MmSAI..90..270S}. The planning future projects like {eASTROGAM} \citep[see e.g.][]{ASTROGAM18} are aiming to increase the sensitivity in the MeV band by one or two orders of magnitude comparing to {COMPTEL GRO} and therefore will provide a new look into the star-forming sites in the Galactic Center, Cygnus X and the others.
 An overview of gamma-ray observations in the starburst regions was given by \citet{ohm16}.
In the Sect. \ref{S:CTA} we take a deeper look on  {\sl Cherenkov Telescope Array} (CTA), one of the nearest and the most ambitious projects, which will provide broad opportunities for studying non-thermal radiation from SFRs.
\subsection {X-ray  facilities}
\label{subsection:Xrayfac}
Up to date, X-ray investigations of the SFRs, including revealing relativistic particle acceleration mechanisms, detecting non-thermal X-ray emission and revising current models of superbubbles, were provided by {\sl Chandra, XMM-Newton and Suzaku} observatories. In the nearest future more advanced instruments are about to come.

\citet{Kavanagh20} reviewed the future prospects of the X-ray 
observations of SFRs, in particular superbubbles and massive clusters, discussed above. In the nearest future two new X-ray observatories will provide new data: {\sl Spectrum Roentgen Gamma} (SRG) \citep{2015SPIE.9603E..0CP,2016SPIE.9905E..1KP}, launched in 2019 and {\sl The X-Ray Imaging Spectroscopy Mission} (XRISM) \citep{Tashiro2018}, which launch is planned in 2022. The X-ray telescopes {eROSITA} and {ART-XC} aboard the SRG observatory are expected to provide detailed observations of LMC and superbubbles therein and also allow thorough spectral studies of large Galactic superbubbles. The {XRISM} observatory has two instruments onboard: soft X-ray spectrometer {\sl Resolve} and soft X-ray imager {\sl Xtend}, and shows the great perspective in high resolution X-ray spectrometry \citep[see ][and the references therein]{Kavanagh20}.

New generation X-ray instruments are projected to launch in the next decades. Among them {\sl Athena} observatory \citep{Nandra2013}, planned to launch in 2031. The spectral and spatial resolution of its instruments are expected to be the qualitative leap in the development of X-ray facilities, enabling observing the X-ray faint superbubbles. Another proposed mission {\sl Lynx} \citep{Gaskin2018} with an estimated launch in 2038, can become the most powerful X-ray observatory with order-of-magnitude advances in capability over {\sl Chandra} and {\sl Athena} would provide unique information on the non-thermal components in SFRs.
\subsection{The Cherenkov Telescope Array}
\label{S:CTA}
Unveiling the physics of star-forming regions in terms of the interplay between cosmic ray production and star formation is one of the Key Science Projects (KSPs) of the upcoming {\sl Cherenkov Telescope Array} (CTA). 
CTA is the largest array of Imaging Cherenkov Telescopes ever conceived (see \cite{ctascience} for detailed information): it will be made of more than 100 Cherenkov telescopes of three different sizes, located in two different sites. The two sites, one in the northern hemisphere, on the island of La Palma, and one in the southern hemisphere, near Paranal, in Chile, are meant to guarantee full sky coverage, while the very large number of telescopes and the different telescope sizes are aimed at obtaining unprecedented sensitivity and an excellent energy and spatial resolution over a very wide range of photon energies. CTA will observe gamma-ray photons with energies between 20 GeV and 300 TeV, and over most of this energy interval, its sensitivity will be one order of magnitude better than existing facilities, which promises to increase by a factor of several the number of detected very high-energy gamma-ray sources. Such a jump in sensitivity will be paralleled by an analogous improvement in terms of spatial and spectral resolution. CTA angular and energy resolution will be better than $0.05^\circ$ and  better than 1\%, respectively. Excellent spatial identification and spectral characterisation are essential for the science performance of the instrument, which, for the first time at these energies, will be affected by source confusion. 

These qualities are also essential to make progress on the main subject of this article, namely assessing the non-thermal aspects of star formation. The topic has several aspects of the outmost scientific relevance and this is why it has been chosen as one of CTA KSPs: about 700 hours of observing time are expected to be devoted to SFRs, with the purpose of learning about the mutual feedback between cosmic ray acceleration and star formation. A number of questions are expected to be answered by CTA observations, as we discuss below.

\subsubsection{Unveiling the nature of the main sources of CRs in the Galaxy}
As mentioned earlier, in the past few years the suspicion has been growing in the CR community, that SNRs might not be the main sources of these particles in the Galaxy. Doubts about the long-standing paradigm that associates CRs of galactic origin to shock acceleration in the supernova blast wave mainly come from the difficulty at explaining energies as high as the CR {\it knee}, $\sim$ 1 PeV, as a result of this process. In order to reach very high energies in shock acceleration, efficient magnetic field amplification is needed  \citep [e.g.][]{blasi2013,amato14}. The most efficient mechanism to this purpose is commonly accepted to be generation of turbulence via the non-resonant streaming instability induced by the particles escaping the system \citep{Bell2004}. The resulting level of field amplification and, as a consequence, the resulting maximum achievable energy, depends on the efficiency of particle acceleration and on the particle spectrum \citep{SB13}. Recent evidence suggests that the latter must be steep: this inference comes both from gamma-ray observations of SNRs and from models of CR propagation through the Galaxy, supplied with the findings by AMS-02 \citep{AMS02BC16,AMS02sec18} about the slope of the diffusion coefficient (see e.g. \cite{2018AdSpR..62.2731A} for a review). Putting together all the pieces of our current knowledge about shock acceleration and magnetic field amplification PeV energies only seem achievable in rare explosions with an extraordinary energy release (e.g. \cite{cardillo15}). 

Parallel to the difficulties faced by isolated SNRs to act as PeVatrons, evidence is growing in favour of star-forming regions as powerful CR sources. First of all, the class of SNRs that might be able to reach PeV energies, if any, consists of remnants of very energetic type II explosions, expanding in the wind of the progenitor star \citep{cardillo15}, mostly located in SFRs. In addition, as amply discussed in the previous sections, the winds of young massive stars are themselves gaining credit as important sources of CRs and the colliding shock waves found in SFRs might be what is needed to enhance the efficiency of the acceleration process and increase the maximum achievable energy.

We earlier discussed (\S~\ref{section:overview}) how gamma-ray observations of three prominent SFRs have shown emission profiles that are consistent with ongoing particle acceleration. These are the OB association Cygnus OB2 (\S~\ref{section:OBstars}), the massive star cluster Wd1 (\S~\ref{section:GRYMSC}) and the Central Molecular Zone (CMZ) towards the Galactic Center (\S~\ref{section:CMZ}). Each of these regions contains a noticeable amount of mechanical energy available for conversion into particle acceleration, between few and several percent of the total energy released in the Galaxy by SN explosions. In addition, in all three cases, the detected gamma-ray spectrum is $\propto E^{-2.3}$, fully consistent with what we expect for CR injection in the ISM based on the CR spectrum detected at the Earth and the most updated modeling of CR propagation through the Galaxy. Finally, and maybe most importantly, no high-energy cut-off has so far been detected up to 100 TeV, suggesting the presence of PeV protons. 

In $\sim$ 2 yrs of observations, CTA will be able to fully constrain the high-energy spectrum of Cygnus OB2, providing an energy dependent morphology, and allowing us to disentangle the contribution of individual sources and determine the maximum particle energy in most of the individual objects \citep{ctascience}. The same is true for Wd1 and for the CMZ.

\subsubsection{Non-linearities in CR transport and CR feedback on star formation}
CTA will start operations at a time when recent and upcoming major facilities are expected to provide us with very high quality data at radio, millimeter and sub-millimeter wavelengths: while ALMA, APEX, IRAM will measure the ionization of molecular clouds due to CRs, SKA (and its pathfinders, such as ASKAP, before it) will provide the most detailed survey ever of the 3-dimensional distribution of HI and CO in SFRs. This information will be used to deconvolve gamma-ray data and obtain direct information on the distribution of CRs.

Especially interesting to this purpose is the case of Westerlund 1: within the first year of observation with CTA, we expect to be able to gather a much better insight in the physics of this object, constraining the acceleration and propagation of cosmic rays within the cluster and its surroundings, by high resolution spectro-morphological analysis of gamma-ray emission throughout an extended range of energies. In fact, the emissions detected by {\sl Fermi} and H.E.S.S. show only partial overlap in space. \cite{Wd1Ohm13} suggest complex transport processes at work and the possible existence of multiple contributing sources of different nature, including SNRs, stellar winds and Pulsar Wind Nebulae (PWNe). The presence in the region of bubbles of HI gas, that partly overlap with the TeV emission detected by H.E.S.S., provides a unique opportunity to study in detail CR propagation in the close vicinity of their acceleration sites. In particular this seems the ideal place where to test the effects of self-generated waves on CR transport. Indeed, a much reduced diffusion coefficient has been implied for the region based on available gamma-ray data. The suppression appears to be even larger (a factor $\sim$ 1000) than that deduced for the so-called TeV halos \citep{hawcgeminga} detected around old Pulsar Wind Nebulae (a factor $\sim$ 100). The study of this region with CTA has the potential to provide, for the first time, direct evidence of non-linear CR transport, lending support to a framework that changes CR propagation not only in the vicinity of their sources, but also on much larger scales, from propagation through the Galaxy \citep{12BABreaks}, to escape from the Galaxy \citep{19BAEscGal} and spreading into the Intergalactic medium \citep{15BAIGM}.

A region of the Galaxy that is not part of the CTA KSP on SFRs, but will however be studied in great detail is the Central Molecular Zone. This region is a bright gamma-ray source (see \S~\ref{section:CMZ}) and contains a large amount of molecular gas in addition to several potential CR sources, including SGR $A^*$ and the winds of young massive stars and SNRs in the Arches cluster. As mentioned above, also here current instruments detect a $E^{-2.3}$ spectrum with no hint of a high-energy cut-off, at least up to 20 TeV. CTA will provide spatially resolved maps of the region, allowing us to distinguish between emission associated to diffuse cosmic rays and specific features. In addition, spectro-morphological analysis of CTA data will provide information on the maximum energy produced by specific sources (stellar winds, SNRs, PWNe and SGR A East) and on the processes governing CR transport in the region: in particular this is the ideal place for the study of advection with the galactic wind and CR penetration in dense clouds. The latter phenomenon is especially important for its back reaction on star formation \citep{2015ARA&A..53..199G}.

\subsection{How CRs impact the overall properties of SFRs at all scales}
One final, extremely relevant question concerning the interplay between star formation and energetic particles is whether there are scaling relations that can be used to extrapolate what we learn locally from detailed observations of nearby systems, to much larger scales, those of star-forming galaxies, starburst galaxies and ULIRGs. CTA is expected to devote $\sim$ 500 h of observing time to address this question. 

Outside of the Milky Way, the only star-forming galaxy for which CTA will be able to resolve the high-energy source population is the LMC. This will be the subject of a long exposure which will also provide us with data on two star-forming regions outside our Galaxy: 30 Doradus, the most active star-forming region in the local group of galaxies, and the star cluster R136, hosting an exceptionally large number of massive O-type stars and Wolf-Rayet stars. The only other {\it normal} (from the point of view of star formation) galaxy with an expected TeV flux above CTA sensitivity is Andromeda, the nearest spiral galaxy, with a well-known star formation rate. The GeV spectrum of Andromeda is flatter than for the Milky Way \citep{2010FermiM31} and this might be a result of different dominant sources of cosmic rays or different transport. A long exposure with CTA should enable us to distinguish between large scale diffuse emission and emission associated with individual sources, providing us with essential clues on the properties of particle transport.

Moving towards higher star formation rates, CTA will observe the two starburst galaxies NGC253 and M82, potentially resolving their core. GeV and TeV emission has been observed from both galaxies by {\it Fermi} \citep{2010FermiM82} and VERITAS \citep{2009VERITASM82} respectively. In these galaxies, powerful winds, possibly driven by CRs themselves, are found to be the main responsible for particle transport at low energies, whereas transport could be predominantly diffusive for particles with energies above several tens of TeV: this could produce a spectral break detectable by CTA. Additional spectral features could be produced by individual sources. The combination of high spatial and spectral resolution could allow us to distinguish between the diffuse starburst emission and the contribution of individual sources, such as PWNe, which are particularly abundant in starburst galaxies \citep{Mannheim12,13OhmPWNe}.

Finally, CTA should observe, with a deep exposure, the ULIRG Arp220, where massive stars are continuously born and SN explode at a rate of 2 per year, as compared with the rate of $\sim$ 1 per century in the Milky Way. This would be the first TeV detection of a galaxy with a star formation rate as high as a few hundreds $M_\odot/yr$ and would put a data-point of fundamental importance on the plot representing the relation between gamma-ray emission and star formation rate.

\section{Summary} \label{sec:sum}
Non-thermal radiation, magnetic fields and cosmic rays 
can play an essential role in the star formation processes providing an important feedback in the galactic environment. Being a product of complicated plasma processes during a gravitational energy release in both protostellar objects and collapse of young massive stars, cosmic rays, their hard radiation and fluctuating magnetic fields provide gas ionization and redistribute the released energy on large scales in the star-forming sites.   

Theoretical models and some gamma-ray observations discussed above give evidences for the young massive star clusters (e.g. Westerlund 1, 2, Cl*1806–20, RSGC 1,2,3 etc)  to be efficient CR sources contributing  to galactic CRs up to $\gsim$ 100 PeV energy regime.     

The specific features of cosmic rays accelerated in young massive star clusters can help in understanding the long-standing questions on the sources of cosmic rays above PeV energies as well as the composition anomalies like the observed $^{22}$Ne/$^{20}$Ne isotopic ratio. The compact clusters of young massive stars can accelerate CRs with high efficiency  
with a quasi-stationary CR distribution around cluster over millions of years interrupted by relatively short periods of supernova explosions when the highest energy CRs can be accelerated.

Given the high fluxes of CRs from the clustered massive stars and supernovae, they may escape from their accelerators producing the regions of enhanced CR-driven turbulence in the vicinity of the sources. This effect may suppress the diffusion around the sources and build CR halos.

 While some aspects of CR origin and magnetic turbulence characteristics remain uncertain, non-thermal radiation and neutrinos produced in the star-forming regions can provide unique information also about the matter distribution in these regions. There are very bright prospects for studies of the non-thermal phenomena in star-forming regions with future multi-messenger observations from radio to gamma-ray energies, gravitational wave and neutrino observatories and we discussed above some of the relevant issues.



\begin{acknowledgements}
A.M.B., A.M. and J.M.D.K. thank the staff of ISSI for their generous hospitality and creating the fruitful cooperation. A.M.B. and M.E.K. were supported by the RSF grant 16-12-10225. E.A. acknowledges support by INAF and ASI through grants PRIN SKA-CTA 2016, INAF-MAINSTREAM 2018 and ASI/INAF n. 2017-14- H.O.
J.M.D.K. gratefully acknowledges funding from the German Research Foundation (DFG) in the form of an Emmy Noether Research Group (grant number KR4801/1-1) and a DFG Sachbeihilfe Grant (grant number KR4801/2-1), from the European Research Council (ERC) under the European Union's Horizon 2020 research and innovation programme via the ERC Starting Grant MUSTANG (grant agreement number 714907), and from Sonderforschungsbereich SFB 881 ``The Milky Way System'' (subproject B2) of the DFG. Some of the modeling was performed at the ``Tornado'' subsystem of the St.~Petersburg Polytechnic University supercomputing center and at the JSCC RAS. AB and EW acknowledge hospitality of the RAS Presidium at the workshop "High-energy processes in space objects: fundamental physics and new detection technolgies" (Moscow, 16-20 Sep 2019).
\end{acknowledgements}


\bibliographystyle{aps-nameyear3}

\begin{thebibliography}{227}
\ifx \bisbn   \undefined \def \bisbn  #1{ISBN #1}\fi
\ifx \binits  \undefined \def \binits#1{#1} \fi
\ifx \bauthor  \undefined \def \bauthor#1{#1} \fi
\ifx \bjtitle  \undefined \def \bjtitle#1{\textrm{#1}}\fi
\ifx \batitle  \undefined \def \batitle#1{#1} \fi
\ifx \bctitle  \undefined \def \bctitle#1{#1} \fi
\ifx \bvolume  \undefined \def \bvolume#1{\textbf{#1}}\fi
\ifx \byear  \undefined \def \byear#1{#1} \fi
\ifx \bissue  \undefined \def \bissue#1{#1} \fi
\ifx \bfpage  \undefined \def \bfpage#1{#1} \fi
\ifx \blpage  \undefined \def \blpage #1{#1} \fi
\ifx \burl  \undefined \def \burl#1{#1} \fi
\ifx \doiurl  \undefined \def \doiurl#1{#1} \fi
\ifx \betal  \undefined \def \betal{et al.} \fi
\ifx \binstitute  \undefined \def \binstitute#1{#1} \fi
\ifx \beditor  \undefined \def \beditor#1{#1} \fi
\ifx \bpublisher  \undefined \def \bpublisher#1{#1} \fi
\ifx \bbtitle  \undefined \def \bbtitle#1{\textit{#1}} \fi
\ifx \bedition  \undefined \def \bedition#1{#1} \fi
\ifx \bseriesno  \undefined \def \bseriesno#1{#1} \fi
\ifx \blocation  \undefined \def \blocation#1{#1} \fi
\ifx \bsertitle  \undefined \def \bsertitle#1{#1} \fi
\ifx \bsnm \undefined \def \bsnm#1{#1} \fi
\ifx \bsuffix \undefined \def \bsuffix#1{#1} \fi
\ifx \bparticle \undefined \def \bparticle#1{#1} \fi
\ifx \barticle \undefined \def \barticle#1{#1} \fi
\ifx \botherref \undefined \def \botherref #1{#1} \fi
\ifx \url \undefined \def \url#1{#1} \fi
\ifx \bchapter \undefined \def \bchapter#1{#1} \fi
\ifx \bbook \undefined \def \bbook#1{#1} \fi
\ifx \bcomment \undefined \def \bcomment#1{#1} \fi
\ifx \oauthor \undefined \def \oauthor#1{#1} \fi
\ifx \citeauthoryear \undefined \def \citeauthoryear#1{#1} \fi
\ifx \texttildelow  \undefined \def \texttildelow{\symbol{126}} \fi
\def \endbibitem {}
\ifx \bconflocation  \undefined \def \bconflocation#1{#1} \fi

\bibitem[\protect\citeauthoryear{{Aartsen} et~al.}{2019}]{IceCube7years}
\begin{barticle}
\bauthor{\binits{M.G.} \bsnm{{Aartsen}}},
\bauthor{\binits{M.} \bsnm{{Ackermann}}},
\bauthor{\binits{J.} \bsnm{{Adams}}}, \betal,
\batitle{{Search for Sources of Astrophysical Neutrinos Using Seven Years of
  IceCube Cascade Events}}.
\bjtitle{\apj}
\bvolume{886}(\bissue{1}),
\bfpage{12}
(\byear{2019}).
doi:\doiurl{10.3847/1538-4357/ab4ae2}
\end{barticle}
\endbibitem

\bibitem[\protect\citeauthoryear{{Abazajian} et~al.}{2014}]{abazajian14}
\begin{barticle}
\bauthor{\binits{K.N.} \bsnm{{Abazajian}}},
\bauthor{\binits{N.} \bsnm{{Canac}}},
\bauthor{\binits{S.} \bsnm{{Horiuchi}}}, \betal,
\batitle{{Astrophysical and dark matter interpretations of extended gamma-ray
  emission from the Galactic Center}}.
\bjtitle{\prd}
\bvolume{90}(\bissue{2}),
\bfpage{023526}
(\byear{2014}).
doi:\doiurl{10.1103/PhysRevD.90.023526}
\end{barticle}
\endbibitem

\bibitem[\protect\citeauthoryear{{Abdo} et~al.}{2010a}]{2010FermiM82}
\begin{barticle}
\bauthor{\binits{A.A.} \bsnm{{Abdo}}},
\bauthor{\binits{M.} \bsnm{{Ackermann}}},
\bauthor{\binits{M.} \bsnm{{Ajello}}}, \betal,
\batitle{{Detection of Gamma-Ray Emission from the Starburst Galaxies M82 and
  NGC 253 with the Large Area Telescope on Fermi}}.
\bjtitle{\apjl}
\bvolume{709}(\bissue{2}),
\bfpage{152}--\blpage{157}
(\byear{2010}a).
doi:\doiurl{10.1088/2041-8205/709/2/L152}
\end{barticle}
\endbibitem

\bibitem[\protect\citeauthoryear{{Abdo} et~al.}{2010b}]{2010FermiM31}
\begin{barticle}
\bauthor{\binits{A.A.} \bsnm{{Abdo}}},
\bauthor{\binits{M.} \bsnm{{Ackermann}}},
\bauthor{\binits{M.} \bsnm{{Ajello}}}, \betal,
\batitle{{Fermi Large Area Telescope observations of Local Group galaxies:
  detection of M 31 and search for M 33}}.
\bjtitle{\aap}
\bvolume{523},
\bfpage{2}
(\byear{2010}b).
doi:\doiurl{10.1051/0004-6361/201015759}
\end{barticle}
\endbibitem

\bibitem[\protect\citeauthoryear{{Abeysekara} et~al.}{2017}]{hawcgeminga}
\begin{barticle}
\bauthor{\binits{A.U.} \bsnm{{Abeysekara}}},
\bauthor{\binits{A.} \bsnm{{Albert}}},
\bauthor{\binits{R.} \bsnm{{Alfaro}}}, \betal,
\batitle{{Extended gamma-ray sources around pulsars constrain the origin of the
  positron flux at Earth}}.
\bjtitle{Science}
\bvolume{358}(\bissue{6365}),
\bfpage{911}--\blpage{914}
(\byear{2017}).
doi:\doiurl{10.1126/science.aan4880}
\end{barticle}
\endbibitem

\bibitem[\protect\citeauthoryear{{Acero} et~al.}{2009}]{2009Sci...326.1080A}
\begin{barticle}
\bauthor{\binits{F.} \bsnm{{Acero}}},
\bauthor{\binits{F.} \bsnm{{Aharonian}}},
\bauthor{\binits{A.G.} \bsnm{{Akhperjanian}}}, \betal,
\batitle{{Detection of Gamma Rays from a Starburst Galaxy}}.
\bjtitle{Science}
\bvolume{326}(\bissue{5956}),
\bfpage{1080}
(\byear{2009}).
doi:\doiurl{10.1126/science.1178826}
\end{barticle}
\endbibitem

\bibitem[\protect\citeauthoryear{{Ackermann} et~al.}{2011}]{AckermannSB2011}
\begin{barticle}
\bauthor{\binits{M.} \bsnm{{Ackermann}}},
\bauthor{\binits{M.} \bsnm{{Ajello}}},
\bauthor{\binits{A.} \bsnm{{Allafort}}}, \betal,
\batitle{{A Cocoon of Freshly Accelerated Cosmic Rays Detected by Fermi in the
  Cygnus Superbubble}}.
\bjtitle{Science}
\bvolume{334},
\bfpage{1103}
(\byear{2011}).
doi:\doiurl{10.1126/science.1210311}
\end{barticle}
\endbibitem

\bibitem[\protect\citeauthoryear{{Ackermann}
  et~al.}{2012}]{2012ApJ...755..164A}
\begin{barticle}
\bauthor{\binits{M.} \bsnm{{Ackermann}}},
\bauthor{\binits{M.} \bsnm{{Ajello}}},
\bauthor{\binits{A.} \bsnm{{Allafort}}}, \betal,
\batitle{{GeV Observations of Star-forming Galaxies with the Fermi Large Area
  Telescope}}.
\bjtitle{\apj}
\bvolume{755}(\bissue{2}),
\bfpage{164}
(\byear{2012}).
doi:\doiurl{10.1088/0004-637X/755/2/164}
\end{barticle}
\endbibitem

\bibitem[\protect\citeauthoryear{{Adamo} et~al.}{2015}]{adamo15}
\begin{barticle}
\bauthor{\binits{A.} \bsnm{{Adamo}}},
\bauthor{\binits{J.M.D.} \bsnm{{Kruijssen}}},
\bauthor{\binits{N.} \bsnm{{Bastian}}}, \betal,
\batitle{{Probing the role of the galactic environment in the formation of
  stellar clusters, using M83 as a test bench}}.
\bjtitle{\mnras}
\bvolume{452},
\bfpage{246}--\blpage{260}
(\byear{2015}).
doi:\doiurl{10.1093/mnras/stv1203}
\end{barticle}
\endbibitem

\bibitem[\protect\citeauthoryear{{Adamo} and
  {Bastian}}{2018}]{2018ASSL..424...91A}
\begin{bchapter}
\bauthor{\binits{A.} \bsnm{{Adamo}}},
\bauthor{\binits{N.} \bsnm{{Bastian}}},
\bctitle{{The Lifecycle of Clusters in Galaxies}},
in \bbtitle{Astrophysics and Space Science Library},
vol. \bseriesno{424},
ed. by \beditor{\binits{S.} \bsnm{{Stahler}}}
\byear{2018},
p. \bfpage{91}.

\end{bchapter}
\endbibitem

\bibitem[\protect\citeauthoryear{{Aguilar} et~al.}{2016}]{AMS02BC16}
\begin{barticle}
\bauthor{\binits{M.} \bsnm{{Aguilar}}},
\bauthor{\binits{L.} \bsnm{{Ali Cavasonza}}},
\bauthor{\binits{G.} \bsnm{{Ambrosi}}}, \betal,
\batitle{{Precision Measurement of the Boron to Carbon Flux Ratio in Cosmic
  Rays from 1.9 GV to 2.6 TV with the Alpha Magnetic Spectrometer on the
  International Space Station}}.
\bjtitle{\prl}
\bvolume{117}(\bissue{23}),
\bfpage{231102}
(\byear{2016}).
doi:\doiurl{10.1103/PhysRevLett.117.231102}
\end{barticle}
\endbibitem

\bibitem[\protect\citeauthoryear{{Aguilar} et~al.}{2018}]{AMS02sec18}
\begin{barticle}
\bauthor{\binits{M.} \bsnm{{Aguilar}}},
\bauthor{\binits{L.} \bsnm{{Ali Cavasonza}}},
\bauthor{\binits{G.} \bsnm{{Ambrosi}}}, \betal,
\batitle{{Observation of New Properties of Secondary Cosmic Rays Lithium,
  Beryllium, and Boron by the Alpha Magnetic Spectrometer on the International
  Space Station}}.
\bjtitle{\prl}
\bvolume{120}(\bissue{2}),
\bfpage{021101}
(\byear{2018}).
doi:\doiurl{10.1103/PhysRevLett.120.021101}
\end{barticle}
\endbibitem

\bibitem[\protect\citeauthoryear{{Aharonian} and
  {Neronov}}{2005}]{Aharonian2005}
\begin{barticle}
\bauthor{\binits{F.} \bsnm{{Aharonian}}},
\bauthor{\binits{A.} \bsnm{{Neronov}}},
\batitle{{High-Energy Gamma Rays from the Massive Black Hole in the Galactic
  Center}}.
\bjtitle{\apj}
\bvolume{619},
\bfpage{306}--\blpage{313}
(\byear{2005}).
doi:\doiurl{10.1086/426426}
\end{barticle}
\endbibitem

\bibitem[\protect\citeauthoryear{{Aharonian} et~al.}{2019}]{Aharonian2019}
\begin{barticle}
\bauthor{\binits{F.} \bsnm{{Aharonian}}},
\bauthor{\binits{R.} \bsnm{{Yang}}},
\bauthor{\binits{E.} \bsnm{{de O{\~n}a Wilhelmi}}},
\batitle{{Massive stars as major factories of Galactic cosmic rays}}.
\bjtitle{Nature Astronomy}
(\byear{2019}).
doi:\doiurl{10.1038/s41550-019-0724-0}
\end{barticle}
\endbibitem

\bibitem[\protect\citeauthoryear{{Aharonian} et~al.}{2002}]{AharonianEtal2002}
\begin{barticle}
\bauthor{\binits{F.} \bsnm{{Aharonian}}},
\bauthor{\binits{A.} \bsnm{{Akhperjanian}}},
\bauthor{\binits{M.} \bsnm{{Beilicke}}}, \betal,
\batitle{{An unidentified TeV source in the vicinity of Cygnus OB2}}.
\bjtitle{\aap}
\bvolume{393},
\bfpage{37}--\blpage{40}
(\byear{2002}).
doi:\doiurl{10.1051/0004-6361:20021171}
\end{barticle}
\endbibitem

\bibitem[\protect\citeauthoryear{{Aharonian} et~al.}{2006}]{aharonian06}
\begin{barticle}
\bauthor{\binits{F.} \bsnm{{Aharonian}}},
\bauthor{\binits{A.G.} \bsnm{{Akhperjanian}}},
\bauthor{\binits{A.R.} \bsnm{{Bazer-Bachi}}}, \betal,
\batitle{{Discovery of very-high-energy {\ensuremath{\gamma}}-rays from the
  Galactic Centre ridge}}.
\bjtitle{\nat}
\bvolume{439}(\bissue{7077}),
\bfpage{695}--\blpage{698}
(\byear{2006}).
doi:\doiurl{10.1038/nature04467}
\end{barticle}
\endbibitem

\bibitem[\protect\citeauthoryear{{Ahlers} and
  {Halzen}}{2014}]{2014PhRvD..90d3005A}
\begin{barticle}
\bauthor{\binits{M.} \bsnm{{Ahlers}}},
\bauthor{\binits{F.} \bsnm{{Halzen}}},
\batitle{{Pinpointing extragalactic neutrino sources in light of recent IceCube
  observations}}.
\bjtitle{\prd}
\bvolume{90}(\bissue{4}),
\bfpage{043005}
(\byear{2014}).
doi:\doiurl{10.1103/PhysRevD.90.043005}
\end{barticle}
\endbibitem

\bibitem[\protect\citeauthoryear{{Alves} and
  {Bouy}}{2012}]{2012A&A...547A..97A}
\begin{barticle}
\bauthor{\binits{J.} \bsnm{{Alves}}},
\bauthor{\binits{H.} \bsnm{{Bouy}}},
\batitle{{Orion revisited. I. The massive cluster in front of the Orion nebula
  cluster}}.
\bjtitle{\aap}
\bvolume{547},
\bfpage{97}
(\byear{2012}).
doi:\doiurl{10.1051/0004-6361/201220119}
\end{barticle}
\endbibitem

\bibitem[\protect\citeauthoryear{{Amato}}{2014}]{amato14}
\begin{barticle}
\bauthor{\binits{E.} \bsnm{{Amato}}},
\batitle{{The origin of galactic cosmic rays}}.
\bjtitle{International Journal of Modern Physics D}
\bvolume{23},
\bfpage{30013}
(\byear{2014}).
doi:\doiurl{10.1142/S0218271814300134}
\end{barticle}
\endbibitem

\bibitem[\protect\citeauthoryear{{Amato} and
  {Blasi}}{2018}]{2018AdSpR..62.2731A}
\begin{barticle}
\bauthor{\binits{E.} \bsnm{{Amato}}},
\bauthor{\binits{P.} \bsnm{{Blasi}}},
\batitle{{Cosmic ray transport in the Galaxy: A review}}.
\bjtitle{\asr}
\bvolume{62}(\bissue{10}),
\bfpage{2731}--\blpage{2749}
(\byear{2018}).
doi:\doiurl{10.1016/j.asr.2017.04.019}
\end{barticle}
\endbibitem

\bibitem[\protect\citeauthoryear{Ammazzalorso
  et~al.}{2020}]{PhysRevLett.124.101102}
\begin{barticle}
\bauthor{\binits{S.} \bsnm{Ammazzalorso}},
\bauthor{\binits{D.} \bsnm{Gruen}},
\bauthor{\binits{M.} \bsnm{Regis}}, \betal,
\batitle{Detection of cross-correlation between gravitational lensing and
  $\ensuremath{\gamma}$ rays}.
\bjtitle{Phys. Rev. Lett.}
\bvolume{124},
\bfpage{101102}
(\byear{2020}).
doi:\doiurl{10.1103/PhysRevLett.124.101102}.
\burl{https://link.aps.org/doi/10.1103/PhysRevLett.124.101102}
\end{barticle}
\endbibitem

\bibitem[\protect\citeauthoryear{Anchordoqui}{2018}]{PhysRevD.97.063010}
\begin{barticle}
\bauthor{\binits{L.A.} \bsnm{Anchordoqui}},
\batitle{Acceleration of ultrahigh-energy cosmic rays in starburst superwinds}.
\bjtitle{Phys. Rev. D}
\bvolume{97},
\bfpage{063010}
(\byear{2018}).
doi:\doiurl{10.1103/PhysRevD.97.063010}.
\burl{https://link.aps.org/doi/10.1103/PhysRevD.97.063010}
\end{barticle}
\endbibitem

\bibitem[\protect\citeauthoryear{{Ao} et~al.}{2013}]{ao13}
\begin{barticle}
\bauthor{\binits{Y.} \bsnm{{Ao}}},
\bauthor{\binits{C.} \bsnm{{Henkel}}},
\bauthor{\binits{K.M.} \bsnm{{Menten}}}, \betal,
\batitle{{The thermal state of molecular clouds in the Galactic center:
  evidence for non-photon-driven heating}}.
\bjtitle{\aap}
\bvolume{550},
\bfpage{135}
(\byear{2013}).
doi:\doiurl{10.1051/0004-6361/201220096}
\end{barticle}
\endbibitem

\bibitem[\protect\citeauthoryear{{Axford}}{1981}]{Axford81}
\begin{bchapter}
\bauthor{\binits{W.I.} \bsnm{{Axford}}},
\bctitle{{The acceleration of galactic cosmic rays}},
in \bbtitle{Origin of Cosmic Rays},
ed. by \beditor{\binits{G.} \bsnm{{Setti}}},
\beditor{\binits{G.} \bsnm{{Spada}}},
\beditor{\binits{A.W.} \bsnm{{Wolfendale}}}
\bsertitle{IAU Symposium},
vol. \bseriesno{94},
\byear{1981},
pp. \bfpage{339}--\blpage{358}
\end{bchapter}
\endbibitem

\bibitem[\protect\citeauthoryear{{Axford} et~al.}{1977}]{1977ICRC...11..132A}
\begin{bchapter}
\bauthor{\binits{W.I.} \bsnm{{Axford}}},
\bauthor{\binits{E.} \bsnm{{Leer}}},
\bauthor{\binits{G.} \bsnm{{Skadron}}},
\bctitle{{The Acceleration of Cosmic Rays by Shock Waves}},
in \bbtitle{International Cosmic Ray Conference}.
\bsertitle{International Cosmic Ray Conference},
vol. \bseriesno{11},
\byear{1977},
p. \bfpage{132}
\end{bchapter}
\endbibitem

\bibitem[\protect\citeauthoryear{Baade and Zwicky}{1934}]{BZ34}
\begin{barticle}
\bauthor{\binits{W.} \bsnm{Baade}},
\bauthor{\binits{F.} \bsnm{Zwicky}},
\batitle{Remarks on super-novae and cosmic rays}.
\bjtitle{Phys. Rev.}
\bvolume{46},
\bfpage{76}--\blpage{77}
(\byear{1934}).
doi:\doiurl{10.1103/PhysRev.46.76.2}.
\burl{http://link.aps.org/doi/10.1103/PhysRev.46.76.2}
\end{barticle}
\endbibitem

\bibitem[\protect\citeauthoryear{{Bai} et~al.}{2019}]{2019arXiv190502773B}
\begin{botherref}
\oauthor{\binits{X.} \bsnm{{Bai}}},
\oauthor{\binits{B.Y.} \bsnm{{Bi}}},
\oauthor{\binits{X.J.} \bsnm{{Bi}}}, et al.,
{The Large High Altitude Air Shower Observatory (LHAASO) Science White Paper}.
arXiv e-prints,
1905--02773
(2019)
\end{botherref}
\endbibitem

\bibitem[\protect\citeauthoryear{{Barnes} et~al.}{2017}]{barnes17}
\begin{barticle}
\bauthor{\binits{A.T.} \bsnm{{Barnes}}},
\bauthor{\binits{S.N.} \bsnm{{Longmore}}},
\bauthor{\binits{C.} \bsnm{{Battersby}}}, \betal,
\batitle{{Star formation rates and efficiencies in the Galactic Centre}}.
\bjtitle{\mnras}
\bvolume{469},
\bfpage{2263}--\blpage{2285}
(\byear{2017}).
doi:\doiurl{10.1093/mnras/stx941}
\end{barticle}
\endbibitem

\bibitem[\protect\citeauthoryear{{Bartels} et~al.}{2016}]{bartels16}
\begin{barticle}
\bauthor{\binits{R.} \bsnm{{Bartels}}},
\bauthor{\binits{S.} \bsnm{{Krishnamurthy}}},
\bauthor{\binits{C.} \bsnm{{Weniger}}},
\batitle{{Strong Support for the Millisecond Pulsar Origin of the Galactic
  Center GeV Excess}}.
\bjtitle{\prl}
\bvolume{116}(\bissue{5}),
\bfpage{051102}
(\byear{2016}).
doi:\doiurl{10.1103/PhysRevLett.116.051102}
\end{barticle}
\endbibitem

\bibitem[\protect\citeauthoryear{{Bell}}{1978a}]{Bell1978b}
\begin{barticle}
\bauthor{\binits{A.R.} \bsnm{{Bell}}},
\batitle{{The acceleration of cosmic rays in shock fronts - II.}}
\bjtitle{\mnras}
\bvolume{182},
\bfpage{443}--\blpage{455}
(\byear{1978}a).
doi:\doiurl{10.1093/mnras/182.3.443}
\end{barticle}
\endbibitem

\bibitem[\protect\citeauthoryear{{Bell}}{1978b}]{Bell78a}
\begin{barticle}
\bauthor{\binits{A.R.} \bsnm{{Bell}}},
\batitle{{The acceleration of cosmic rays in shock fronts. I}}.
\bjtitle{\mnras}
\bvolume{182},
\bfpage{147}--\blpage{156}
(\byear{1978}b)
\end{barticle}
\endbibitem

\bibitem[\protect\citeauthoryear{{Bell}}{2004}]{Bell2004}
\begin{barticle}
\bauthor{\binits{A.R.} \bsnm{{Bell}}},
\batitle{{Turbulent amplification of magnetic field and diffusive shock
  acceleration of cosmic rays}}.
\bjtitle{\mnras}
\bvolume{353},
\bfpage{550}--\blpage{558}
(\byear{2004})
\end{barticle}
\endbibitem

\bibitem[\protect\citeauthoryear{{Bell} et~al.}{2013}]{BellEtal2013}
\begin{barticle}
\bauthor{\binits{A.R.} \bsnm{{Bell}}},
\bauthor{\binits{K.M.} \bsnm{{Schure}}},
\bauthor{\binits{B.} \bsnm{{Reville}}}, \betal,
\batitle{{Cosmic-ray acceleration and escape from supernova remnants}}.
\bjtitle{\mnras}
\bvolume{431},
\bfpage{415}--\blpage{429}
(\byear{2013})
\end{barticle}
\endbibitem

\bibitem[\protect\citeauthoryear{{Berezhko} et~al.}{1994}]{Berezhko1994}
\begin{barticle}
\bauthor{\binits{E.G.} \bsnm{{Berezhko}}},
\bauthor{\binits{V.K.} \bsnm{{Yelshin}}},
\bauthor{\binits{L.T.} \bsnm{{Ksenofontov}}},
\batitle{{Numerical investigation of cosmic ray acceleration in supernova
  remnants}}.
\bjtitle{\aspp}
\bvolume{2}(\bissue{2}),
\bfpage{215}--\blpage{227}
(\byear{1994}).
doi:\doiurl{10.1016/0927-6505(94)90043-4}
\end{barticle}
\endbibitem

\bibitem[\protect\citeauthoryear{{Berezinskii} et~al.}{1990}]{Berezinski90}
\begin{bbook}
\bauthor{\binits{V.S.} \bsnm{{Berezinskii}}},
\bauthor{\binits{S.V.} \bsnm{{Bulanov}}},
\bauthor{\binits{V.A.} \bsnm{{Dogiel}}}, \betal,
\bbtitle{{Astrophysics of cosmic rays, North-Holland, Amsterdam}}
\byear{1990}
\end{bbook}
\endbibitem

\bibitem[\protect\citeauthoryear{{Berlanas} et~al.}{2018}]{2018A&A...612A..50B}
\begin{barticle}
\bauthor{\binits{S.R.} \bsnm{{Berlanas}}},
\bauthor{\binits{A.} \bsnm{{Herrero}}},
\bauthor{\binits{F.} \bsnm{{Comer{\'o}n}}}, \betal,
\batitle{{New massive members of Cygnus OB2}}.
\bjtitle{\aap}
\bvolume{612},
\bfpage{50}
(\byear{2018}).
doi:\doiurl{10.1051/0004-6361/201731856}
\end{barticle}
\endbibitem

\bibitem[\protect\citeauthoryear{{Berlanas} et~al.}{2019}]{2019MNRAS.484.1838B}
\begin{barticle}
\bauthor{\binits{S.R.} \bsnm{{Berlanas}}},
\bauthor{\binits{N.J.} \bsnm{{Wright}}},
\bauthor{\binits{A.} \bsnm{{Herrero}}}, \betal,
\batitle{{Disentangling the spatial substructure of Cygnus OB2 from Gaia DR2}}.
\bjtitle{\mnras}
\bvolume{484}(\bissue{2}),
\bfpage{1838}--\blpage{1842}
(\byear{2019}).
doi:\doiurl{10.1093/mnras/stz117}
\end{barticle}
\endbibitem

\bibitem[\protect\citeauthoryear{{Binns} et~al.}{2005}]{Binns2005}
\begin{barticle}
\bauthor{\binits{W.R.} \bsnm{{Binns}}},
\bauthor{\binits{M.E.} \bsnm{{Wiedenbeck}}},
\bauthor{\binits{M.} \bsnm{{Arnould}}}, \betal,
\batitle{{Cosmic-Ray Neon, Wolf-Rayet Stars, and the Superbubble Origin of
  Galactic Cosmic Rays}}.
\bjtitle{\apj}
\bvolume{634}(\bissue{1}),
\bfpage{351}--\blpage{364}
(\byear{2005}).
doi:\doiurl{10.1086/496959}
\end{barticle}
\endbibitem

\bibitem[\protect\citeauthoryear{{Blandford} and {Eichler}}{1987}]{BE87}
\begin{barticle}
\bauthor{\binits{R.} \bsnm{{Blandford}}},
\bauthor{\binits{D.} \bsnm{{Eichler}}},
\batitle{{Particle acceleration at astrophysical shocks: A theory of cosmic ray
  origin}}.
\bjtitle{Phys. Rep.}
\bvolume{154},
\bfpage{1}--\blpage{75}
(\byear{1987})
\end{barticle}
\endbibitem

\bibitem[\protect\citeauthoryear{{Blandford} and
  {Ostriker}}{1978}]{Blandford1978}
\begin{barticle}
\bauthor{\binits{R.D.} \bsnm{{Blandford}}},
\bauthor{\binits{J.P.} \bsnm{{Ostriker}}},
\batitle{{Particle acceleration by astrophysical shocks.}}
\bjtitle{\apjl}
\bvolume{221},
\bfpage{29}--\blpage{32}
(\byear{1978}).
doi:\doiurl{10.1086/182658}
\end{barticle}
\endbibitem

\bibitem[\protect\citeauthoryear{{Blandford} and {Funk}}{2007}]{Blandford2007}
\begin{bchapter}
\bauthor{\binits{R.} \bsnm{{Blandford}}},
\bauthor{\binits{S.} \bsnm{{Funk}}},
\bctitle{{The Magnetic Bootstrap}},
in \bbtitle{The First GLAST Symposium},
ed. by \beditor{\binits{S.} \bsnm{{Ritz}}},
\beditor{\binits{P.} \bsnm{{Michelson}}},
\beditor{\binits{C.A.} \bsnm{{Meegan}}}
\bsertitle{American Institute of Physics Conference Series},
vol. \bseriesno{921},
\byear{2007},
pp. \bfpage{62}--\blpage{64}.
doi:\doiurl{10.1063/1.2757268}
\end{bchapter}
\endbibitem

\bibitem[\protect\citeauthoryear{{Blasi}}{2013}]{blasi2013}
\begin{barticle}
\bauthor{\binits{P.} \bsnm{{Blasi}}},
\batitle{{The origin of galactic cosmic rays}}.
\bjtitle{\aapr}
\bvolume{21},
\bfpage{70}
(\byear{2013}).
doi:\doiurl{10.1007/s00159-013-0070-7}
\end{barticle}
\endbibitem

\bibitem[\protect\citeauthoryear{{Blasi} and {Amato}}{2019}]{19BAEscGal}
\begin{barticle}
\bauthor{\binits{P.} \bsnm{{Blasi}}},
\bauthor{\binits{E.} \bsnm{{Amato}}},
\batitle{{Escape of Cosmic Rays from the Galaxy and Effects on the
  Circumgalactic Medium}}.
\bjtitle{\prl}
\bvolume{122}(\bissue{5}),
\bfpage{051101}
(\byear{2019}).
doi:\doiurl{10.1103/PhysRevLett.122.051101}
\end{barticle}
\endbibitem

\bibitem[\protect\citeauthoryear{{Blasi} et~al.}{2015}]{15BAIGM}
\begin{barticle}
\bauthor{\binits{P.} \bsnm{{Blasi}}},
\bauthor{\binits{E.} \bsnm{{Amato}}},
\bauthor{\binits{M.} \bsnm{{D'Angelo}}},
\batitle{{High-Energy Cosmic Ray Self-Confinement Close to Extra-Galactic
  Sources}}.
\bjtitle{\prl}
\bvolume{115}(\bissue{12}),
\bfpage{121101}
(\byear{2015}).
doi:\doiurl{10.1103/PhysRevLett.115.121101}
\end{barticle}
\endbibitem

\bibitem[\protect\citeauthoryear{{Blasi} et~al.}{2012}]{12BABreaks}
\begin{barticle}
\bauthor{\binits{P.} \bsnm{{Blasi}}},
\bauthor{\binits{E.} \bsnm{{Amato}}},
\bauthor{\binits{P.D.} \bsnm{{Serpico}}},
\batitle{{Spectral Breaks as a Signature of Cosmic Ray Induced Turbulence in
  the Galaxy}}.
\bjtitle{\prl}
\bvolume{109}(\bissue{6}),
\bfpage{061101}
(\byear{2012}).
doi:\doiurl{10.1103/PhysRevLett.109.061101}
\end{barticle}
\endbibitem

\bibitem[\protect\citeauthoryear{{Bouy} and
  {Alves}}{2015}]{2015A&A...584A..26B}
\begin{barticle}
\bauthor{\binits{H.} \bsnm{{Bouy}}},
\bauthor{\binits{J.} \bsnm{{Alves}}},
\batitle{{Cosmography of OB stars in the solar neighbourhood}}.
\bjtitle{\aap}
\bvolume{584},
\bfpage{26}
(\byear{2015}).
doi:\doiurl{10.1051/0004-6361/201527058}
\end{barticle}
\endbibitem

\bibitem[\protect\citeauthoryear{{Brahimi} et~al.}{2020}]{Brahimi20}
\begin{barticle}
\bauthor{\binits{L.} \bsnm{{Brahimi}}},
\bauthor{\binits{A.} \bsnm{{Marcowith}}},
\bauthor{\binits{V.S.} \bsnm{{Ptuskin}}},
\batitle{{Nonlinear diffusion of cosmic rays escaping from supernova remnants:
  Cold partially neutral atomic and molecular phases}}.
\bjtitle{\aap}
\bvolume{633},
\bfpage{72}
(\byear{2020}).
doi:\doiurl{10.1051/0004-6361/201936166}
\end{barticle}
\endbibitem

\bibitem[\protect\citeauthoryear{{Brandt} and {Kocsis}}{2015}]{brandt15}
\begin{barticle}
\bauthor{\binits{T.D.} \bsnm{{Brandt}}},
\bauthor{\binits{B.} \bsnm{{Kocsis}}},
\batitle{{Disrupted Globular Clusters Can Explain the Galactic Center Gamma-Ray
  Excess}}.
\bjtitle{\apj}
\bvolume{812}(\bissue{1}),
\bfpage{15}
(\byear{2015}).
doi:\doiurl{10.1088/0004-637X/812/1/15}
\end{barticle}
\endbibitem

\bibitem[\protect\citeauthoryear{{Breitschwerdt}
  et~al.}{1991}]{1991A&A...245...79B}
\begin{barticle}
\bauthor{\binits{D.} \bsnm{{Breitschwerdt}}},
\bauthor{\binits{J.F.} \bsnm{{McKenzie}}},
\bauthor{\binits{H.J.} \bsnm{{Voelk}}},
\batitle{{Galactic winds. I. Cosmic ray and wave-driven winds from the
  galaxy.}}
\bjtitle{\aap}
\bvolume{245},
\bfpage{79}
(\byear{1991})
\end{barticle}
\endbibitem

\bibitem[\protect\citeauthoryear{{Bruno} and
  {Carbone}}{2016}]{2016LNP...928.....B}
\begin{bbook}
\bauthor{\binits{R.} \bsnm{{Bruno}}},
\bauthor{\binits{V.} \bsnm{{Carbone}}},
\bbtitle{{Turbulence in the Solar Wind}},
vol. \bseriesno{928}
\byear{2016}.
doi:\doiurl{10.1007/978-3-319-43440-7}
\end{bbook}
\endbibitem

\bibitem[\protect\citeauthoryear{{Bykov}}{2001}]{Bykov2001}
\begin{barticle}
\bauthor{\binits{A.M.} \bsnm{{Bykov}}},
\batitle{{Particle Acceleration and Nonthermal Phenomena in Superbubbles}}.
\bjtitle{\ssr}
\bvolume{99},
\bfpage{317}--\blpage{326}
(\byear{2001})
\end{barticle}
\endbibitem

\bibitem[\protect\citeauthoryear{{Bykov}}{2014}]{Bykov2014}
\begin{barticle}
\bauthor{\binits{A.M.} \bsnm{{Bykov}}},
\batitle{{Nonthermal particles and photons in starburst regions and
  superbubbles}}.
\bjtitle{\aapr}
\bvolume{22},
\bfpage{77}
(\byear{2014}).
doi:\doiurl{10.1007/s00159-014-0077-8}
\end{barticle}
\endbibitem

\bibitem[\protect\citeauthoryear{{Bykov} and {Fleishman}}{1992}]{Bykov92}
\begin{barticle}
\bauthor{\binits{A.M.} \bsnm{{Bykov}}},
\bauthor{\binits{G.D.} \bsnm{{Fleishman}}},
\batitle{{On non-thermal particle generation in superbubbles.}}
\bjtitle{\mnras}
\bvolume{255},
\bfpage{269}--\blpage{275}
(\byear{1992}).
doi:\doiurl{10.1093/mnras/255.2.269}
\end{barticle}
\endbibitem

\bibitem[\protect\citeauthoryear{{Bykov} and
  {Toptygin}}{1987}]{1987Ap&SS.138..341B}
\begin{barticle}
\bauthor{\binits{A.M.} \bsnm{{Bykov}}},
\bauthor{\binits{I.N.} \bsnm{{Toptygin}}},
\batitle{{Effect of Shocks on Interstellar Turbulence and Cosmic-Ray
  Dynamics}}.
\bjtitle{\apss}
\bvolume{138}(\bissue{2}),
\bfpage{341}--\blpage{354}
(\byear{1987}).
doi:\doiurl{10.1007/BF00637855}
\end{barticle}
\endbibitem

\bibitem[\protect\citeauthoryear{{Bykov} and {Toptygin}}{1993}]{BT93}
\begin{barticle}
\bauthor{\binits{A.M.} \bsnm{{Bykov}}},
\bauthor{\binits{I.N.} \bsnm{{Toptygin}}},
\batitle{{Particle kinetics in highly turbulent plasmas (renormalization and
  self-consistent field methods)}}.
\bjtitle{Phys. Usp.}
\bvolume{36},
\bfpage{1020}--\blpage{1052}
(\byear{1993})
\end{barticle}
\endbibitem

\bibitem[\protect\citeauthoryear{{Bykov} and
  {Toptygin}}{2001}]{2001AstL...27..625B}
\begin{barticle}
\bauthor{\binits{A.M.} \bsnm{{Bykov}}},
\bauthor{\binits{I.N.} \bsnm{{Toptygin}}},
\batitle{{A Model of Particle Acceleration to High Energies by Multiple
  Supernova Explosions in OB Associations}}.
\bjtitle{Astronomy Letters}
\bvolume{27}(\bissue{10}),
\bfpage{625}--\blpage{633}
(\byear{2001}).
doi:\doiurl{10.1134/1.1404456}
\end{barticle}
\endbibitem

\bibitem[\protect\citeauthoryear{{Bykov} et~al.}{2013}]{bbmo13}
\begin{barticle}
\bauthor{\binits{A.M.} \bsnm{{Bykov}}},
\bauthor{\binits{A.} \bsnm{{Brandenburg}}},
\bauthor{\binits{M.A.} \bsnm{{Malkov}}}, \betal,
\batitle{{Microphysics of Cosmic Ray Driven Plasma Instabilities}}.
\bjtitle{\ssr}
\bvolume{178},
\bfpage{201}--\blpage{232}
(\byear{2013}).
doi:\doiurl{10.1007/s11214-013-9988-3}
\end{barticle}
\endbibitem

\bibitem[\protect\citeauthoryear{{Bykov} et~al.}{2015}]{BEGO2015MNRAS}
\begin{barticle}
\bauthor{\binits{A.M.} \bsnm{{Bykov}}},
\bauthor{\binits{D.C.} \bsnm{{Ellison}}},
\bauthor{\binits{P.E.} \bsnm{{Gladilin}}}, \betal,
\batitle{{Ultrahard spectra of PeV neutrinos from supernovae in compact star
  clusters}}.
\bjtitle{\mnras}
\bvolume{453},
\bfpage{113}--\blpage{121}
(\byear{2015}).
doi:\doiurl{10.1093/mnras/stv1606}
\end{barticle}
\endbibitem

\bibitem[\protect\citeauthoryear{{Bykov} et~al.}{2018}]{2018SSRv..214...41B}
\begin{barticle}
\bauthor{\binits{A.M.} \bsnm{{Bykov}}},
\bauthor{\binits{D.C.} \bsnm{{Ellison}}},
\bauthor{\binits{A.} \bsnm{{Marcowith}}}, \betal,
\batitle{{Cosmic Ray Production in Supernovae}}.
\bjtitle{\ssr}
\bvolume{214}(\bissue{1}),
\bfpage{41}
(\byear{2018}).
doi:\doiurl{10.1007/s11214-018-0479-4}
\end{barticle}
\endbibitem

\bibitem[\protect\citeauthoryear{{Bykov} et~al.}{2019}]{Bykov2019}
\begin{barticle}
\bauthor{\binits{A.M.} \bsnm{{Bykov}}},
\bauthor{\binits{M.E.} \bsnm{{Kalyashova}}},
\bauthor{\binits{D.C.} \bsnm{{Ellison}}}, \betal,
\batitle{{High-energy cosmic rays from compact galactic star clusters: Particle
  fluxes and anisotropy}}.
\bjtitle{\asr}
\bvolume{64}(\bissue{12}),
\bfpage{2439}--\blpage{2444}
(\byear{2019}).
doi:\doiurl{10.1016/j.asr.2019.06.005}
\end{barticle}
\endbibitem

\bibitem[\protect\citeauthoryear{{Cardillo} et~al.}{2015}]{cardillo15}
\begin{barticle}
\bauthor{\binits{M.} \bsnm{{Cardillo}}},
\bauthor{\binits{E.} \bsnm{{Amato}}},
\bauthor{\binits{P.} \bsnm{{Blasi}}},
\batitle{{On the cosmic ray spectrum from type II supernovae expanding in their
  red giant presupernova wind}}.
\bjtitle{Astroparticle Physics}
\bvolume{69},
\bfpage{1}--\blpage{10}
(\byear{2015}).
doi:\doiurl{10.1016/j.astropartphys.2015.03.002}
\end{barticle}
\endbibitem

\bibitem[\protect\citeauthoryear{{Casse} and {Paul}}{1982}]{Casse1982}
\begin{barticle}
\bauthor{\binits{M.} \bsnm{{Casse}}},
\bauthor{\binits{J.A.} \bsnm{{Paul}}},
\batitle{{On the stellar origin of the Ne-22 excess in cosmic rays}}.
\bjtitle{\apj}
\bvolume{258},
\bfpage{860}--\blpage{863}
(\byear{1982}).
doi:\doiurl{10.1086/160132}
\end{barticle}
\endbibitem

\bibitem[\protect\citeauthoryear{{Cesarsky} and {Montmerle}}{1983}]{cm83}
\begin{barticle}
\bauthor{\binits{C.J.} \bsnm{{Cesarsky}}},
\bauthor{\binits{T.} \bsnm{{Montmerle}}},
\batitle{{Gamma rays from active regions in the galaxy - The possible
  contribution of stellar winds}}.
\bjtitle{\ssr}
\bvolume{36},
\bfpage{173}--\blpage{193}
(\byear{1983}).
doi:\doiurl{10.1007/BF00167503}
\end{barticle}
\endbibitem

\bibitem[\protect\citeauthoryear{{Cesarsky} and {Volk}}{1978}]{Cesarsky78}
\begin{barticle}
\bauthor{\binits{C.J.} \bsnm{{Cesarsky}}},
\bauthor{\binits{H.J.} \bsnm{{Volk}}},
\batitle{{Cosmic Ray Penetration into Molecular Clouds}}.
\bjtitle{\aap}
\bvolume{70},
\bfpage{367}
(\byear{1978})
\end{barticle}
\endbibitem

\bibitem[\protect\citeauthoryear{{Chernyakova} et~al.}{2011}]{chernyakova11}
\begin{barticle}
\bauthor{\binits{M.} \bsnm{{Chernyakova}}},
\bauthor{\binits{D.} \bsnm{{Malyshev}}},
\bauthor{\binits{F.A.} \bsnm{{Aharonian}}}, \betal,
\batitle{{The High-energy, Arcminute-scale Galactic Center Gamma-ray Source}}.
\bjtitle{\apj}
\bvolume{726}(\bissue{2}),
\bfpage{60}
(\byear{2011}).
doi:\doiurl{10.1088/0004-637X/726/2/60}
\end{barticle}
\endbibitem

\bibitem[\protect\citeauthoryear{{Chevalier} and {Clegg}}{1985}]{chev_clegg85}
\begin{barticle}
\bauthor{\binits{R.A.} \bsnm{{Chevalier}}},
\bauthor{\binits{A.W.} \bsnm{{Clegg}}},
\batitle{{Wind from a starburst galaxy nucleus}}.
\bjtitle{\nat}
\bvolume{317},
\bfpage{44}
(\byear{1985}).
doi:\doiurl{10.1038/317044a0}
\end{barticle}
\endbibitem

\bibitem[\protect\citeauthoryear{{Chevance} et~al.}{2020}]{chevance20}
\begin{barticle}
\bauthor{\binits{M.} \bsnm{{Chevance}}},
\bauthor{\binits{J.M.D.} \bsnm{{Kruijssen}}},
\bauthor{\binits{A.P.S.} \bsnm{{Hygate}}}, \betal,
\batitle{{The lifecycle of molecular clouds in nearby star-forming disc
  galaxies}}.
\bjtitle{\mnras}
\bvolume{493}(\bissue{2}),
\bfpage{2872}--\blpage{2909}
(\byear{2020}).
doi:\doiurl{10.1093/mnras/stz3525}
\end{barticle}
\endbibitem

\bibitem[\protect\citeauthoryear{{Cioffi} et~al.}{1988}]{Cioffi1988}
\begin{barticle}
\bauthor{\binits{D.F.} \bsnm{{Cioffi}}},
\bauthor{\binits{C.F.} \bsnm{{McKee}}},
\bauthor{\binits{E.} \bsnm{{Bertschinger}}},
\batitle{{Dynamics of Radiative Supernova Remnants}}.
\bjtitle{\apj}
\bvolume{334},
\bfpage{252}
(\byear{1988}).
doi:\doiurl{10.1086/166834}
\end{barticle}
\endbibitem

\bibitem[\protect\citeauthoryear{{Comer{\'o}n} et~al.}{2016}]{ComeronEtal2016}
\begin{barticle}
\bauthor{\binits{F.} \bsnm{{Comer{\'o}n}}},
\bauthor{\binits{A.A.} \bsnm{{Djupvik}}},
\bauthor{\binits{N.} \bsnm{{Schneider}}}, \betal,
\batitle{{Red supergiants and the past of Cygnus OB2}}.
\bjtitle{\aap}
\bvolume{586},
\bfpage{46}
(\byear{2016}).
doi:\doiurl{10.1051/0004-6361/201527517}
\end{barticle}
\endbibitem

\bibitem[\protect\citeauthoryear{{Commer{\c{c}}on} et~al.}{2019}]{Commercon19}
\begin{barticle}
\bauthor{\binits{B.} \bsnm{{Commer{\c{c}}on}}},
\bauthor{\binits{A.} \bsnm{{Marcowith}}},
\bauthor{\binits{Y.} \bsnm{{Dubois}}},
\batitle{{Cosmic-ray propagation in the bi-stable interstellar medium. I.
  Conditions for cosmic-ray trapping}}.
\bjtitle{\aap}
\bvolume{622},
\bfpage{143}
(\byear{2019}).
doi:\doiurl{10.1051/0004-6361/201833809}
\end{barticle}
\endbibitem

\bibitem[\protect\citeauthoryear{{Condon}}{1992}]{Condon1992ARAA}
\begin{barticle}
\bauthor{\binits{J.J.} \bsnm{{Condon}}},
\batitle{{Radio emission from normal galaxies.}}
\bjtitle{\araa}
\bvolume{30},
\bfpage{575}--\blpage{611}
(\byear{1992}).
doi:\doiurl{10.1146/annurev.aa.30.090192.003043}
\end{barticle}
\endbibitem

\bibitem[\protect\citeauthoryear{{Cox}}{2005}]{2005ARA&A..43..337C}
\begin{barticle}
\bauthor{\binits{D.P.} \bsnm{{Cox}}},
\batitle{{The Three-Phase Interstellar Medium Revisited}}.
\bjtitle{\araa}
\bvolume{43}(\bissue{1}),
\bfpage{337}--\blpage{385}
(\byear{2005}).
doi:\doiurl{10.1146/annurev.astro.43.072103.150615}
\end{barticle}
\endbibitem

\bibitem[\protect\citeauthoryear{{Crocker}}{2012}]{crocker12}
\begin{barticle}
\bauthor{\binits{R.M.} \bsnm{{Crocker}}},
\batitle{{Non-thermal insights on mass and energy flows through the Galactic
  Centre and into the Fermi bubbles}}.
\bjtitle{\mnras}
\bvolume{423}(\bissue{4}),
\bfpage{3512}--\blpage{3539}
(\byear{2012}).
doi:\doiurl{10.1111/j.1365-2966.2012.21149.x}
\end{barticle}
\endbibitem

\bibitem[\protect\citeauthoryear{{Crocker} et~al.}{2010}]{crocker10}
\begin{barticle}
\bauthor{\binits{R.M.} \bsnm{{Crocker}}},
\bauthor{\binits{D.I.} \bsnm{{Jones}}},
\bauthor{\binits{F.} \bsnm{{Melia}}}, \betal,
\batitle{{A lower limit of 50 microgauss for the magnetic field near the
  Galactic Centre}}.
\bjtitle{\nat}
\bvolume{463}(\bissue{7277}),
\bfpage{65}--\blpage{67}
(\byear{2010}).
doi:\doiurl{10.1038/nature08635}
\end{barticle}
\endbibitem

\bibitem[\protect\citeauthoryear{{CTA consortium}}{2019}]{ctascience}
\begin{bbook}
\bauthor{\bsnm{{CTA consortium}}},
\bbtitle{Science with the Cherenkov Telescope Array}
(\bpublisher{WORLD SCIENTIFIC}, \blocation{???}, \byear{2019}).
doi:\doiurl{10.1142/10986}.
\burl{https://www.worldscientific.com/doi/abs/10.1142/10986}
\end{bbook}
\endbibitem

\bibitem[\protect\citeauthoryear{{D'Angelo} et~al.}{2016}]{Dangelo2016}
\begin{barticle}
\bauthor{\binits{M.} \bsnm{{D'Angelo}}},
\bauthor{\binits{P.} \bsnm{{Blasi}}},
\bauthor{\binits{E.} \bsnm{{Amato}}},
\batitle{{Grammage of cosmic rays around Galactic supernova remnants}}.
\bjtitle{\prd}
\bvolume{94}(\bissue{8}),
\bfpage{083003}
(\byear{2016}).
doi:\doiurl{10.1103/PhysRevD.94.083003}
\end{barticle}
\endbibitem

\bibitem[\protect\citeauthoryear{{D'Angelo} et~al.}{2018}]{Dangelo2018}
\begin{barticle}
\bauthor{\binits{M.} \bsnm{{D'Angelo}}},
\bauthor{\binits{G.} \bsnm{{Morlino}}},
\bauthor{\binits{E.} \bsnm{{Amato}}}, \betal,
\batitle{{Diffuse gamma-ray emission from self-confined cosmic rays around
  Galactic sources}}.
\bjtitle{\mnras}
\bvolume{474}(\bissue{2}),
\bfpage{1944}--\blpage{1954}
(\byear{2018}).
doi:\doiurl{10.1093/mnras/stx2828}
\end{barticle}
\endbibitem

\bibitem[\protect\citeauthoryear{{de Angelis} et~al.}{2018}]{ASTROGAM18}
\begin{barticle}
\bauthor{\binits{A.} \bsnm{{de Angelis}}},
\bauthor{\binits{V.} \bsnm{{Tatischeff}}},
\bauthor{\binits{I.A.} \bsnm{{Grenier}}}, \betal,
\batitle{{Science with e-ASTROGAM. A space mission for MeV-GeV gamma-ray
  astrophysics}}.
\bjtitle{Journal of High Energy Astrophysics}
\bvolume{19},
\bfpage{1}--\blpage{106}
(\byear{2018}).
doi:\doiurl{10.1016/j.jheap.2018.07.001}
\end{barticle}
\endbibitem

\bibitem[\protect\citeauthoryear{{De Becker} and
  {Raucq}}{2013}]{2013A&A...558A..28D}
\begin{barticle}
\bauthor{\binits{M.} \bsnm{{De Becker}}},
\bauthor{\binits{F.} \bsnm{{Raucq}}},
\batitle{{Catalogue of particle-accelerating colliding-wind binaries}}.
\bjtitle{\aap}
\bvolume{558},
\bfpage{28}
(\byear{2013}).
doi:\doiurl{10.1051/0004-6361/201322074}
\end{barticle}
\endbibitem

\bibitem[\protect\citeauthoryear{{Drury}}{2011}]{Drury2011}
\begin{barticle}
\bauthor{\binits{L.O.} \bsnm{{Drury}}},
\batitle{{Escaping the accelerator: how, when and in what numbers do cosmic
  rays get out of supernova remnants?}}
\bjtitle{\mnras}
\bvolume{415}(\bissue{2}),
\bfpage{1807}--\blpage{1814}
(\byear{2011}).
doi:\doiurl{10.1111/j.1365-2966.2011.18824.x}
\end{barticle}
\endbibitem

\bibitem[\protect\citeauthoryear{{Drury} et~al.}{2001}]{Drury2001}
\begin{barticle}
\bauthor{\binits{L.O.} \bsnm{{Drury}}},
\bauthor{\binits{D.E.} \bsnm{{Ellison}}},
\bauthor{\binits{F.A.} \bsnm{{Aharonian}}}, \betal,
\batitle{{Test of galactic cosmic-ray source models - Working Group Report}}.
\bjtitle{\ssr}
\bvolume{99},
\bfpage{329}--\blpage{352}
(\byear{2001})
\end{barticle}
\endbibitem

\bibitem[\protect\citeauthoryear{{Efremov} and {Elmegreen}}{1998}]{efremov98}
\begin{barticle}
\bauthor{\binits{Y.N.} \bsnm{{Efremov}}},
\bauthor{\binits{B.G.} \bsnm{{Elmegreen}}},
\batitle{{Hierarchical star formation from the time-space distribution of star
  clusters in the Large Magellanic Cloud}}.
\bjtitle{\mnras}
\bvolume{299},
\bfpage{588}--\blpage{594}
(\byear{1998}).
doi:\doiurl{10.1046/j.1365-8711.1998.01819.x}
\end{barticle}
\endbibitem

\bibitem[\protect\citeauthoryear{{Ekstr{\"o}m} et~al.}{2012}]{Ekstrom2012}
\begin{barticle}
\bauthor{\binits{S.} \bsnm{{Ekstr{\"o}m}}},
\bauthor{\binits{C.} \bsnm{{Georgy}}},
\bauthor{\binits{P.} \bsnm{{Eggenberger}}}, \betal,
\batitle{{Grids of stellar models with rotation. I. Models from 0.8 to 120
  M$_{\odot}$ at solar metallicity (Z = 0.014)}}.
\bjtitle{\aap}
\bvolume{537},
\bfpage{146}
(\byear{2012}).
doi:\doiurl{10.1051/0004-6361/201117751}
\end{barticle}
\endbibitem

\bibitem[\protect\citeauthoryear{{Farber} et~al.}{2018}]{2018ApJ...856..112F}
\begin{barticle}
\bauthor{\binits{R.} \bsnm{{Farber}}},
\bauthor{\binits{M.} \bsnm{{Ruszkowski}}},
\bauthor{\binits{H.-Y.K.} \bsnm{{Yang}}}, \betal,
\batitle{{Impact of Cosmic-Ray Transport on Galactic Winds}}.
\bjtitle{\apj}
\bvolume{856}(\bissue{2}),
\bfpage{112}
(\byear{2018}).
doi:\doiurl{10.3847/1538-4357/aab26d}
\end{barticle}
\endbibitem

\bibitem[\protect\citeauthoryear{{Farmer} and {Goldreich}}{2004}]{Farmer2004}
\begin{barticle}
\bauthor{\binits{A.J.} \bsnm{{Farmer}}},
\bauthor{\binits{P.} \bsnm{{Goldreich}}},
\batitle{{Wave Damping by Magnetohydrodynamic Turbulence and Its Effect on
  Cosmic-Ray Propagation in the Interstellar Medium}}.
\bjtitle{\apj}
\bvolume{604}(\bissue{2}),
\bfpage{671}--\blpage{674}
(\byear{2004}).
doi:\doiurl{10.1086/382040}
\end{barticle}
\endbibitem

\bibitem[\protect\citeauthoryear{{Garcia-Munoz} et~al.}{1979}]{Garcia1979}
\begin{barticle}
\bauthor{\binits{M.} \bsnm{{Garcia-Munoz}}},
\bauthor{\binits{J.A.} \bsnm{{Simpson}}},
\bauthor{\binits{J.P.} \bsnm{{Wefel}}},
\batitle{{The isotopes of neon in the galactic cosmic rays}}.
\bjtitle{\apjl}
\bvolume{232},
\bfpage{95}--\blpage{99}
(\byear{1979}).
doi:\doiurl{10.1086/183043}
\end{barticle}
\endbibitem

\bibitem[\protect\citeauthoryear{{Gaskin} et~al.}{2018}]{Gaskin2018}
\begin{bchapter}
\bauthor{\binits{J.A.} \bsnm{{Gaskin}}},
\bauthor{\binits{A.} \bsnm{{Dominguez}}},
\bauthor{\binits{K.} \bsnm{{Gelmis}}}, \betal,
\bctitle{{The Lynx X-ray Observatory: concept study overview and status}},
in \bbtitle{\procspie}.
\bsertitle{Society of Photo-Optical Instrumentation Engineers (SPIE) Conference
  Series},
vol. \bseriesno{10699},
\byear{2018},
p. \bfpage{106990}.
doi:\doiurl{10.1117/12.2314149}
\end{bchapter}
\endbibitem

\bibitem[\protect\citeauthoryear{{Georgy} et~al.}{2012}]{Georgy2012}
\begin{barticle}
\bauthor{\binits{C.} \bsnm{{Georgy}}},
\bauthor{\binits{S.} \bsnm{{Ekstr{\"o}m}}},
\bauthor{\binits{G.} \bsnm{{Meynet}}}, \betal,
\batitle{{Grids of stellar models with rotation. II. WR populations and
  supernovae/GRB progenitors at Z = 0.014}}.
\bjtitle{\aap}
\bvolume{542},
\bfpage{29}
(\byear{2012}).
doi:\doiurl{10.1051/0004-6361/201118340}
\end{barticle}
\endbibitem

\bibitem[\protect\citeauthoryear{{Getman} et~al.}{2018}]{Getman2018}
\begin{barticle}
\bauthor{\binits{K.V.} \bsnm{{Getman}}},
\bauthor{\binits{M.A.} \bsnm{{Kuhn}}},
\bauthor{\binits{E.D.} \bsnm{{Feigelson}}}, \betal,
\batitle{{Young star clusters in nearby molecular clouds}}.
\bjtitle{\mnras}
\bvolume{477}(\bissue{1}),
\bfpage{298}--\blpage{324}
(\byear{2018}).
doi:\doiurl{10.1093/mnras/sty473}
\end{barticle}
\endbibitem

\bibitem[\protect\citeauthoryear{{Gieseler} et~al.}{2000}]{Gieseler2000}
\begin{barticle}
\bauthor{\binits{U.D.J.} \bsnm{{Gieseler}}},
\bauthor{\binits{T.W.} \bsnm{{Jones}}},
\bauthor{\binits{H.} \bsnm{{Kang}}},
\batitle{{Time dependent cosmic-ray shock acceleration with self-consistent
  injection}}.
\bjtitle{\aap}
\bvolume{364},
\bfpage{911}--\blpage{922}
(\byear{2000})
\end{barticle}
\endbibitem

\bibitem[\protect\citeauthoryear{{Ginsburg} et~al.}{2016}]{ginsburg16}
\begin{barticle}
\bauthor{\binits{A.} \bsnm{{Ginsburg}}},
\bauthor{\binits{C.} \bsnm{{Henkel}}},
\bauthor{\binits{Y.} \bsnm{{Ao}}}, \betal,
\batitle{{Dense gas in the Galactic central molecular zone is warm and heated
  by turbulence}}.
\bjtitle{\aap}
\bvolume{586},
\bfpage{50}
(\byear{2016}).
doi:\doiurl{10.1051/0004-6361/201526100}
\end{barticle}
\endbibitem

\bibitem[\protect\citeauthoryear{{Ginzburg} and
  {Syrovatskii}}{1964}]{Ginzburg1964}
\begin{bbook}
\bauthor{\binits{V.L.} \bsnm{{Ginzburg}}},
\bauthor{\binits{S.I.} \bsnm{{Syrovatskii}}},
\bbtitle{{The Origin of Cosmic Rays}}
\byear{1964}
\end{bbook}
\endbibitem

\bibitem[\protect\citeauthoryear{{Greiner} et~al.}{2012}]{GRIPS12}
\begin{barticle}
\bauthor{\binits{J.} \bsnm{{Greiner}}},
\bauthor{\binits{K.} \bsnm{{Mannheim}}},
\bauthor{\binits{F.} \bsnm{{Aharonian}}}, \betal,
\batitle{{GRIPS - Gamma-Ray Imaging, Polarimetry and Spectroscopy}}.
\bjtitle{Experimental Astronomy}
\bvolume{34}(\bissue{2}),
\bfpage{551}--\blpage{582}
(\byear{2012}).
doi:\doiurl{10.1007/s10686-011-9255-0}
\end{barticle}
\endbibitem

\bibitem[\protect\citeauthoryear{{Grenier} et~al.}{2015}]{2015ARA&A..53..199G}
\begin{barticle}
\bauthor{\binits{I.A.} \bsnm{{Grenier}}},
\bauthor{\binits{J.H.} \bsnm{{Black}}},
\bauthor{\binits{A.W.} \bsnm{{Strong}}},
\batitle{{The Nine Lives of Cosmic Rays in Galaxies}}.
\bjtitle{\araa}
\bvolume{53},
\bfpage{199}--\blpage{246}
(\byear{2015}).
doi:\doiurl{10.1146/annurev-astro-082214-122457}
\end{barticle}
\endbibitem

\bibitem[\protect\citeauthoryear{{Grimaldo} et~al.}{2019}]{2019ApJ...871...55G}
\begin{barticle}
\bauthor{\binits{E.} \bsnm{{Grimaldo}}},
\bauthor{\binits{A.} \bsnm{{Reimer}}},
\bauthor{\binits{R.} \bsnm{{Kissmann}}}, \betal,
\batitle{{Proton Acceleration in Colliding Stellar Wind Binaries}}.
\bjtitle{\apj}
\bvolume{871}(\bissue{1}),
\bfpage{55}
(\byear{2019}).
doi:\doiurl{10.3847/1538-4357/aaf6ee}
\end{barticle}
\endbibitem

\bibitem[\protect\citeauthoryear{{Guo} et~al.}{2017}]{GuoEtal2017}
\begin{barticle}
\bauthor{\binits{Y.-Q.} \bsnm{{Guo}}},
\bauthor{\binits{Z.} \bsnm{{Tian}}},
\bauthor{\binits{Z.} \bsnm{{Wang}}}, \betal,
\batitle{{The Galactic Center: A Petaelectronvolt Cosmic-ray Acceleration
  Factory}}.
\bjtitle{\apj}
\bvolume{836},
\bfpage{233}
(\byear{2017}).
doi:\doiurl{10.3847/1538-4357/aa5f58}
\end{barticle}
\endbibitem

\bibitem[\protect\citeauthoryear{{Gupta} et~al.}{2020}]{Gupta2019}
\begin{barticle}
\bauthor{\binits{S.} \bsnm{{Gupta}}},
\bauthor{\binits{B.B.} \bsnm{{Nath}}},
\bauthor{\binits{P.} \bsnm{{Sharma}}}, \betal,
\batitle{{Realistic modelling of wind and supernovae shocks in star clusters:
  addressing $^{22}$Ne/$^{20}$Ne and other problems in Galactic cosmic rays}}.
\bjtitle{\mnras}
\bvolume{493}(\bissue{3}),
\bfpage{3159}--\blpage{3177}
(\byear{2020}).
doi:\doiurl{10.1093/mnras/staa286}
\end{barticle}
\endbibitem

\bibitem[\protect\citeauthoryear{{Guti{\'e}rrez} et~al.}{2020}]{NGC253_20}
\begin{botherref}
\oauthor{\binits{E.M.} \bsnm{{Guti{\'e}rrez}}},
\oauthor{\binits{G.E.} \bsnm{{Romero}}},
\oauthor{\binits{F.L.} \bsnm{{Vieyro}}},
{Cosmic rays from the nearby starburst galaxy NGC 253: the effect of a low
  luminosity active galactic nucleus}.
arXiv e-prints,
2003--09410
(2020)
\end{botherref}
\endbibitem

\bibitem[\protect\citeauthoryear{{Han}}{2017}]{Han2017}
\begin{barticle}
\bauthor{\binits{J.L.} \bsnm{{Han}}},
\batitle{{Observing Interstellar and Intergalactic Magnetic Fields}}.
\bjtitle{\araa}
\bvolume{55}(\bissue{1}),
\bfpage{111}--\blpage{157}
(\byear{2017}).
doi:\doiurl{10.1146/annurev-astro-091916-055221}
\end{barticle}
\endbibitem

\bibitem[\protect\citeauthoryear{{Haverkorn}
  et~al.}{2008}]{2008ApJ...680..362H}
\begin{barticle}
\bauthor{\binits{M.} \bsnm{{Haverkorn}}},
\bauthor{\binits{J.C.} \bsnm{{Brown}}},
\bauthor{\binits{B.M.} \bsnm{{Gaensler}}}, \betal,
\batitle{{The Outer Scale of Turbulence in the Magnetoionized Galactic
  Interstellar Medium}}.
\bjtitle{\apj}
\bvolume{680}(\bissue{1}),
\bfpage{362}--\blpage{370}
(\byear{2008}).
doi:\doiurl{10.1086/587165}
\end{barticle}
\endbibitem

\bibitem[\protect\citeauthoryear{{Heiles}}{1990}]{1990ApJ...354..483H}
\begin{barticle}
\bauthor{\binits{C.} \bsnm{{Heiles}}},
\batitle{{Clustered Supernovae versus the Gaseous Disk and Halo}}.
\bjtitle{\apj}
\bvolume{354},
\bfpage{483}
(\byear{1990}).
doi:\doiurl{10.1086/168709}
\end{barticle}
\endbibitem

\bibitem[\protect\citeauthoryear{{Henshaw} et~al.}{2016}]{henshaw16}
\begin{barticle}
\bauthor{\binits{J.D.} \bsnm{{Henshaw}}},
\bauthor{\binits{S.N.} \bsnm{{Longmore}}},
\bauthor{\binits{J.M.D.} \bsnm{{Kruijssen}}}, \betal,
\batitle{{Molecular gas kinematics within the central 250 pc of the Milky
  Way}}.
\bjtitle{\mnras}
\bvolume{457},
\bfpage{2675}--\blpage{2702}
(\byear{2016}).
doi:\doiurl{10.1093/mnras/stw121}
\end{barticle}
\endbibitem

\bibitem[\protect\citeauthoryear{{Herczeg} and
  {Hillenbrand}}{2015}]{2015ApJ...808...23H}
\begin{barticle}
\bauthor{\binits{G.J.} \bsnm{{Herczeg}}},
\bauthor{\binits{L.A.} \bsnm{{Hillenbrand}}},
\batitle{{Empirical Isochrones for Low Mass Stars in Nearby Young
  Associations}}.
\bjtitle{\apj}
\bvolume{808}(\bissue{1}),
\bfpage{23}
(\byear{2015}).
doi:\doiurl{10.1088/0004-637X/808/1/23}
\end{barticle}
\endbibitem

\bibitem[\protect\citeauthoryear{{H.E.S.S. Collaboration}
  et~al.}{2016}]{HESS2016}
\begin{barticle}
\bauthor{\bsnm{{H.E.S.S. Collaboration}}},
\bauthor{\binits{A.} \bsnm{{Abramowski}}},
\bauthor{\binits{F.} \bsnm{{Aharonian}}}, \betal,
\batitle{{Acceleration of petaelectronvolt protons in the Galactic Centre}}.
\bjtitle{\nat}
\bvolume{531},
\bfpage{476}--\blpage{479}
(\byear{2016}).
doi:\doiurl{10.1038/nature17147}
\end{barticle}
\endbibitem

\bibitem[\protect\citeauthoryear{{H.E.S.S. Collaboration}
  et~al.}{2018a}]{2018A&A...612A...9H}
\begin{barticle}
\bauthor{\bsnm{{H.E.S.S. Collaboration}}},
\bauthor{\binits{H.} \bsnm{{Abdalla}}},
\bauthor{\binits{A.} \bsnm{{Abramowski}}}, \betal,
\batitle{{Characterising the VHE diffuse emission in the central 200 parsecs of
  our Galaxy with H.E.S.S.}}
\bjtitle{\aap}
\bvolume{612},
\bfpage{9}
(\byear{2018}a).
doi:\doiurl{10.1051/0004-6361/201730824}
\end{barticle}
\endbibitem

\bibitem[\protect\citeauthoryear{{H.E.S.S. Collaboration}
  et~al.}{2018b}]{2018A&A...612A..11H}
\begin{barticle}
\bauthor{\bsnm{{H.E.S.S. Collaboration}}},
\bauthor{\binits{H.} \bsnm{{Abdalla}}},
\bauthor{\binits{A.} \bsnm{{Abramowski}}}, \betal,
\batitle{{Extended VHE {\ensuremath{\gamma}}-ray emission towards SGR1806-20,
  LBV 1806-20, and stellar cluster Cl* 1806-20}}.
\bjtitle{\aap}
\bvolume{612},
\bfpage{11}
(\byear{2018}b).
doi:\doiurl{10.1051/0004-6361/201628695}
\end{barticle}
\endbibitem

\bibitem[\protect\citeauthoryear{{H.E.S.S. Collaboration}
  et~al.}{2018c}]{2018A&A...612A...1H}
\begin{barticle}
\bauthor{\bsnm{{H.E.S.S. Collaboration}}},
\bauthor{\binits{H.} \bsnm{{Abdalla}}},
\bauthor{\binits{A.} \bsnm{{Abramowski}}}, \betal,
\batitle{{The H.E.S.S. Galactic plane survey}}.
\bjtitle{\aap}
\bvolume{612},
\bfpage{1}
(\byear{2018}c).
doi:\doiurl{10.1051/0004-6361/201732098}
\end{barticle}
\endbibitem

\bibitem[\protect\citeauthoryear{{H.E.S.S. Collaboration}
  et~al.}{2018d}]{NGC253_HESS_Fermi18}
\begin{barticle}
\bauthor{\bsnm{{H.E.S.S. Collaboration}}},
\bauthor{\binits{H.} \bsnm{{Abdalla}}},
\bauthor{\binits{F.} \bsnm{{Aharonian}}}, \betal,
\batitle{{The starburst galaxy NGC 253 revisited by H.E.S.S. and Fermi-LAT}}.
\bjtitle{\aap}
\bvolume{617},
\bfpage{73}
(\byear{2018}d).
doi:\doiurl{10.1051/0004-6361/201833202}
\end{barticle}
\endbibitem

\bibitem[\protect\citeauthoryear{{Higdon} and
  {Lingenfelter}}{2003}]{Higdon2003}
\begin{barticle}
\bauthor{\binits{J.C.} \bsnm{{Higdon}}},
\bauthor{\binits{R.E.} \bsnm{{Lingenfelter}}},
\batitle{{The Superbubble Origin of $^{22}$Ne in Cosmic Rays}}.
\bjtitle{\apj}
\bvolume{590},
\bfpage{822}--\blpage{832}
(\byear{2003}).
doi:\doiurl{10.1086/375192}
\end{barticle}
\endbibitem

\bibitem[\protect\citeauthoryear{{Holguin} et~al.}{2019}]{2019MNRAS.490.1271H}
\begin{barticle}
\bauthor{\binits{F.} \bsnm{{Holguin}}},
\bauthor{\binits{M.} \bsnm{{Ruszkowski}}},
\bauthor{\binits{A.} \bsnm{{Lazarian}}}, \betal,
\batitle{{Role of cosmic-ray streaming and turbulent damping in driving
  galactic winds}}.
\bjtitle{\mnras}
\bvolume{490}(\bissue{1}),
\bfpage{1271}--\blpage{1282}
(\byear{2019}).
doi:\doiurl{10.1093/mnras/stz2568}
\end{barticle}
\endbibitem

\bibitem[\protect\citeauthoryear{{Hona} et~al.}{2019}]{HAWC_Cygnus_19}
\begin{bchapter}
\bauthor{\binits{B.} \bsnm{{Hona}}},
\bauthor{\binits{H.} \bsnm{{Fleischhack}}},
\bauthor{\binits{P.} \bsnm{{Huentemeyer}}},
\bctitle{{Testing the Limits of Particle Acceleration in Cygnus OB2 with
  HAWC}},
in \bbtitle{36th International Cosmic Ray Conference (ICRC2019)}.
\bsertitle{International Cosmic Ray Conference},
vol. \bseriesno{36},
\byear{2019},
p. \bfpage{699}
\end{bchapter}
\endbibitem

\bibitem[\protect\citeauthoryear{{Hooper} and {Goodenough}}{2011}]{hooper11}
\begin{barticle}
\bauthor{\binits{D.} \bsnm{{Hooper}}},
\bauthor{\binits{L.} \bsnm{{Goodenough}}},
\batitle{{Dark matter annihilation in the Galactic Center as seen by the Fermi
  Gamma Ray Space Telescope}}.
\bjtitle{Phys. Lett. B}
\bvolume{697}(\bissue{5}),
\bfpage{412}--\blpage{428}
(\byear{2011}).
doi:\doiurl{10.1016/j.physletb.2011.02.029}
\end{barticle}
\endbibitem

\bibitem[\protect\citeauthoryear{{Hooper} and {Linden}}{2011}]{hooper11b}
\begin{barticle}
\bauthor{\binits{D.} \bsnm{{Hooper}}},
\bauthor{\binits{T.} \bsnm{{Linden}}},
\batitle{{Origin of the gamma rays from the Galactic Center}}.
\bjtitle{\prd}
\bvolume{84}(\bissue{12}),
\bfpage{123005}
(\byear{2011}).
doi:\doiurl{10.1103/PhysRevD.84.123005}
\end{barticle}
\endbibitem

\bibitem[\protect\citeauthoryear{{Hopkins} et~al.}{2019}]{2019MNRAS.tmp.2993H}
\begin{botherref}
\oauthor{\binits{P.F.} \bsnm{{Hopkins}}},
\oauthor{\binits{T.K.} \bsnm{{Chan}}},
\oauthor{\binits{S.} \bsnm{{Garrison-Kimmel}}}, et al.,
{But What About Cosmic Rays, Magnetic Fields, Conduction, \&amp; Viscosity in
  Galaxy Formation}.
\mnras,
2993
(2019).
doi:\doiurl{10.1093/mnras/stz3321}
\end{botherref}
\endbibitem

\bibitem[\protect\citeauthoryear{{Hosek} et~al.}{2019}]{Hosek2019}
\begin{barticle}
\bauthor{\binits{M.W.} \bsnm{{Hosek}} \bsuffix{Jr.}},
\bauthor{\binits{J.R.} \bsnm{{Lu}}},
\bauthor{\binits{J.} \bsnm{{Anderson}}}, \betal,
\batitle{{The Unusual Initial Mass Function of the Arches Cluster}}.
\bjtitle{\apj}
\bvolume{870},
\bfpage{44}
(\byear{2019}).
doi:\doiurl{10.3847/1538-4357/aaef90}
\end{barticle}
\endbibitem

\bibitem[\protect\citeauthoryear{{Howarth} et~al.}{1997}]{Howarth1997}
\begin{botherref}
\oauthor{\binits{I.D.} \bsnm{{Howarth}}},
\oauthor{\binits{K.W.} \bsnm{{Siebert}}},
\oauthor{\binits{G.A.J.} \bsnm{{Hussain}}}, et al.,
{VizieR Online Data Catalog: Rotational Velocities of 373 OB stars (Howarth+
  1997)}.
VizieR Online Data Catalog,
284--265
(1997)
\end{botherref}
\endbibitem

\bibitem[\protect\citeauthoryear{{IceCube Collaboration}
  et~al.}{2019}]{2019arXiv191008488I}
\begin{botherref}
\oauthor{\bsnm{{IceCube Collaboration}}},
\oauthor{\binits{M.G.} \bsnm{{Aartsen}}},
\oauthor{\binits{M.} \bsnm{{Ackermann}}}, et al.,
{Time-integrated Neutrino Source Searches with 10 years of IceCube Data}.
arXiv e-prints,
1910--08488
(2019)
\end{botherref}
\endbibitem

\bibitem[\protect\citeauthoryear{{Inoue}}{2019}]{Inoue19}
\begin{barticle}
\bauthor{\binits{T.} \bsnm{{Inoue}}},
\batitle{{Bell-instability-mediated Spectral Modulation of Hadronic Gamma-Rays
  from a Supernova Remnant Interacting with a Molecular Cloud}}.
\bjtitle{\apj}
\bvolume{872}(\bissue{1}),
\bfpage{46}
(\byear{2019}).
doi:\doiurl{10.3847/1538-4357/aafb70}
\end{barticle}
\endbibitem

\bibitem[\protect\citeauthoryear{{Ipavich}}{1975}]{ipavich1975}
\begin{barticle}
\bauthor{\binits{F.M.} \bsnm{{Ipavich}}},
\batitle{{Galactic winds driven by cosmic rays.}}
\bjtitle{\apj}
\bvolume{196},
\bfpage{107}--\blpage{120}
(\byear{1975}).
doi:\doiurl{10.1086/153397}
\end{barticle}
\endbibitem

\bibitem[\protect\citeauthoryear{{Jansson} and {Farrar}}{2012}]{Jansson1}
\begin{barticle}
\bauthor{\binits{R.} \bsnm{{Jansson}}},
\bauthor{\binits{G.R.} \bsnm{{Farrar}}},
\batitle{{A New Model of the Galactic Magnetic Field}}.
\bjtitle{\apj}
\bvolume{757},
\bfpage{14}
(\byear{2012}).
doi:\doiurl{10.1088/0004-637X/757/1/14}
\end{barticle}
\endbibitem

\bibitem[\protect\citeauthoryear{{Joubaud} et~al.}{2019}]{OrionEridanusAA19}
\begin{barticle}
\bauthor{\binits{T.} \bsnm{{Joubaud}}},
\bauthor{\binits{I.A.} \bsnm{{Grenier}}},
\bauthor{\binits{J.} \bsnm{{Ballet}}}, \betal,
\batitle{{Gas shells and magnetic fields in the Orion-Eridanus superbubble}}.
\bjtitle{\aap}
\bvolume{631},
\bfpage{52}
(\byear{2019}).
doi:\doiurl{10.1051/0004-6361/201936239}
\end{barticle}
\endbibitem

\bibitem[\protect\citeauthoryear{{Joubaud} et~al.}{2020}]{OrionEridanusCR}
\begin{barticle}
\bauthor{\binits{T.} \bsnm{{Joubaud}}},
\bauthor{\binits{I.A.} \bsnm{{Grenier}}},
\bauthor{\binits{J.M.} \bsnm{{Casandjian}}}, \betal,
\batitle{The cosmic-ray content of the orion-eridanus superbubble}.
\bjtitle{A\&A}
\bvolume{635},
\bfpage{96}
(\byear{2020}).
doi:\doiurl{10.1051/0004-6361/201937205}.
\burl{https://doi.org/10.1051/0004-6361/201937205}
\end{barticle}
\endbibitem

\bibitem[\protect\citeauthoryear{{Jouvin} et~al.}{2017}]{2017MNRAS.467.4622J}
\begin{barticle}
\bauthor{\binits{L.} \bsnm{{Jouvin}}},
\bauthor{\binits{A.} \bsnm{{Lemi{\`e}re}}},
\bauthor{\binits{R.} \bsnm{{Terrier}}},
\batitle{{Does the SN rate explain the very high energy cosmic rays in the
  central 200 pc of our Galaxy?}}
\bjtitle{\mnras}
\bvolume{467}(\bissue{4}),
\bfpage{4622}--\blpage{4630}
(\byear{2017}).
doi:\doiurl{10.1093/mnras/stx361}
\end{barticle}
\endbibitem

\bibitem[\protect\citeauthoryear{{Katsuta} et~al.}{2017}]{Katsuta2017}
\begin{barticle}
\bauthor{\binits{J.} \bsnm{{Katsuta}}},
\bauthor{\binits{Y.} \bsnm{{Uchiyama}}},
\bauthor{\binits{S.} \bsnm{{Funk}}},
\batitle{{Extended Gamma-Ray Emission from the G25.0+0.0 Region: A Star-forming
  Region Powered by the Newly Found OB Association?}}
\bjtitle{\apj}
\bvolume{839},
\bfpage{129}
(\byear{2017}).
doi:\doiurl{10.3847/1538-4357/aa6aa3}
\end{barticle}
\endbibitem

\bibitem[\protect\citeauthoryear{{Kauffmann} et~al.}{2017}]{kauffmann17}
\begin{barticle}
\bauthor{\binits{J.} \bsnm{{Kauffmann}}},
\bauthor{\binits{T.} \bsnm{{Pillai}}},
\bauthor{\binits{Q.} \bsnm{{Zhang}}}, \betal,
\batitle{{The Galactic Center Molecular Cloud Survey. I. A steep linewidth-size
  relation and suppression of star formation}}.
\bjtitle{\aap}
\bvolume{603},
\bfpage{89}
(\byear{2017}).
doi:\doiurl{10.1051/0004-6361/201628088}
\end{barticle}
\endbibitem

\bibitem[\protect\citeauthoryear{{Kavanagh}}{2020}]{Kavanagh20}
\begin{barticle}
\bauthor{\binits{P.J.} \bsnm{{Kavanagh}}},
\batitle{{Thermal and non-thermal X-ray emission from stellar clusters and
  superbubbles}}.
\bjtitle{\apss}
\bvolume{365}(\bissue{1}),
\bfpage{6}
(\byear{2020}).
doi:\doiurl{10.1007/s10509-019-3719-5}
\end{barticle}
\endbibitem

\bibitem[\protect\citeauthoryear{{Kawamura} et~al.}{2009}]{kawamura09}
\begin{barticle}
\bauthor{\binits{A.} \bsnm{{Kawamura}}},
\bauthor{\binits{Y.} \bsnm{{Mizuno}}},
\bauthor{\binits{T.} \bsnm{{Minamidani}}}, \betal,
\batitle{{The Second Survey of the Molecular Clouds in the Large Magellanic
  Cloud by NANTEN. II. Star Formation}}.
\bjtitle{\apjs}
\bvolume{184},
\bfpage{1}--\blpage{17}
(\byear{2009}).
doi:\doiurl{10.1088/0067-0049/184/1/1}
\end{barticle}
\endbibitem

\bibitem[\protect\citeauthoryear{{Krause} et~al.}{2013}]{2013A&A...550A..49K}
\begin{barticle}
\bauthor{\binits{M.} \bsnm{{Krause}}},
\bauthor{\binits{K.} \bsnm{{Fierlinger}}},
\bauthor{\binits{R.} \bsnm{{Diehl}}}, \betal,
\batitle{{Feedback by massive stars and the emergence of superbubbles. I.
  Energy efficiency and Vishniac instabilities}}.
\bjtitle{\aap}
\bvolume{550},
\bfpage{49}
(\byear{2013}).
doi:\doiurl{10.1051/0004-6361/201220060}
\end{barticle}
\endbibitem

\bibitem[\protect\citeauthoryear{{Krause} et~al.}{2014}]{2014A&A...566A..94K}
\begin{barticle}
\bauthor{\binits{M.} \bsnm{{Krause}}},
\bauthor{\binits{R.} \bsnm{{Diehl}}},
\bauthor{\binits{H.} \bsnm{{B{\"o}hringer}}}, \betal,
\batitle{{Feedback by massive stars and the emergence of superbubbles. II.
  X-ray properties}}.
\bjtitle{\aap}
\bvolume{566},
\bfpage{94}
(\byear{2014}).
doi:\doiurl{10.1051/0004-6361/201423871}
\end{barticle}
\endbibitem

\bibitem[\protect\citeauthoryear{{Krieger} et~al.}{2017}]{krieger17}
\begin{barticle}
\bauthor{\binits{N.} \bsnm{{Krieger}}},
\bauthor{\binits{J.} \bsnm{{Ott}}},
\bauthor{\binits{H.} \bsnm{{Beuther}}}, \betal,
\batitle{{The Survey of Water and Ammonia in the Galactic Center (SWAG):
  Molecular Cloud Evolution in the Central Molecular Zone}}.
\bjtitle{\apj}
\bvolume{850},
\bfpage{77}
(\byear{2017}).
doi:\doiurl{10.3847/1538-4357/aa951c}
\end{barticle}
\endbibitem

\bibitem[\protect\citeauthoryear{{Kruijssen} and
  {Longmore}}{2013}]{kruijssen13}
\begin{barticle}
\bauthor{\binits{J.M.D.} \bsnm{{Kruijssen}}},
\bauthor{\binits{S.N.} \bsnm{{Longmore}}},
\batitle{{Comparing molecular gas across cosmic time-scales: the Milky Way as
  both a typical spiral galaxy and a high-redshift galaxy analogue}}.
\bjtitle{\mnras}
\bvolume{435},
\bfpage{2598}--\blpage{2603}
(\byear{2013}).
doi:\doiurl{10.1093/mnras/stt1634}
\end{barticle}
\endbibitem

\bibitem[\protect\citeauthoryear{{Kruijssen} et~al.}{2015}]{kruijssen15}
\begin{barticle}
\bauthor{\binits{J.M.D.} \bsnm{{Kruijssen}}},
\bauthor{\binits{J.E.} \bsnm{{Dale}}},
\bauthor{\binits{S.N.} \bsnm{{Longmore}}},
\batitle{{The dynamical evolution of molecular clouds near the Galactic Centre
  - I. Orbital structure and evolutionary timeline}}.
\bjtitle{\mnras}
\bvolume{447},
\bfpage{1059}--\blpage{1079}
(\byear{2015}).
doi:\doiurl{10.1093/mnras/stu2526}
\end{barticle}
\endbibitem

\bibitem[\protect\citeauthoryear{{Kruijssen} et~al.}{2014}]{kruijssen14}
\begin{barticle}
\bauthor{\binits{J.M.D.} \bsnm{{Kruijssen}}},
\bauthor{\binits{S.N.} \bsnm{{Longmore}}},
\bauthor{\binits{B.G.} \bsnm{{Elmegreen}}}, \betal,
\batitle{{What controls star formation in the central 500 pc of the Galaxy?}}
\bjtitle{\mnras}
\bvolume{440},
\bfpage{3370}--\blpage{3391}
(\byear{2014}).
doi:\doiurl{10.1093/mnras/stu494}
\end{barticle}
\endbibitem

\bibitem[\protect\citeauthoryear{{Kruijssen} et~al.}{2019}]{kruijssen19}
\begin{barticle}
\bauthor{\binits{J.M.D.} \bsnm{{Kruijssen}}},
\bauthor{\binits{A.} \bsnm{{Schruba}}},
\bauthor{\binits{M.} \bsnm{{Chevance}}}, \betal,
\batitle{{Fast and inefficient star formation due to short-lived molecular
  clouds and rapid feedback}}.
\bjtitle{\nat}
\bvolume{569},
\bfpage{519}
(\byear{2019})
\end{barticle}
\endbibitem

\bibitem[\protect\citeauthoryear{{Kruijssen}}{2012}]{DK12}
\begin{barticle}
\bauthor{\binits{J.M.D.} \bsnm{{Kruijssen}}},
\batitle{{On the fraction of star formation occurring in bound stellar
  clusters}}.
\bjtitle{\mnras}
\bvolume{426}(\bissue{4}),
\bfpage{3008}--\blpage{3040}
(\byear{2012}).
doi:\doiurl{10.1111/j.1365-2966.2012.21923.x}
\end{barticle}
\endbibitem

\bibitem[\protect\citeauthoryear{{Krumholz} et~al.}{2017}]{krumholz17}
\begin{barticle}
\bauthor{\binits{M.R.} \bsnm{{Krumholz}}},
\bauthor{\binits{J.M.D.} \bsnm{{Kruijssen}}},
\bauthor{\binits{R.M.} \bsnm{{Crocker}}},
\batitle{{A dynamical model for gas flows, star formation and nuclear winds in
  galactic centres}}.
\bjtitle{\mnras}
\bvolume{466},
\bfpage{1213}--\blpage{1233}
(\byear{2017}).
doi:\doiurl{10.1093/mnras/stw3195}
\end{barticle}
\endbibitem

\bibitem[\protect\citeauthoryear{{Krumholz} and
  {Federrath}}{2019}]{2019FrASS...6....7K}
\begin{barticle}
\bauthor{\binits{M.R.} \bsnm{{Krumholz}}},
\bauthor{\binits{C.} \bsnm{{Federrath}}},
\batitle{{The Role of Magnetic Fields in Setting the Star Formation Rate and
  the Initial Mass Function}}.
\bjtitle{Frontiers in Astronomy and Space Sciences}
\bvolume{6},
\bfpage{7}
(\byear{2019}).
doi:\doiurl{10.3389/fspas.2019.00007}
\end{barticle}
\endbibitem

\bibitem[\protect\citeauthoryear{{Krymskii}}{1977}]{Kry77}
\begin{barticle}
\bauthor{\binits{G.F.} \bsnm{{Krymskii}}},
\batitle{{A regular mechanism for the acceleration of charged particles on the
  front of a shock wave}}.
\bjtitle{Akademiia Nauk SSSR Doklady}
\bvolume{234},
\bfpage{1306}--\blpage{1308}
(\byear{1977})
\end{barticle}
\endbibitem

\bibitem[\protect\citeauthoryear{{Kuhn} et~al.}{2010}]{Kuhn}
\begin{barticle}
\bauthor{\binits{M.A.} \bsnm{{Kuhn}}},
\bauthor{\binits{K.V.} \bsnm{{Getman}}},
\bauthor{\binits{E.D.} \bsnm{{Feigelson}}}, \betal,
\batitle{{A Chandra Observation of the Obscured Star-forming Complex W40}}.
\bjtitle{\apj}
\bvolume{725}(\bissue{2}),
\bfpage{2485}--\blpage{2506}
(\byear{2010}).
doi:\doiurl{10.1088/0004-637X/725/2/2485}
\end{barticle}
\endbibitem

\bibitem[\protect\citeauthoryear{{Kulsrud}}{2005}]{2005ppfa.book.....K}
\begin{bbook}
\bauthor{\binits{R.M.} \bsnm{{Kulsrud}}},
\bbtitle{{Plasma physics for astrophysics, Princeton University Press}}
\byear{2005}
\end{bbook}
\endbibitem

\bibitem[\protect\citeauthoryear{{Lazarian}}{2016}]{Lazarian2016}
\begin{barticle}
\bauthor{\binits{A.} \bsnm{{Lazarian}}},
\batitle{{Damping of Alfv{\'e}n Waves by Turbulence and Its Consequences: From
  Cosmic-ray Streaming to Launching Winds}}.
\bjtitle{\apj}
\bvolume{833}(\bissue{2}),
\bfpage{131}
(\byear{2016}).
doi:\doiurl{10.3847/1538-4357/833/2/131}
\end{barticle}
\endbibitem

\bibitem[\protect\citeauthoryear{{Lemoine}}{2018}]{2018NPPP..297..267L}
\begin{barticle}
\bauthor{\binits{M.} \bsnm{{Lemoine}}},
\batitle{{On the origin of ultra-high rigidity cosmic rays}}.
\bjtitle{Nuclear and Particle Physics Proceedings}
\bvolume{297-299},
\bfpage{267}--\blpage{276}
(\byear{2018}).
doi:\doiurl{10.1016/j.nuclphysbps.2018.07.037}
\end{barticle}
\endbibitem

\bibitem[\protect\citeauthoryear{{Lemoine}}{2019}]{2019PhRvD..99h3006L}
\begin{barticle}
\bauthor{\binits{M.} \bsnm{{Lemoine}}},
\batitle{{Generalized Fermi acceleration}}.
\bjtitle{\prd}
\bvolume{99}(\bissue{8}),
\bfpage{083006}
(\byear{2019}).
doi:\doiurl{10.1103/PhysRevD.99.083006}
\end{barticle}
\endbibitem

\bibitem[\protect\citeauthoryear{{Lim} et~al.}{2013}]{Lim2013}
\begin{barticle}
\bauthor{\binits{B.} \bsnm{{Lim}}},
\bauthor{\binits{M.-Y.} \bsnm{{Chun}}},
\bauthor{\binits{H.} \bsnm{{Sung}}}, \betal,
\batitle{{The Starburst Cluster Westerlund 1: The Initial Mass Function and
  Mass Segregation}}.
\bjtitle{\aj}
\bvolume{145},
\bfpage{46}
(\byear{2013}).
doi:\doiurl{10.1088/0004-6256/145/2/46}
\end{barticle}
\endbibitem

\bibitem[\protect\citeauthoryear{{Limongi} and {Chieffi}}{2018}]{Limongi2018}
\begin{barticle}
\bauthor{\binits{M.} \bsnm{{Limongi}}},
\bauthor{\binits{A.} \bsnm{{Chieffi}}},
\batitle{{Presupernova Evolution and Explosive Nucleosynthesis of Rotating
  Massive Stars in the Metallicity Range -3 {\ensuremath{\leq}} [Fe/H]
  {\ensuremath{\leq}} 0}}.
\bjtitle{\apjs}
\bvolume{237}(\bissue{1}),
\bfpage{13}
(\byear{2018}).
doi:\doiurl{10.3847/1538-4365/aacb24}
\end{barticle}
\endbibitem

\bibitem[\protect\citeauthoryear{{Lingenfelter}}{2018}]{Lingenfelter2018}
\begin{barticle}
\bauthor{\binits{R.E.} \bsnm{{Lingenfelter}}},
\batitle{{Cosmic rays from supernova remnants and superbubbles}}.
\bjtitle{\asr}
\bvolume{62},
\bfpage{2750}--\blpage{2763}
(\byear{2018}).
doi:\doiurl{10.1016/j.asr.2017.04.006}
\end{barticle}
\endbibitem

\bibitem[\protect\citeauthoryear{{Lingenfelter}}{2019}]{2019ApJS..245...30L}
\begin{barticle}
\bauthor{\binits{R.E.} \bsnm{{Lingenfelter}}},
\batitle{{The Origin of Cosmic Rays: How Their Composition Defines Their
  Sources and Sites and the Processes of Their Mixing, Injection, and
  Acceleration}}.
\bjtitle{\apjs}
\bvolume{245}(\bissue{2}),
\bfpage{30}
(\byear{2019}).
doi:\doiurl{10.3847/1538-4365/ab4b58}
\end{barticle}
\endbibitem

\bibitem[\protect\citeauthoryear{{Lodders}}{2003}]{Lodders2003}
\begin{barticle}
\bauthor{\binits{K.} \bsnm{{Lodders}}},
\batitle{{Solar System Abundances and Condensation Temperatures of the
  Elements}}.
\bjtitle{\apj}
\bvolume{591},
\bfpage{1220}--\blpage{1247}
(\byear{2003}).
doi:\doiurl{10.1086/375492}
\end{barticle}
\endbibitem

\bibitem[\protect\citeauthoryear{{Loeb} and {Waxman}}{2006}]{LW06}
\begin{barticle}
\bauthor{\binits{A.} \bsnm{{Loeb}}},
\bauthor{\binits{E.} \bsnm{{Waxman}}},
\batitle{{The cumulative background of high energy neutrinos from starburst
  galaxies}}.
\bjtitle{\jcap}
\bvolume{2006}(\bissue{5}),
\bfpage{003}
(\byear{2006}).
doi:\doiurl{10.1088/1475-7516/2006/05/003}
\end{barticle}
\endbibitem

\bibitem[\protect\citeauthoryear{{Longmore} et~al.}{2013}]{longmore13}
\begin{barticle}
\bauthor{\binits{S.N.} \bsnm{{Longmore}}},
\bauthor{\binits{J.} \bsnm{{Bally}}},
\bauthor{\binits{L.} \bsnm{{Testi}}}, \betal,
\batitle{{Variations in the Galactic star formation rate and density thresholds
  for star formation}}.
\bjtitle{\mnras}
\bvolume{429},
\bfpage{987}--\blpage{1000}
(\byear{2013}).
doi:\doiurl{10.1093/mnras/sts376}
\end{barticle}
\endbibitem

\bibitem[\protect\citeauthoryear{{Longmore} et~al.}{2014}]{longmore14}
\begin{botherref}
\oauthor{\binits{S.N.} \bsnm{{Longmore}}},
\oauthor{\binits{J.M.D.} \bsnm{{Kruijssen}}},
\oauthor{\binits{N.} \bsnm{{Bastian}}}, et al.,
{The Formation and Early Evolution of Young Massive Clusters}.
Protostars and Planets VI,
291--314
(2014).
doi:\doiurl{10.2458/azu\_uapress\_9780816531240-ch013}
\end{botherref}
\endbibitem

\bibitem[\protect\citeauthoryear{{Longmore} and
  {Kruijssen}}{2018}]{2018Galax...6...55L}
\begin{barticle}
\bauthor{\binits{S.} \bsnm{{Longmore}}},
\bauthor{\binits{J.M.D.} \bsnm{{Kruijssen}}},
\batitle{{Constraints on the Distribution of Gas and Young Stars in the
  Galactic Centre in the Context of Interpreting Gamma Ray Emission Features}}.
\bjtitle{Galaxies}
\bvolume{6}(\bissue{2}),
\bfpage{55}
(\byear{2018}).
doi:\doiurl{10.3390/galaxies6020055}
\end{barticle}
\endbibitem

\bibitem[\protect\citeauthoryear{{Lu}}{2018}]{2018ASSL..424...69L}
\begin{bchapter}
\bauthor{\binits{J.R.} \bsnm{{Lu}}},
\bctitle{{Massive Young Clusters Near the Galactic Center}},
in \bbtitle{Astrophysics and Space Science Library},
vol. \bseriesno{424},
ed. by \beditor{\binits{S.} \bsnm{{Stahler}}}
\byear{2018},
p. \bfpage{69}.
\end{bchapter}
\endbibitem

\bibitem[\protect\citeauthoryear{{Lukasiak} et~al.}{1994}]{Lukasiak1994}
\begin{barticle}
\bauthor{\binits{A.} \bsnm{{Lukasiak}}},
\bauthor{\binits{P.} \bsnm{{Ferrando}}},
\bauthor{\binits{F.B.} \bsnm{{McDonald}}}, \betal,
\batitle{{Cosmic-ray isotopic composition of C, N, O, Ne, Mg, SI nuclei in the
  energy range 50-200 MeV per nucleon measured by the Voyager spacecraft during
  the solar minimum period}}.
\bjtitle{\apj}
\bvolume{426},
\bfpage{366}--\blpage{372}
(\byear{1994}).
doi:\doiurl{10.1086/174072}
\end{barticle}
\endbibitem

\bibitem[\protect\citeauthoryear{{Mac Low} and
  {McCray}}{1988}]{1988ApJ...324..776M}
\begin{barticle}
\bauthor{\binits{M.-M.} \bsnm{{Mac Low}}},
\bauthor{\binits{R.} \bsnm{{McCray}}},
\batitle{{Superbubbles in Disk Galaxies}}.
\bjtitle{\apj}
\bvolume{324},
\bfpage{776}
(\byear{1988}).
doi:\doiurl{10.1086/165936}
\end{barticle}
\endbibitem

\bibitem[\protect\citeauthoryear{{Malkov} et~al.}{2013}]{Malkov2013}
\begin{barticle}
\bauthor{\binits{M.A.} \bsnm{{Malkov}}},
\bauthor{\binits{P.H.} \bsnm{{Diamond}}},
\bauthor{\binits{R.Z.} \bsnm{{Sagdeev}}}, \betal,
\batitle{{Analytic Solution for Self-regulated Collective Escape of Cosmic Rays
  from Their Acceleration Sites}}.
\bjtitle{\apj}
\bvolume{768}(\bissue{1}),
\bfpage{73}
(\byear{2013}).
doi:\doiurl{10.1088/0004-637X/768/1/73}
\end{barticle}
\endbibitem

\bibitem[\protect\citeauthoryear{{Mannheim} et~al.}{2012}]{Mannheim12}
\begin{barticle}
\bauthor{\binits{K.} \bsnm{{Mannheim}}},
\bauthor{\binits{D.} \bsnm{{Els{\"a}sser}}},
\bauthor{\binits{O.} \bsnm{{Tibolla}}},
\batitle{{Gamma-rays from pulsar wind nebulae in starburst galaxies}}.
\bjtitle{Astroparticle Physics}
\bvolume{35}(\bissue{12}),
\bfpage{797}--\blpage{800}
(\byear{2012}).
doi:\doiurl{10.1016/j.astropartphys.2012.02.009}
\end{barticle}
\endbibitem

\bibitem[\protect\citeauthoryear{{Marcowith} et~al.}{2016}]{Marcowith2016}
\begin{barticle}
\bauthor{\binits{A.} \bsnm{{Marcowith}}},
\bauthor{\binits{A.} \bsnm{{Bret}}},
\bauthor{\binits{A.} \bsnm{{Bykov}}}, \betal,
\batitle{{The microphysics of collisionless shock waves}}.
\bjtitle{\rpp}
\bvolume{79}(\bissue{4}),
\bfpage{046901}
(\byear{2016}).
doi:\doiurl{10.1088/0034-4885/79/4/046901}
\end{barticle}
\endbibitem

\bibitem[\protect\citeauthoryear{{Martin} et~al.}{2009}]{2009A&A...506..703M}
\begin{barticle}
\bauthor{\binits{P.} \bsnm{{Martin}}},
\bauthor{\binits{J.} \bsnm{{Kn{\"o}dlseder}}},
\bauthor{\binits{R.} \bsnm{{Diehl}}}, \betal,
\batitle{{New estimates of the gamma-ray line emission of the Cygnus region
  from INTEGRAL/SPI observations}}.
\bjtitle{\aap}
\bvolume{506}(\bissue{2}),
\bfpage{703}--\blpage{710}
(\byear{2009}).
doi:\doiurl{10.1051/0004-6361/200912178}
\end{barticle}
\endbibitem

\bibitem[\protect\citeauthoryear{{Martin} et~al.}{2010}]{Martinea10}
\begin{barticle}
\bauthor{\binits{P.} \bsnm{{Martin}}},
\bauthor{\binits{J.} \bsnm{{Kn{\"o}dlseder}}},
\bauthor{\binits{G.} \bsnm{{Meynet}}}, \betal,
\batitle{{Predicted gamma-ray line emission from the Cygnus complex}}.
\bjtitle{\aap}
\bvolume{511},
\bfpage{86}
(\byear{2010}).
doi:\doiurl{10.1051/0004-6361/200912864}
\end{barticle}
\endbibitem

\bibitem[\protect\citeauthoryear{{Maurin} et~al.}{2016}]{Maurin2016}
\begin{barticle}
\bauthor{\binits{G.} \bsnm{{Maurin}}},
\bauthor{\binits{A.} \bsnm{{Marcowith}}},
\bauthor{\binits{N.} \bsnm{{Komin}}}, \betal,
\batitle{{Embedded star clusters as sources of high-energy cosmic rays .
  Modelling and constraints}}.
\bjtitle{\aap}
\bvolume{591},
\bfpage{71}
(\byear{2016}).
doi:\doiurl{10.1051/0004-6361/201628465}
\end{barticle}
\endbibitem

\bibitem[\protect\citeauthoryear{{Medina} et~al.}{2014}]{2014MNRAS.445.1797M}
\begin{barticle}
\bauthor{\binits{S.-N.X.} \bsnm{{Medina}}},
\bauthor{\binits{S.J.} \bsnm{{Arthur}}},
\bauthor{\binits{W.J.} \bsnm{{Henney}}}, \betal,
\batitle{{Turbulence in simulated H II regions}}.
\bjtitle{\mnras}
\bvolume{445}(\bissue{2}),
\bfpage{1797}--\blpage{1819}
(\byear{2014}).
doi:\doiurl{10.1093/mnras/stu1862}
\end{barticle}
\endbibitem

\bibitem[\protect\citeauthoryear{{Meynet} and
  {Maeder}}{2005}]{2005A&A...429..581M}
\begin{barticle}
\bauthor{\binits{G.} \bsnm{{Meynet}}},
\bauthor{\binits{A.} \bsnm{{Maeder}}},
\batitle{{Stellar evolution with rotation. XI. Wolf-Rayet star populations at
  different metallicities}}.
\bjtitle{\aap}
\bvolume{429},
\bfpage{581}--\blpage{598}
(\byear{2005}).
doi:\doiurl{10.1051/0004-6361:20047106}
\end{barticle}
\endbibitem

\bibitem[\protect\citeauthoryear{{Miville-Deschenes}
  et~al.}{1995}]{1995ApJ...454..316M}
\begin{barticle}
\bauthor{\binits{M.-A.} \bsnm{{Miville-Deschenes}}},
\bauthor{\binits{G.} \bsnm{{Joncas}}},
\bauthor{\binits{D.} \bsnm{{Durand }}},
\batitle{{The H II Region Sharpless 170: A Multiscale Analysis of the H alpha
  Velocity Field}}.
\bjtitle{\apj}
\bvolume{454},
\bfpage{316}
(\byear{1995}).
doi:\doiurl{10.1086/176484}
\end{barticle}
\endbibitem

\bibitem[\protect\citeauthoryear{{Molinari} et~al.}{2011}]{molinari11}
\begin{barticle}
\bauthor{\binits{S.} \bsnm{{Molinari}}},
\bauthor{\binits{J.} \bsnm{{Bally}}},
\bauthor{\binits{A.} \bsnm{{Noriega-Crespo}}}, \betal,
\batitle{{A 100 pc Elliptical and Twisted Ring of Cold and Dense Molecular
  Clouds Revealed by Herschel Around the Galactic Center}}.
\bjtitle{\apjl}
\bvolume{735},
\bfpage{33}
(\byear{2011}).
doi:\doiurl{10.1088/2041-8205/735/2/L33}
\end{barticle}
\endbibitem

\bibitem[\protect\citeauthoryear{{Monin} and
  {Yaglom}}{1971}]{1971sfmm.book.....M}
\begin{bbook}
\bauthor{\binits{A.S.} \bsnm{{Monin}}},
\bauthor{\binits{A.M.} \bsnm{{Yaglom}}},
\bbtitle{{Statistical fluid mechanics; mechanics of turbulence, The MIT Press}}
\byear{1971}
\end{bbook}
\endbibitem

\bibitem[\protect\citeauthoryear{{Morris} and {Serabyn}}{1996}]{morris96}
\begin{barticle}
\bauthor{\binits{M.} \bsnm{{Morris}}},
\bauthor{\binits{E.} \bsnm{{Serabyn}}},
\batitle{{The Galactic Center Environment}}.
\bjtitle{\araa}
\bvolume{34},
\bfpage{645}--\blpage{702}
(\byear{1996}).
doi:\doiurl{10.1146/annurev.astro.34.1.645}
\end{barticle}
\endbibitem

\bibitem[\protect\citeauthoryear{{Nandra} et~al.}{2013}]{Nandra2013}
\begin{botherref}
\oauthor{\binits{K.} \bsnm{{Nandra}}},
\oauthor{\binits{D.} \bsnm{{Barret}}},
\oauthor{\binits{X.} \bsnm{{Barcons}}}, et al.,
{The Hot and Energetic Universe: A White Paper presenting the science theme
  motivating the Athena+ mission}.
arXiv e-prints,
1306--2307
(2013)
\end{botherref}
\endbibitem

\bibitem[\protect\citeauthoryear{{Nava} and {Gabici}}{2013}]{Nava2013}
\begin{barticle}
\bauthor{\binits{L.} \bsnm{{Nava}}},
\bauthor{\binits{S.} \bsnm{{Gabici}}},
\batitle{{Anisotropic cosmic ray diffusion and gamma-ray production close to
  supernova remnants, with an application to W28}}.
\bjtitle{\mnras}
\bvolume{429}(\bissue{2}),
\bfpage{1643}--\blpage{1651}
(\byear{2013}).
doi:\doiurl{10.1093/mnras/sts450}
\end{barticle}
\endbibitem

\bibitem[\protect\citeauthoryear{{Nava} et~al.}{2016}]{Nava2016}
\begin{barticle}
\bauthor{\binits{L.} \bsnm{{Nava}}},
\bauthor{\binits{S.} \bsnm{{Gabici}}},
\bauthor{\binits{A.} \bsnm{{Marcowith}}}, \betal,
\batitle{{Non-linear diffusion of cosmic rays escaping from supernova remnants
  - I. The effect of neutrals}}.
\bjtitle{\mnras}
\bvolume{461}(\bissue{4}),
\bfpage{3552}--\blpage{3562}
(\byear{2016}).
doi:\doiurl{10.1093/mnras/stw1592}
\end{barticle}
\endbibitem

\bibitem[\protect\citeauthoryear{{Nava} et~al.}{2019}]{Nava2019}
\begin{barticle}
\bauthor{\binits{L.} \bsnm{{Nava}}},
\bauthor{\binits{S.} \bsnm{{Recchia}}},
\bauthor{\binits{S.} \bsnm{{Gabici}}}, \betal,
\batitle{{Non-linear diffusion of cosmic rays escaping from supernova remnants
  - II. Hot ionized media}}.
\bjtitle{\mnras}
\bvolume{484}(\bissue{2}),
\bfpage{2684}--\blpage{2691}
(\byear{2019}).
doi:\doiurl{10.1093/mnras/stz137}
\end{barticle}
\endbibitem

\bibitem[\protect\citeauthoryear{{Nogueras-Lara}
  et~al.}{2019}]{2019NatAs.tmp....4N}
\begin{botherref}
\oauthor{\binits{F.} \bsnm{{Nogueras-Lara}}},
\oauthor{\binits{R.} \bsnm{{Sch{\"o}del}}},
\oauthor{\binits{A.T.} \bsnm{{Gallego-Calvente}}}, et al.,
{Early formation and recent starburst activity in the nuclear disk of the Milky
  Way}.
Nature Astronomy,
4
(2019).
doi:\doiurl{10.1038/s41550-019-0967-9}
\end{botherref}
\endbibitem

\bibitem[\protect\citeauthoryear{{Oey}}{2007}]{2007IAUS..237..106O}
\begin{bchapter}
\bauthor{\binits{M.S.} \bsnm{{Oey}}},
\bctitle{{Towards resolving the evolution of multi-supernova superbubbles}},
in \bbtitle{Triggered Star Formation in a Turbulent ISM},
ed. by \beditor{\binits{B.G.} \bsnm{{Elmegreen}}},
\beditor{\binits{J.} \bsnm{{Palous}}}
\bsertitle{IAU Symposium},
vol. \bseriesno{237},
\byear{2007},
pp. \bfpage{106}--\blpage{113}.
doi:\doiurl{10.1017/S1743921307001305}
\end{bchapter}
\endbibitem

\bibitem[\protect\citeauthoryear{{Oey} and
  {Clarke}}{1997}]{1997MNRAS.289..570O}
\begin{barticle}
\bauthor{\binits{M.S.} \bsnm{{Oey}}},
\bauthor{\binits{C.J.} \bsnm{{Clarke}}},
\batitle{{The superbubble size distribution in the interstellar medium of
  galaxies.}}
\bjtitle{\mnras}
\bvolume{289},
\bfpage{570}--\blpage{588}
(\byear{1997}).
doi:\doiurl{10.1093/mnras/289.3.570}
\end{barticle}
\endbibitem

\bibitem[\protect\citeauthoryear{{Ohm}}{2016}]{ohm16}
\begin{barticle}
\bauthor{\binits{S.} \bsnm{{Ohm}}},
\batitle{{Starburst galaxies as seen by gamma-ray telescopes}}.
\bjtitle{C R Phys.}
\bvolume{17},
\bfpage{585}--\blpage{593}
(\byear{2016}).
doi:\doiurl{10.1016/j.crhy.2016.04.003}
\end{barticle}
\endbibitem

\bibitem[\protect\citeauthoryear{{Ohm} and {Hinton}}{2013}]{13OhmPWNe}
\begin{barticle}
\bauthor{\binits{S.} \bsnm{{Ohm}}},
\bauthor{\binits{J.A.} \bsnm{{Hinton}}},
\batitle{{Non-thermal emission from pulsar-wind nebulae in starburst
  galaxies.}}
\bjtitle{\mnras}
\bvolume{429},
\bfpage{70}--\blpage{74}
(\byear{2013}).
doi:\doiurl{10.1093/mnrasl/sls025}
\end{barticle}
\endbibitem

\bibitem[\protect\citeauthoryear{{Ohm} et~al.}{2013}]{Wd1Ohm13}
\begin{barticle}
\bauthor{\binits{S.} \bsnm{{Ohm}}},
\bauthor{\binits{J.A.} \bsnm{{Hinton}}},
\bauthor{\binits{R.} \bsnm{{White}}},
\batitle{{{\ensuremath{\gamma}}-ray emission from the Westerlund 1 region}}.
\bjtitle{\mnras}
\bvolume{434}(\bissue{3}),
\bfpage{2289}--\blpage{2294}
(\byear{2013}).
doi:\doiurl{10.1093/mnras/stt1170}
\end{barticle}
\endbibitem

\bibitem[\protect\citeauthoryear{{Olive} and {Schramm}}{1982}]{Olive1982}
\begin{barticle}
\bauthor{\binits{K.A.} \bsnm{{Olive}}},
\bauthor{\binits{D.N.} \bsnm{{Schramm}}},
\batitle{{OB associations and the nonuniversality of the cosmic abundances -
  Implications for cosmic rays and meteorites}}.
\bjtitle{\apj}
\bvolume{257},
\bfpage{276}--\blpage{282}
(\byear{1982}).
doi:\doiurl{10.1086/159986}
\end{barticle}
\endbibitem

\bibitem[\protect\citeauthoryear{{Owen} et~al.}{2019}]{2019A&A...626A..85O}
\begin{barticle}
\bauthor{\binits{E.R.} \bsnm{{Owen}}},
\bauthor{\binits{K.} \bsnm{{Wu}}},
\bauthor{\binits{X.} \bsnm{{Jin}}}, \betal,
\batitle{{Starburst and post-starburst high-redshift protogalaxies. The
  feedback impact of high energy cosmic rays}}.
\bjtitle{\aap}
\bvolume{626},
\bfpage{85}
(\byear{2019}).
doi:\doiurl{10.1051/0004-6361/201834350}
\end{barticle}
\endbibitem

\bibitem[\protect\citeauthoryear{{Padovani} et~al.}{2020}]{2020SSRv..216...29P}
\begin{barticle}
\bauthor{\binits{M.} \bsnm{{Padovani}}},
\bauthor{\binits{A.V.} \bsnm{{Ivlev}}},
\bauthor{\binits{D.} \bsnm{{Galli}}}, \betal,
\batitle{{Impact of Low-Energy Cosmic Rays on Star Formation}}.
\bjtitle{\ssr}
\bvolume{216}(\bissue{2}),
\bfpage{29}
(\byear{2020}).
doi:\doiurl{10.1007/s11214-020-00654-1}
\end{barticle}
\endbibitem

\bibitem[\protect\citeauthoryear{{Parizot} et~al.}{2004}]{Parizot2004}
\begin{barticle}
\bauthor{\binits{E.} \bsnm{{Parizot}}},
\bauthor{\binits{A.} \bsnm{{Marcowith}}},
\bauthor{\binits{E.} \bsnm{{van der Swaluw}}}, \betal,
\batitle{{Superbubbles and energetic particles in the Galaxy. I. Collective
  effects of particle acceleration}}.
\bjtitle{\aap}
\bvolume{424},
\bfpage{747}--\blpage{760}
(\byear{2004}).
doi:\doiurl{10.1051/0004-6361:20041269}
\end{barticle}
\endbibitem

\bibitem[\protect\citeauthoryear{{Pavlinsky}
  et~al.}{2015}]{2015SPIE.9603E..0CP}
\begin{bbook}
\bauthor{\binits{M.} \bsnm{{Pavlinsky}}},
\bauthor{\binits{V.} \bsnm{{Akimov}}},
\bauthor{\binits{V.} \bsnm{{Levin}}}, \betal,
\bctitle{{Status of ART-XC / SRG instrument}},
in \bbtitle{Society of Photo-Optical Instrumentation Engineers (SPIE)
  Conference Series},
vol. \bseriesno{9603}
\byear{2015},
p. \bfpage{96030}.
doi:\doiurl{10.1117/12.2190184}
\end{bbook}
\endbibitem

\bibitem[\protect\citeauthoryear{{Penny}}{1996}]{Penny1996}
\begin{barticle}
\bauthor{\binits{L.R.} \bsnm{{Penny}}},
\batitle{{Projected Rotational Velocities of O-Type Stars}}.
\bjtitle{\apj}
\bvolume{463},
\bfpage{737}
(\byear{1996}).
doi:\doiurl{10.1086/177286}
\end{barticle}
\endbibitem

\bibitem[\protect\citeauthoryear{{Peretti} et~al.}{2020}]{Peretti2020}
\begin{barticle}
\bauthor{\binits{E.} \bsnm{{Peretti}}},
\bauthor{\binits{P.} \bsnm{{Blasi}}},
\bauthor{\binits{F.} \bsnm{{Aharonian}}}, \betal,
\batitle{{Contribution of starburst nuclei to the diffuse gamma-ray and
  neutrino flux}}.
\bjtitle{\mnras}
(\byear{2020}).
doi:\doiurl{10.1093/mnras/staa698}
\end{barticle}
\endbibitem

\bibitem[\protect\citeauthoryear{{Pittard} and
  {Parkin}}{2010}]{2010MNRAS.403.1657P}
\begin{barticle}
\bauthor{\binits{J.M.} \bsnm{{Pittard}}},
\bauthor{\binits{E.R.} \bsnm{{Parkin}}},
\batitle{{3D models of radiatively driven colliding winds in massive O + O star
  binaries - III. Thermal X-ray emission}}.
\bjtitle{\mnras}
\bvolume{403}(\bissue{4}),
\bfpage{1657}--\blpage{1683}
(\byear{2010}).
doi:\doiurl{10.1111/j.1365-2966.2010.15776.x}
\end{barticle}
\endbibitem

\bibitem[\protect\citeauthoryear{{Pittard} et~al.}{2019}]{2019arXiv191205299P}
\begin{botherref}
\oauthor{\binits{J.M.} \bsnm{{Pittard}}},
\oauthor{\binits{G.S.} \bsnm{{Vila}}},
\oauthor{\binits{G.E.} \bsnm{{Romero}}},
{Diffusive shock acceleration and the resulting non-thermal emission from
  colliding-wind binary systems}.
arXiv e-prints,
1912--05299
(2019)
\end{botherref}
\endbibitem

\bibitem[\protect\citeauthoryear{{Prantzos}}{2012}]{Prantzos2012}
\begin{barticle}
\bauthor{\binits{N.} \bsnm{{Prantzos}}},
\batitle{{On the origin and composition of Galactic cosmic rays}}.
\bjtitle{\aap}
\bvolume{538},
\bfpage{80}
(\byear{2012}).
doi:\doiurl{10.1051/0004-6361/201117448}
\end{barticle}
\endbibitem

\bibitem[\protect\citeauthoryear{{Predehl} et~al.}{2016}]{2016SPIE.9905E..1KP}
\begin{bbook}
\bauthor{\binits{P.} \bsnm{{Predehl}}},
\bauthor{\binits{R.} \bsnm{{Andritschke}}},
\bauthor{\binits{V.} \bsnm{{Babyshkin}}}, \betal,
\bctitle{{eROSITA on SRG}},
in \bbtitle{Society of Photo-Optical Instrumentation Engineers (SPIE)
  Conference Series},
vol. \bseriesno{9905}
\byear{2016},
p. \bfpage{99051}.
doi:\doiurl{10.1117/12.2235092}
\end{bbook}
\endbibitem

\bibitem[\protect\citeauthoryear{{Pshirkov} et~al.}{2011}]{Pshirkov}
\begin{barticle}
\bauthor{\binits{M.S.} \bsnm{{Pshirkov}}},
\bauthor{\binits{P.G.} \bsnm{{Tinyakov}}},
\bauthor{\binits{P.P.} \bsnm{{Kronberg}}}, \betal,
\batitle{{Deriving the Global Structure of the Galactic Magnetic Field from
  Faraday Rotation Measures of Extragalactic Sources}}.
\bjtitle{\apj}
\bvolume{738},
\bfpage{192}
(\byear{2011}).
doi:\doiurl{10.1088/0004-637X/738/2/192}
\end{barticle}
\endbibitem

\bibitem[\protect\citeauthoryear{{Ptuskin} et~al.}{2008}]{Ptuskin2008}
\begin{barticle}
\bauthor{\binits{V.S.} \bsnm{{Ptuskin}}},
\bauthor{\binits{V.N.} \bsnm{{Zirakashvili}}},
\bauthor{\binits{A.A.} \bsnm{{Plesser}}},
\batitle{{Non-linear diffusion of cosmic rays}}.
\bjtitle{\asr}
\bvolume{42}(\bissue{3}),
\bfpage{486}--\blpage{490}
(\byear{2008}).
doi:\doiurl{10.1016/j.asr.2007.12.007}
\end{barticle}
\endbibitem

\bibitem[\protect\citeauthoryear{{Reeves}}{1978}]{Reeves1978}
\begin{bchapter}
\bauthor{\binits{H.} \bsnm{{Reeves}}},
\bctitle{{The 'big bang' theory of the origin of the solar system}},
in \bbtitle{IAU Colloq. 52: Protostars and Planets},
ed. by \beditor{\binits{T.} \bsnm{{Gehrels}}},
\beditor{\binits{M.S.} \bsnm{{Matthews}}},
\byear{1978},
pp. \bfpage{399}--\blpage{423}
\end{bchapter}
\endbibitem

\bibitem[\protect\citeauthoryear{{Reina-Campos} and
  {Kruijssen}}{2017}]{reinacampos17}
\begin{barticle}
\bauthor{\binits{M.} \bsnm{{Reina-Campos}}},
\bauthor{\binits{J.M.D.} \bsnm{{Kruijssen}}},
\batitle{{A unified model for the maximum mass scales of molecular clouds,
  stellar clusters and high-redshift clumps}}.
\bjtitle{\mnras}
\bvolume{469},
\bfpage{1282}--\blpage{1298}
(\byear{2017}).
doi:\doiurl{10.1093/mnras/stx790}
\end{barticle}
\endbibitem

\bibitem[\protect\citeauthoryear{{Romero}}{2019}]{2019RLSFN.tmp....3R}
\begin{botherref}
\oauthor{\binits{G.E.} \bsnm{{Romero}}},
{Gamma rays from colliding winds in massive binaries}.
Rend. Fis. Acc. Lincei,
3
(2019).
doi:\doiurl{10.1007/s12210-019-00763-2}
\end{botherref}
\endbibitem

\bibitem[\protect\citeauthoryear{{Romero} and
  {M{\"u}ller}}{2019}]{2019arXiv191207969R}
\begin{botherref}
\oauthor{\binits{G.E.} \bsnm{{Romero}}},
\oauthor{\binits{A.L.} \bsnm{{M{\"u}ller}}},
{Gamma Rays from Large-Scale Outflows in Starburst Galaxies}.
arXiv e-prints,
1912--07969
(2019)
\end{botherref}
\endbibitem

\bibitem[\protect\citeauthoryear{{Ruszkowski}
  et~al.}{2017}]{2017ApJ...834..208R}
\begin{barticle}
\bauthor{\binits{M.} \bsnm{{Ruszkowski}}},
\bauthor{\binits{H.-Y.K.} \bsnm{{Yang}}},
\bauthor{\binits{E.} \bsnm{{Zweibel}}},
\batitle{{Global Simulations of Galactic Winds Including Cosmic-ray
  Streaming}}.
\bjtitle{\apj}
\bvolume{834}(\bissue{2}),
\bfpage{208}
(\byear{2017}).
doi:\doiurl{10.3847/1538-4357/834/2/208}
\end{barticle}
\endbibitem

\bibitem[\protect\citeauthoryear{{Schlickeiser}}{2002}]{Schlickeiser02}
\begin{bbook}
\bauthor{\binits{R.} \bsnm{{Schlickeiser}}},
\bbtitle{{Cosmic Ray Astrophysics, Springer}}
\byear{2002}
\end{bbook}
\endbibitem

\bibitem[\protect\citeauthoryear{{Schure} and {Bell}}{2013}]{SB13}
\begin{barticle}
\bauthor{\binits{K.M.} \bsnm{{Schure}}},
\bauthor{\binits{A.R.} \bsnm{{Bell}}},
\batitle{{Cosmic ray acceleration in young supernova remnants}}.
\bjtitle{\mnras}
\bvolume{435},
\bfpage{1174}--\blpage{1185}
(\byear{2013}).
doi:\doiurl{10.1093/mnras/stt1371}
\end{barticle}
\endbibitem

\bibitem[\protect\citeauthoryear{{Schure} et~al.}{2012}]{SchureEtal2012}
\begin{barticle}
\bauthor{\binits{K.M.} \bsnm{{Schure}}},
\bauthor{\binits{A.R.} \bsnm{{Bell}}},
\bauthor{\binits{L.} \bsnm{{O'C Drury}}}, \betal,
\batitle{{Diffusive Shock Acceleration and Magnetic Field Amplification}}.
\bjtitle{\ssr}
\bvolume{173},
\bfpage{491}--\blpage{519}
(\byear{2012})
\end{barticle}
\endbibitem

\bibitem[\protect\citeauthoryear{{Seo} et~al.}{2018}]{Seo2018}
\begin{barticle}
\bauthor{\binits{J.} \bsnm{{Seo}}},
\bauthor{\binits{H.} \bsnm{{Kang}}},
\bauthor{\binits{D.} \bsnm{{Ryu}}},
\batitle{{The Contribution of Stellar Winds to Cosmic Ray Production}}.
\bjtitle{Journal of Korean Astronomical Society}
\bvolume{51},
\bfpage{37}--\blpage{48}
(\byear{2018}).
doi:\doiurl{10.5303/JKAS.2018.51.2.37}
\end{barticle}
\endbibitem

\bibitem[\protect\citeauthoryear{{Seta} and {Beck}}{2019}]{2019Galax...7...45S}
\begin{barticle}
\bauthor{\binits{A.} \bsnm{{Seta}}},
\bauthor{\binits{R.} \bsnm{{Beck}}},
\batitle{{Revisiting the Equipartition Assumption in Star-Forming Galaxies}}.
\bjtitle{Galaxies}
\bvolume{7}(\bissue{2}),
\bfpage{45}
(\byear{2019}).
doi:\doiurl{10.3390/galaxies7020045}
\end{barticle}
\endbibitem

\bibitem[\protect\citeauthoryear{{Shara} et~al.}{2017}]{Shara2017}
\begin{barticle}
\bauthor{\binits{M.M.} \bsnm{{Shara}}},
\bauthor{\binits{S.M.} \bsnm{{Crawford}}},
\bauthor{\binits{D.} \bsnm{{Vanbeveren}}}, \betal,
\batitle{{The spin rates of O stars in WR + O binaries - I. Motivation,
  methodology, and first results from SALT}}.
\bjtitle{\mnras}
\bvolume{464}(\bissue{2}),
\bfpage{2066}--\blpage{2074}
(\byear{2017}).
doi:\doiurl{10.1093/mnras/stw2450}
\end{barticle}
\endbibitem

\bibitem[\protect\citeauthoryear{{Siegert}}{2019}]{2019MmSAI..90..270S}
\begin{barticle}
\bauthor{\binits{T.} \bsnm{{Siegert}}},
\batitle{{INTEGRAL contributions to {\ensuremath{\gamma}}-ray line studies}}.
\bjtitle{\memsai}
\bvolume{90},
\bfpage{270}
(\byear{2019})
\end{barticle}
\endbibitem

\bibitem[\protect\citeauthoryear{{Sormani} et~al.}{2018}]{sormani18}
\begin{barticle}
\bauthor{\binits{M.C.} \bsnm{{Sormani}}},
\bauthor{\binits{R.G.} \bsnm{{Tre{\ss}}}},
\bauthor{\binits{M.} \bsnm{{Ridley}}}, \betal,
\batitle{{A theoretical explanation for the Central Molecular Zone asymmetry}}.
\bjtitle{\mnras}
\bvolume{475},
\bfpage{2383}--\blpage{2402}
(\byear{2018}).
doi:\doiurl{10.1093/mnras/stx3258}
\end{barticle}
\endbibitem

\bibitem[\protect\citeauthoryear{{Sormani} and {Barnes}}{2019}]{sormani19}
\begin{barticle}
\bauthor{\binits{M.C.} \bsnm{{Sormani}}},
\bauthor{\binits{A.T.} \bsnm{{Barnes}}},
\batitle{{Mass inflow rate into the Central Molecular Zone: observational
  determination and evidence of episodic accretion}}.
\bjtitle{\mnras}
\bvolume{484}(\bissue{1}),
\bfpage{1213}--\blpage{1219}
(\byear{2019}).
doi:\doiurl{10.1093/mnras/stz046}
\end{barticle}
\endbibitem

\bibitem[\protect\citeauthoryear{{Stolte} et~al.}{2014}]{stolte14}
\begin{barticle}
\bauthor{\binits{A.} \bsnm{{Stolte}}},
\bauthor{\binits{B.} \bsnm{{Hu{\ss}mann}}},
\bauthor{\binits{M.R.} \bsnm{{Morris}}}, \betal,
\batitle{{The Orbital Motion of the Quintuplet Cluster{\textemdash}A Common
  Origin for the Arches and Quintuplet Clusters?}}
\bjtitle{\apj}
\bvolume{789}(\bissue{2}),
\bfpage{115}
(\byear{2014}).
doi:\doiurl{10.1088/0004-637X/789/2/115}
\end{barticle}
\endbibitem

\bibitem[\protect\citeauthoryear{{Su} et~al.}{2010}]{su10}
\begin{barticle}
\bauthor{\binits{M.} \bsnm{{Su}}},
\bauthor{\binits{T.R.} \bsnm{{Slatyer}}},
\bauthor{\binits{D.P.} \bsnm{{Finkbeiner}}},
\batitle{{Giant Gamma-ray Bubbles from Fermi-LAT: Active Galactic Nucleus
  Activity or Bipolar Galactic Wind?}}
\bjtitle{\apj}
\bvolume{724},
\bfpage{1044}--\blpage{1082}
(\byear{2010}).
doi:\doiurl{10.1088/0004-637X/724/2/1044}
\end{barticle}
\endbibitem

\bibitem[\protect\citeauthoryear{{Tamborra} et~al.}{2014}]{2014JCAP...09..043T}
\begin{barticle}
\bauthor{\binits{I.} \bsnm{{Tamborra}}},
\bauthor{\binits{S.} \bsnm{{Ando}}},
\bauthor{\binits{K.} \bsnm{{Murase}}},
\batitle{{Star-forming galaxies as the origin of diffuse high-energy
  backgrounds: gamma-ray and neutrino connections, and implications for
  starburst history}}.
\bjtitle{\jcap}
\bvolume{2014}(\bissue{9}),
\bfpage{043}
(\byear{2014}).
doi:\doiurl{10.1088/1475-7516/2014/09/043}
\end{barticle}
\endbibitem

\bibitem[\protect\citeauthoryear{{Tashiro} et~al.}{2018}]{Tashiro2018}
\begin{bchapter}
\bauthor{\binits{M.} \bsnm{{Tashiro}}},
\bauthor{\binits{H.} \bsnm{{Maejima}}},
\bauthor{\binits{K.} \bsnm{{Toda}}}, \betal,
\bctitle{{Concept of the X-ray Astronomy Recovery Mission}},
in \bbtitle{\procspie}.
\bsertitle{Society of Photo-Optical Instrumentation Engineers (SPIE) Conference
  Series},
vol. \bseriesno{10699},
\byear{2018},
p. \bfpage{1069922}.
doi:\doiurl{10.1117/12.2309455}
\end{bchapter}
\endbibitem

\bibitem[\protect\citeauthoryear{{Tatischeff} and
  {Gabici}}{2018}]{2018ARNPS..68..377T}
\begin{barticle}
\bauthor{\binits{V.} \bsnm{{Tatischeff}}},
\bauthor{\binits{S.} \bsnm{{Gabici}}},
\batitle{{Particle Acceleration by Supernova Shocks and Spallogenic
  Nucleosynthesis of Light Elements}}.
\bjtitle{\arnps}
\bvolume{68}(\bissue{1}),
\bfpage{377}--\blpage{404}
(\byear{2018}).
doi:\doiurl{10.1146/annurev-nucl-101917-021151}
\end{barticle}
\endbibitem

\bibitem[\protect\citeauthoryear{{Telezhinsky} et~al.}{2012}]{Telezinsky2012}
\begin{barticle}
\bauthor{\binits{I.} \bsnm{{Telezhinsky}}},
\bauthor{\binits{V.V.} \bsnm{{Dwarkadas}}},
\bauthor{\binits{M.} \bsnm{{Pohl}}},
\batitle{{Particle spectra from acceleration at forward and reverse shocks of
  young Type Ia Supernova Remnants}}.
\bjtitle{\aspp}
\bvolume{35}(\bissue{6}),
\bfpage{300}--\blpage{311}
(\byear{2012}).
doi:\doiurl{10.1016/j.astropartphys.2011.10.001}
\end{barticle}
\endbibitem

\bibitem[\protect\citeauthoryear{{Tolksdorf}
  et~al.}{2019}]{2019ApJ...879...66T}
\begin{barticle}
\bauthor{\binits{T.} \bsnm{{Tolksdorf}}},
\bauthor{\binits{I.A.} \bsnm{{Grenier}}},
\bauthor{\binits{T.} \bsnm{{Joubaud}}}, \betal,
\batitle{{Cosmic Rays in Superbubbles}}.
\bjtitle{\apj}
\bvolume{879}(\bissue{2}),
\bfpage{66}
(\byear{2019}).
doi:\doiurl{10.3847/1538-4357/ab24c6}
\end{barticle}
\endbibitem

\bibitem[\protect\citeauthoryear{{Toptygin}}{1985}]{1985crim.book.....T}
\begin{bbook}
\bauthor{\binits{I.N.} \bsnm{{Toptygin}}},
\bbtitle{{Cosmic rays in interplanetary magnetic fields, North Holland}}
\byear{1985}
\end{bbook}
\endbibitem

\bibitem[\protect\citeauthoryear{{Truelove} and {McKee}}{1999}]{Truelove1999}
\begin{barticle}
\bauthor{\binits{J.K.} \bsnm{{Truelove}}},
\bauthor{\binits{C.F.} \bsnm{{McKee}}},
\batitle{{Evolution of Nonradiative Supernova Remnants}}.
\bjtitle{\apjs}
\bvolume{120}(\bissue{2}),
\bfpage{299}--\blpage{326}
(\byear{1999}).
doi:\doiurl{10.1086/313176}
\end{barticle}
\endbibitem

\bibitem[\protect\citeauthoryear{{Tversko{\v{i}}}}{1968}]{1968JETP...26..821T}
\begin{barticle}
\bauthor{\binits{B.A.} \bsnm{{Tversko{\v{i}}}}},
\batitle{{Theory of Turbulent Acceleration of Charged Particles in a Plasma}}.
\bjtitle{\jetp}
\bvolume{26},
\bfpage{821}
(\byear{1968})
\end{barticle}
\endbibitem

\bibitem[\protect\citeauthoryear{{VERITAS Collaboration}
  et~al.}{2009}]{2009VERITASM82}
\begin{barticle}
\bauthor{\bsnm{{VERITAS Collaboration}}},
\bauthor{\binits{V.A.} \bsnm{{Acciari}}},
\bauthor{\binits{E.} \bsnm{{Aliu}}}, \betal,
\batitle{{A connection between star formation activity and cosmic rays in the
  starburst galaxy M82}}.
\bjtitle{\nat}
\bvolume{462}(\bissue{7274}),
\bfpage{770}--\blpage{772}
(\byear{2009}).
doi:\doiurl{10.1038/nature08557}
\end{barticle}
\endbibitem

\bibitem[\protect\citeauthoryear{{Voss} et~al.}{2009}]{2009A&A...504..531V}
\begin{barticle}
\bauthor{\binits{R.} \bsnm{{Voss}}},
\bauthor{\binits{R.} \bsnm{{Diehl}}},
\bauthor{\binits{D.H.} \bsnm{{Hartmann}}}, \betal,
\batitle{{Using population synthesis of massive stars to study the interstellar
  medium near OB associations}}.
\bjtitle{\aap}
\bvolume{504}(\bissue{2}),
\bfpage{531}--\blpage{542}
(\byear{2009}).
doi:\doiurl{10.1051/0004-6361/200912260}
\end{barticle}
\endbibitem

\bibitem[\protect\citeauthoryear{{Voss} et~al.}{2010}]{2010A&A...520A..51V}
\begin{barticle}
\bauthor{\binits{R.} \bsnm{{Voss}}},
\bauthor{\binits{R.} \bsnm{{Diehl}}},
\bauthor{\binits{J.S.} \bsnm{{Vink}}}, \betal,
\batitle{{Probing the evolving massive star population in Orion with kinematic
  and radioactive tracers}}.
\bjtitle{\aap}
\bvolume{520},
\bfpage{51}
(\byear{2010}).
doi:\doiurl{10.1051/0004-6361/201014408}
\end{barticle}
\endbibitem

\bibitem[\protect\citeauthoryear{{Wang} et~al.}{2019}]{2019arXiv191207874W}
\begin{botherref}
\oauthor{\binits{W.} \bsnm{{Wang}}},
\oauthor{\binits{T.} \bsnm{{Siegert}}},
\oauthor{\binits{Z.G.} \bsnm{{Dai}}}, et al.,
{Gamma-ray Emission of 60Fe and 26Al Radioactivities in our Galaxy}.
arXiv e-prints,
1912--07874
(2019)
\end{botherref}
\endbibitem

\bibitem[\protect\citeauthoryear{{Ward} et~al.}{2019}]{ward19}
\begin{botherref}
\oauthor{\binits{J.L.} \bsnm{{Ward}}},
\oauthor{\binits{J.M.D.} \bsnm{{Kruijssen}}},
\oauthor{\binits{H.-W.} \bsnm{{Rix}}},
{Not all stars form in clusters -- $Gaia$-DR2 uncovers the origin of OB
  associations}.
\mnras~submitted,
1910--06974
(2019)
\end{botherref}
\endbibitem

\bibitem[\protect\citeauthoryear{{Waxman}}{2017}]{2017nacs.book...33W}
\begin{botherref}
\oauthor{\binits{E.} \bsnm{{Waxman}}},
{The Origin of IceCube's Neutrinos: Cosmic Ray Accelerators Embedded in Star
  Forming Calorimeters},
ed. by T. {Gaisser}, A. {Karle}
2017,
pp. 33--45.
\end{botherref}
\endbibitem

\bibitem[\protect\citeauthoryear{{Wiedenbeck} and
  {Greiner}}{1981}]{Wiedenbeck1981}
\begin{barticle}
\bauthor{\binits{M.E.} \bsnm{{Wiedenbeck}}},
\bauthor{\binits{D.E.} \bsnm{{Greiner}}},
\batitle{{Isotopic anomalies in the galactic cosmic-ray source}}.
\bjtitle{\prl}
\bvolume{46},
\bfpage{682}--\blpage{685}
(\byear{1981}).
doi:\doiurl{10.1103/PhysRevLett.46.682}
\end{barticle}
\endbibitem

\bibitem[\protect\citeauthoryear{{Woosley} and {Weaver}}{1981}]{Woosley1981}
\begin{barticle}
\bauthor{\binits{S.E.} \bsnm{{Woosley}}},
\bauthor{\binits{T.A.} \bsnm{{Weaver}}},
\batitle{{Anomalous isotopic composition of cosmic rays}}.
\bjtitle{\apj}
\bvolume{243},
\bfpage{651}--\blpage{659}
(\byear{1981}).
doi:\doiurl{10.1086/158631}
\end{barticle}
\endbibitem

\bibitem[\protect\citeauthoryear{{Xu} et~al.}{2016}]{Xu2016}
\begin{barticle}
\bauthor{\binits{S.} \bsnm{{Xu}}},
\bauthor{\binits{H.} \bsnm{{Yan}}},
\bauthor{\binits{A.} \bsnm{{Lazarian}}},
\batitle{{Damping of Magnetohydrodynamic Turbulence in Partially Ionized
  Plasma: Implications for Cosmic Ray Propagation}}.
\bjtitle{\apj}
\bvolume{826}(\bissue{2}),
\bfpage{166}
(\byear{2016}).
doi:\doiurl{10.3847/0004-637X/826/2/166}
\end{barticle}
\endbibitem

\bibitem[\protect\citeauthoryear{{Yang} et~al.}{2019}]{AharonianLincei2019}
\begin{botherref}
\oauthor{\binits{R.-z.} \bsnm{{Yang}}},
\oauthor{\binits{F.} \bsnm{{Aharonian}}},
\oauthor{\binits{E.} \bsnm{{de O{\~n}a Wilhelmi}}},
{Massive star clusters as the an alternative source population of galactic
  cosmic rays}.
Rend. Fis. Acc. Lincei,
34
(2019).
doi:\doiurl{10.1007/s12210-019-00819-3}
\end{botherref}
\endbibitem

\bibitem[\protect\citeauthoryear{{Yoast-Hull} et~al.}{2017}]{Yoast-Hull2017}
\begin{barticle}
\bauthor{\binits{T.M.} \bsnm{{Yoast-Hull}}},
\bauthor{\binits{J.S.} \bsnm{{Gallagher}}},
\bauthor{\binits{F.} \bsnm{{Halzen}}}, \betal,
\batitle{{Gamma-ray puzzle in Cygnus X: Implications for high-energy
  neutrinos}}.
\bjtitle{\prd}
\bvolume{96}(\bissue{4}),
\bfpage{043011}
(\byear{2017}).
doi:\doiurl{10.1103/PhysRevD.96.043011}
\end{barticle}
\endbibitem

\bibitem[\protect\citeauthoryear{{Yusef-Zadeh} and
  {Wardle}}{2019}]{yusefzadeh19}
\begin{barticle}
\bauthor{\binits{F.} \bsnm{{Yusef-Zadeh}}},
\bauthor{\binits{M.} \bsnm{{Wardle}}},
\batitle{{Cosmic-ray-driven outflow from the Galactic Centre and the origin of
  magnetized radio filaments}}.
\bjtitle{\mnras}
\bvolume{490}(\bissue{1}),
\bfpage{1}--\blpage{5}
(\byear{2019}).
doi:\doiurl{10.1093/mnrasl/slz134}
\end{barticle}
\endbibitem

\bibitem[\protect\citeauthoryear{{Zirakashvili} and {Ptuskin}}{2008}]{Zira08}
\begin{barticle}
\bauthor{\binits{V.N.} \bsnm{{Zirakashvili}}},
\bauthor{\binits{V.S.} \bsnm{{Ptuskin}}},
\batitle{{Diffusive Shock Acceleration with Magnetic Amplification by
  Nonresonant Streaming Instability in Supernova Remnants}}.
\bjtitle{\apj}
\bvolume{678}(\bissue{2}),
\bfpage{939}--\blpage{949}
(\byear{2008}).
doi:\doiurl{10.1086/529580}
\end{barticle}
\endbibitem

\end{thebibliography}

\end{document}